# Real Time Models of the Asynchronous Circuits: the Delay Theory

Serban E. Vlad

Oradea City Hall, P-ta Unirii, Nr. 1, 3700, Oradea, Romania
www.oradea.ro/serban/index.html, serban_e_vlad@yahoo.com

## Table of Contents

**List of Abbreviations**

| | | |
|---|---|---|
| SC | Stability condition | 6.1.1 |
| DC | Delay condition | 6.2.1 |
| $CC_{BDC}$ | Consistency condition (of the bounded delay condition) | 7.1.2 |
| BDC | Bounded delay condition | 7.2.2 |
| FDC | Fixed delay condition | 7.4.2 |
| AIC | Absolute inertial condition | 8.1.2 |
| AIDC | Absolute inertial delay condition | 8.2.1 |
| $CC_{BAIIDC}$ | Consistency condition (of the bounded absolute inertial delay condition) | 8.3.2 |
| BAIDC | Bounded absolute inertial delay condition | 8.4.1 |
| RIC | Relative inertial condition | 9.1.2 |
| RIDC | Relative inertial delay condition | 9.2.1 |
| $CC_{BRIDC}$ | Consistency condition (of the bounded relative inertial delay condition) | 9.3.2 |
| BRIDC | Bounded relative inertial delay condition | 9.4.1 |
| DBRIDC | Deterministic bounded relative inertial delay condition | 9.5.3 |
| BDC', AIC', RIC' | Variants of BDC, AIC, RIC | 10.1 |
| DBRIDC', SDBRIDC' | Variants of DBRIDC | 10.2.2 |

# 1. Introduction

Digital electrical engineering is a non-formalized theory and one of the major causes of this situation consists in the complexity of Mother Nature, things cannot be completely different from those in medicine, for example. We are too restricted to finding quick solutions to the problems that arise in order to take the time to strengthen a sound theoretical foundation of the reasoning that we do. Obviously, the political, military, economical and technological importance of digital electrical engineering is itself an obstacle in the spreading of consolidated theories.

      In fact, the reader of such literature can remark the existing distance from the deductive theories, the way that the mathematicians use them. We reproduce a point of view that we consider to be representative in this direction belonging to L. Rougier: '*Reasoning is deductive or is not at all*'.

The consequences of non-formalization are known. Many researchers do not give the right importance to the scientific language and words like *definition, theorem, proof* are titles of descriptive paragraphs rather than having the meaning that is accepted by the logicians. A fascinating job is, in this context, the translation is a precise mathematical language of what is intuitively, imprecisely explained with natural language by the engineers and this can be done in several ways. Our work has many such examples, let's just mention here the notion of inertia that is important and confusing at the same time. By reading with a ball-point pen in our hand, we infer that the inertia's inertia is not inertia, a paradox that should end the discussion on the validity of a theory. The theoretical construction continues however, without visible implications on the subsequent results, by using the methods of the non-deductive investigations.

The purpose of delay theory is that of writing systems of equations and inequalities with electrical signals, that model the behavior of the asynchronous circuits.

The (*electrical*) *signals* are the functions $f : \mathbf{R} \to \{0,1\}$ where $\mathbf{R}$, the set of the real numbers, is the time set. We ask that they:

- be constant in the interval $(-\infty, 0)$, with the variant that we have used elsewhere: be null in the interval $(-\infty, 0)$, in other words 0 is the initial time instant

- be constant on intervals $[t', t'')$ that are left closed and right open

- have a finite number of discontinuity points (i.e. a finite number of switches) in any bounded interval.

The *asynchronous circuits* (also called *asynchronous systems*, or *asynchronous automata* or *timed automata* in literature) are these electrical devices that can be modeled by using signals.

The fundamental (asynchronous) circuit in delay theory is the *delay circuit*, also called *delay buffer*, the circuit that computes the identity $1_{\{0,1\}}$ and the fundamental notion is that of *delay condition*, or shortly *delay*, the real time model of the delay circuit.

We show the way that the 'inertia's paradox' has been solved. First, the definition of the delays is given. Second, the pure delays are defined. Third, all the delays different from the pure delays are considered to be by definition inertial. Fourth, the serial connection of the delays is their composition, as binary relations. The serial connection of the inertial delays results in an inertial delay, but the type of inertia is likely to differ. The bounded delays have the nice property that, under the serial connection, the delays remain bounded and thus the type of inertia remains the same; the absolute inertial delays are in the same situation. The relative inertial delays are not closed under the serial connection, the 'paradox'.

We shall describe now, informally, the work of the delay circuit.

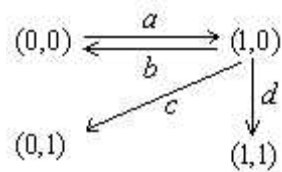

Fig 1

We have noted with $u : \mathbf{R} \to \{0,1\}$ its input and with $x : \mathbf{R} \to \{0,1\}$ its output. Both $u, x$ are signals. In Fig 1, the couples of binary numbers, temporarily called *states*, represent values $(u(t), x(t))$, with $t \in \mathbf{R}$ and $a, b, c, d$ are the *labels* (= the *names*) of the transitions $(u(t'), x(t')) \to (u(t''), x(t''))$. In such transitions, we suppose that $t' < t''$ and that $t'' - t'$ is a small infinitesimal. A suitable notation for this is $t' = t'' - 0$.

The real numbers $0 < d_{r,\min} \leq d_{r,\max}$ are given, the meaning of the index '$r$' being that of *raise* (switch from 0 to 1) of a signal, event symbolized by the validity of the equation
$$\overline{x(t-0)} \cdot x(t) = 1$$
Dually the real numbers $0 < d_{f,\min} \leq d_{f,\max}$ are given, the meaning of the index '$f$' being that of *fall* (switch from 1 to 0) of a signal and that event is symbolized by the validity of the equation
$$x(t-0) \cdot \overline{x(t)} = 1$$
We suppose that at the initial time instant $t_0 \geq 0$ the circuit is in the initial state $(0,0)$:
$$\forall \xi \in (-\infty, t_0), u(\xi) = 0$$
$$\forall \xi \in (-\infty, t_0], x(\xi) = 0$$
This state is *stable*, meaning that the delay circuit could remain indefinitely long there, if the input is 0:
$$\forall h > 0, \forall \xi \in [t_0, t_0 + h), u(\xi) = 0 \Rightarrow x(t_0 + h) = 0$$
A switch of the input takes place at $t_0$
$$\overline{u(t_0-0)} \cdot u(t_0) = 1$$
and the delay circuit follows the trajectory labeled $a$, i.e. $(0,0) \to (1,0)$. The hypothesis states that both the input and the output remain constant in the interval $[t_0, t_1)$
$$\forall \xi \in [t_0, t_1), u(\xi) = 1$$
$$\forall \xi \in [t_0, t_1), x(\xi) = 0$$
and the problem is to describe the behavior of the circuit at $t_1$. Three possibilities exist, those of running one the transitions $b, c, d$, depending on the values of $t_1$ and $u(t_1)$.

$b$:  it is necessarily run at $t_1$ if $t_1 - t_0 < d_{r,\min}$ and if $u$ switches from 1 to 0 at the time instant $t_1$
$$t_0 < t_1 < t_0 + d_{r,\min} \text{ and } u(t_1-0) \cdot \overline{u(t_1)} = 1 \text{ and } \overline{x(t_1-0)} \cdot x(t_1) = 0$$
The interpretation is that the circuit's inertia did not allow a fast switch of $x$ from 0 to 1 happen.

$b,c$: any of them is possible to be run at $t_1$ ($x(t_1) = 0$ for $b$ and $x(t_1) = 1$ for $c$) if $d_{r,\min} \leq t_1 - t_0 < d_{r,\max}$ and if $u$ switches from 1 to 0 at $t_1$
$$t_0 + d_{r,\min} \leq t_1 < t_0 + d_{r,\max} \text{ and } u(t_1-0) \cdot \overline{u(t_1)} = 1 \text{ and } \overline{x(t_1-0)} \cdot x(t_1) = 0$$
$$t_0 + d_{r,\min} \leq t_1 < t_0 + d_{r,\max} \text{ and } u(t_1-0) \cdot \overline{u(t_1)} = 1 \text{ and } \overline{x(t_1-0)} \cdot x(t_1) = 1$$
if $t_1 - t_0 = d_{r,\max}$ and if $u$ switches from 1 to 0 at $t_1$, then $c$ is necessary
$$t_1 = t_0 + d_{r,\max} \text{ and } u(t_1-0) \cdot \overline{u(t_1)} = 1 \text{ and } \overline{x(t_1-0)} \cdot x(t_1) = 1$$

$d$:  if $u(t_1) = 1$, then it is possible at $t_1$ for $d_{r,\min} \leq t_1 - t_0 < d_{r,\max}$ and it is necessary at $t_1$ for $t_1 - t_0 = d_{r,\max}$:
$$t_0 + d_{r,\min} \leq t_1 \leq t_0 + d_{r,\max} \text{ and } \overline{u(t_1-0) \cdot \overline{u(t_1)}} = 0 \text{ and } \overline{x(t_1-0)} \cdot x(t_1) = 1$$

The intuitive description of the circuit continues by asking that the dual statements hold also, as resulted by the replacement of '$r$', 0, 1 with '$f$', 1, 0.

The circuit computes the identity on $\{0,1\}$ because the states $(0,0)$, $(1,1)$ are stable and these are the only stable states of the circuit.

A possible manner of describing the previous facts is given by the system

$$\bigcap_{\xi \in [t-d_{r,\max},t)} u(\xi) \leq x(t) \leq \bigcup_{\xi \in [t-d_{f,\max},t)} u(\xi)$$

$$\overline{x(t-0)} \cdot x(t) \leq \bigcap_{\xi \in [t-d_{r,\min},t)} u(\xi)$$

$$x(t-0) \cdot \overline{x(t)} \leq \bigcap_{\xi \in [t-d_{f,\min},t)} \overline{u(\xi)}$$

and this might seem not quite obvious for the moment.

The chapter is organized in sections, numbered with 1, 2, 3, ... the sections have several paragraphs: 2.1, 3.2, ... and the paragraphs are usually organized in subparagraphs: 2.1.1, 4.5.2, ... The important equations and inequalities have been numbered, as well as all the figures and tables. The notation 3.2 (2) refers to equation or inequality (2) of paragraph 3.2 (that has no subparagraphs, in this case) and the notation 4.1.2 (3) refers to equation or inequality (3) of the subparagraph 2 from the paragraph 4.1.

In Section 2 we give several examples of models for the sake of creating intuition and this is a presentation of our intentions. The theory starting with section 3 is supposed to be self-contained.

In Section 3 we fix some fundamental concepts and notations on the $\boldsymbol{R} \to \{0,1\}$ functions.

Section 4 defines the signals and gives some useful properties on them.

In Section 5 we present the informal definitions of the delays, with long quotations from several authors.

The sections that follow represent the core of this chapter. In Section 6 we define the delays, as well as their determinism, order, time invariance, constancy, symmetry and serial connection. Section 7 is dedicated to the bounded delays and in Sections 8, 9 we define and characterize the absolute and the relative inertial conditions and delays. Section 10 shows some variants of the concepts from Sections 7, 8, 9 and introduces a special form of deterministic delays. Section 11 closes the chapter with new examples and a generalization.

We thank in advance to all those that will want to bring corrections and improvements to our results.

## 2. Motivating Examples

### 2.1 Example 1 The Delay Circuit

The symbol of the delay circuit is the next one

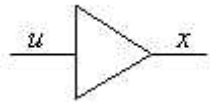

Fig 2

We consider different possibilities of modelation of this circuit, a way to anticipate the facts that will be presented later. $u, x$ are $\boldsymbol{R} \to \{0,1\}$ functions and moreover they are signals, with constant values for any $t < 0$.

**SC** *Stability*[1] (*unbounded delays*) If $u$ is of the form
$$u(t) = u(t) \cdot \chi_{(-\infty,t_0)}(t) \oplus u(t_0) \cdot \chi_{[t_0,\infty)}(t)$$
then $x$ is of the form
$$x(t) = x(t) \cdot \chi_{(-\infty,t_1)}(t) \oplus u(t_0) \cdot \chi_{[t_1,\infty)}(t)$$
where $t_0 \geq 0, t_1 \geq 0$ and $\chi_{(\ )} : \mathbf{R} \to \{0,1\}$ is the *characteristic function*.

**BDC'** *Upper bounded, lower unbounded delays* $d_r > 0, d_f > 0$ exist so that the next system is satisfied:
$$\bigcap_{\xi \in [t-d_r,t)} u(\xi) \leq x(t) \leq \bigcup_{\xi \in [t-d_f,t)} u(\xi)$$

**BDC** *Bounded delays* $0 \leq m_r \leq d_r, 0 \leq m_f \leq d_f$ exist and the system is the next one
$$\bigcap_{\xi \in [t-d_r,t-d_r+m_r]} u(\xi) \leq x(t) \leq \bigcup_{\xi \in [t-d_f,t-d_f+m_f]} u(\xi)$$

**FDC** *Fixed delays* (*ideal delays*) The relation between $u$ and $x$ is, for $d \geq 0$
$$x(t) = u(t-d)$$

**AIC** *Absolute inertia* $\delta_r \geq 0, \delta_f \geq 0$ exist so that $x$ satisfies
$$\overline{x(t-0)} \cdot x(t) \leq \bigcap_{\xi \in [t,t+\delta_r]} x(\xi)$$
$$x(t-0) \cdot \overline{x(t)} \leq \bigcap_{\xi \in [t,t+\delta_f]} \overline{x(\xi)}$$
This inertia condition is added to one of SC, BDC', BDC, FDC.

**RIC** *Relative inertia* $0 \leq \mu_r \leq \delta_r, 0 \leq \mu_f \leq \delta_f$ are given so that
$$\overline{x(t-0)} \cdot x(t) \leq \bigcap_{\xi \in [t-\delta_r,t-\delta_r+\mu_r]} u(\xi)$$
$$x(t-0) \cdot \overline{x(t)} \leq \bigcap_{\xi \in [t-\delta_f,t-\delta_f+\mu_f]} \overline{u(\xi)}$$
are satisfied. Similarly with absolute inertia, relative inertia is a request added to one of SC, BDC', BDC, FDC.

**DBRIDC** *Deterministic bounded relative inertial delays* If in BDC+RIC $\mu_r, \delta_r, \mu_f, \delta_f$ coincide with $m_r, d_r, m_f, d_f$ the system takes the special deterministic form
$$\overline{x(t-0)} \cdot x(t) = \overline{x(t-0)} \cdot \bigcap_{\xi \in [t-d_r,t-d_r+m_r]} u(\xi)$$
$$x(t-0) \cdot \overline{x(t)} = x(t-0) \cdot \bigcap_{\xi \in [t-d_f,t-d_f+m_f]} \overline{u(\xi)}$$

**SDBRIDC'** *Symmetrical deterministic upper bounded, lower unbounded relative inertial delays*, version of DBRIDC consisting in the next equation

---
[1] In the abbreviations that we use: SC, BDC,… the letter 'C' comes from 'condition': stability (condition), bounded delay (condition),…

$$Dx(t) = (x(t-0) \oplus u(t-0)) \cdot \overline{\bigcup_{\xi \in (t-d,t)} Du(\xi)} \cdot \chi_{[d,\infty)}(t)$$

where

$$Dx(t) = \overline{x(t-0)} \cdot x(t) \cup x(t-0) \cdot \overline{x(t)} = x(t-0) \oplus x(t)$$

is the left derivative of $x$.

All the solutions of BDC', BDC, FDC, DBRIDC, SDBRIDC' satisfy $x(0-0) = u(0-0)$ and some of the previous systems satisfy also supplementary *conditions of consistency* (i.e. the existence of a solution $x$ for any $u$).

## 2.2 Example 2 Circuit with Feedback Using a Delay Circuit

In the circuit from 2.1 Fig 2 we suppose that $u = x$ and this corresponds to the next circuit

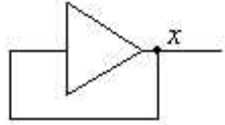

Fig 3

**SC** The satisfaction of SC does not bring any information on $x$, as it consists in a tautology of the form $\neg A \vee A$, where the proposition $A$ is the equation

$$\exists t_0 \geq 0, x(t) = x(t) \cdot \chi_{(-\infty, t_0)}(t) \oplus x(t_0) \cdot \chi_{[t_0, \infty)}(t)$$

Interpretation: the circuit can be stable or unstable.

**BDC'** The system is

$$\bigcap_{\xi \in [t-d_r, t)} x(\xi) \leq x(t) \leq \bigcup_{\xi \in [t-d_f, t)} x(\xi)$$

with $d_r > 0, d_f > 0$. Let $t_0 \geq 0$ so that $\forall t < t_0, x(t) = 0$. Because $\bigcup_{\xi \in [t_0 - d_f, t_0)} x(\xi) = 0$, we get $x(t_0) = 0$. Similarly, let $t_0 \geq 0$ so that $\forall t < t_0, x(t) = 1$. Because $\bigcap_{\xi \in [t_0 - d_r, t_0)} x(\xi) = 1$, we obtain $x(t_0) = 1$. $t_0$ was arbitrary previously, so that the only solutions of BDC' are the constant functions.

On the other hand, the constant functions satisfy any supplementary inertial condition AIC, RIC because $\overline{x(t-0)} \cdot x(t) = x(t-0) \cdot \overline{x(t)} = 0$.

**BDC** We have the system

$$\bigcap_{\xi \in [t-d_r, t-d_r + m_r]} x(\xi) \leq x(t) \leq \bigcup_{\xi \in [t-d_f, t-d_f + m_f]} x(\xi)$$

where $0 \leq m_r \leq d_r, 0 \leq m_f \leq d_f$. Let us suppose in the beginning, when solving it, that $x(0-0) = 0$. The solutions are the next ones.

<u>Case</u> $d_f - m_f > 0$

We can show analogously with BDC' that the only solution is $x(t) = 0$.

<u>Case</u> $d_f - m_f = 0$, $d_r > 0$

As the inequality $x(t) \leq \bigcup_{\xi \in [t-d_f, t]} x(\xi)$ is satisfied by all $x$, BDC has in this case the same

solutions like (in other words: is equivalent with)

$$\bigcap_{\xi \in [t-d_r, t-d_r+m_r]} x(\xi) \leq x(t) \tag{1}$$

and any solution can be written under one of the forms

$$x(t) = 0 \tag{2}$$

$$x(t) = \chi_{[t_0, \infty)}(t) \tag{3}$$

$$x(t) = \chi_{[t_0, t_1)}(t) \oplus \chi_{[t_2, t_3)}(t) \oplus \ldots \oplus \chi_{[t_{2n}, t_{2n+1})}(t) \tag{4}$$

$$x(t) = \chi_{[t_0, t_1)}(t) \oplus \chi_{[t_2, t_3)}(t) \oplus \ldots \oplus \chi_{[t_{2n}, t_{2n+1})}(t) \oplus \chi_{[t_{2n+2}, \infty)}(t) \tag{5}$$

$$x(t) = \chi_{[t_0, t_1)}(t) \oplus \chi_{[t_2, t_3)}(t) \oplus \ldots \oplus \chi_{[t_{2n}, t_{2n+1})}(t) \oplus \ldots \tag{6}$$

(2),…,(6) represent all the signals $x$ with $x(0-0) = 0$, where $0 \leq t_0 < t_1 < t_2 < \ldots$ is unbounded, arbitrary. (2), (3) satisfy (1) without supplementary requests. Because if the term $\chi_{[t_{2k}, t_{2k+1})}$ satisfies $t_{2k+1} - t_{2k} > m_r$ we have that

$$\bigcap_{\xi \in [t-d_r, t-d_r+m_r]} \chi_{[t_{2k}, t_{2k+1})}(\xi) = \chi_{[t_{2k}+d_r, t_{2k+1}+d_r-m_r)}(t)$$

is not null, in order that (4),…,(6) be solutions of (1), the next property should be true for all $k \geq 0$:

$$t_{2k+1} - t_{2k} > m_r \Rightarrow [t_{2k} + d_r, t_{2k+1} + d_r - m_r) \subset supp\ x$$

We have noted $supp\ x = \{t \mid x(t) = 1\}$ the support set of $x$.

A special case of (1) is the one when $m_r = 0$:

$$x(t - d_r) \leq x(t) \tag{7}$$

and then for all $k \geq 0$ the next inclusion

$$[t_{2k} + d_r, t_{2k+1} + d_r) \subset supp\ x$$

is fulfilled. For example, the 'periodical' functions

$$x(t) = \chi_{[t_0, t_1)}(t) \oplus \chi_{[t_0+d_r, t_1+d_r)}(t) \oplus \ldots \oplus \chi_{[t_0+n \cdot d_r, t_1+n \cdot d_r)}(t) \oplus \ldots$$

where $0 \leq t_0 < t_1 \leq t_0 + d_r$ satisfy (7) because

$$x(t - d_r) = \chi_{[t_0+d_r, t_1+d_r)}(t) \oplus \chi_{[t_0+2d_r, t_1+2d_r)}(t) \oplus \ldots \oplus \chi_{[t_0+(n+1) \cdot d_r, t_1+(n+1) \cdot d_r)}(t) \oplus \ldots$$

An interesting situation in BDC+AIC is the special case $\delta_r \geq m_r, \delta_f = 0$ when the inclusion

$$[t_{2k} + d_r, t_{2k+1} + d_r - m_r) \subset supp\ x$$

is true for all $k \geq 0$ and all solutions $x$, the hypothesis $t_{2k+1} - t_{2k} > \delta_r \geq m_r$ being satisfied due to AIC.

Adding RIC in the case $d_f - m_f = 0, d_r > 0$ of BDC, under the form

$$\overline{x(t-0)} \cdot x(t) \leq \bigcap_{\xi \in [t-\delta_r, t-\delta_r+\mu_r]} x(\xi) \tag{8}$$

$$x(t-0) \cdot \overline{x(t)} \leq \bigcap_{\xi \in [t-\delta_f, t-\delta_f+\mu_f]} \overline{x(\xi)} \tag{9}$$

implies if $\delta_r > 0$ that $x(t) = 0$. For $\delta_r = 0$, inequality (8) becomes trivial: $\overline{x(t-0)} \cdot x(t) \leq x(t)$ and then, if $\delta_f > 0$, the restrictions corresponding to RIC on the solutions $x$ of BDC are expressed under the form, see (4),…,(6)

$$\chi_{\{t_1, t_3, \ldots\}}(t) = x(t-0) \cdot \overline{x(t)} \leq$$

$$\leq \bigcap_{\xi \in [t-\delta_f, t-\delta_f+\mu_f]} \overline{x(\xi)} = \chi_{(-\infty, t_0+\delta_f-\mu_f) \vee [t_1+\delta_f, t_2+\delta_f-\mu_f) \vee [t_3+\delta_f, t_4+\delta_f-\mu_f) \vee \ldots}(t) \quad (\text{t})$$

i.e. equivalently
$$\{t_1, t_3, \ldots\} \subset (-\infty, t_0+\delta_f-\mu_f) \vee [t_1+\delta_f, t_2+\delta_f-\mu_f) \vee [t_3+\delta_f, t_4+\delta_f-\mu_f) \vee \ldots$$

$\delta_r = \delta_f = 0$ means triviality for RIC.

<u>Case</u> $d_f - m_f = 0, d_r = 0$

BDC consists in
$$x(t) \leq x(t) \leq \bigcup_{\xi \in [t-d_f, t]} x(\xi)$$

and all the signals $x$ satisfy it.

By duality, the possibility $x(0-0) = 1$ is analyzed. We observe for example that if $d_f - m_f > 0, d_r - m_r > 0$ then the only solutions of BDC are the constant functions.

**FDC** The equation to be solved is
$$x(t) = x(t-d), d \geq 0$$
If $d > 0$, then the solutions are the two constant functions and if $d = 0$ then the solutions are all the signals.

**DBRIDC** The system is
$$\overline{x(t-0)} \cdot x(t) = \overline{x(t-0)} \cdot \bigcap_{\xi \in [t-d_r, t-d_r+m_r]} x(\xi) \quad (10)$$

$$x(t-0) \cdot \overline{x(t)} = x(t-0) \cdot \bigcap_{\xi \in [t-d_f, t-d_f+m_f]} \overline{x(\xi)} \quad (11)$$

and we suppose like before that $x(0-0) = 0$.

<u>Case</u> $d_r > 0$

The only solution is $x(t) = 0$.

<u>Case</u> $d_r = 0, d_f = m_f > 0$

The switch from 0 to 1 is possible, because (10) takes the trivial form $\overline{x(t-0)} \cdot x(t) = \overline{x(t-0)} \cdot x(t)$. From this moment $\bigcap_{\xi \in [t-d_f, t]} \overline{x(\xi)}$ is null, thus the solutions have one of the forms (2), (3).

<u>Case</u> $d_r = 0, d_f = m_f = 0$

All the signals $x$ satisfy the system, (10), (11) being both trivial.

<u>Case</u> $d_r = 0, d_f > m_f \geq 0$

The switch from 0 to 1 seems possible and let $t_0$ be the moment of the first such switch, thus $\overline{x(t_0 - 0)} \cdot x(t_0) = 1$. At the time instant $t_1 > t_0$ characterized by $[t_0, t_1) \subset supp\ x$, (11) becomes
$$\overline{x(t_1)} = \bigcap_{\xi \in [t_1-d_f, t_1-d_f+m_f]} \overline{x(\xi)} \quad (12)$$

For all $t_1 - d_f + m_f < t_0$, i.e. if $0 < t_1 - t_0 < d_f - m_f$, the right member of (12) is 1 and the switch of $x$ from 1 to 0 necessary. We have reached a contradiction showing that DBRIDC has no solution $x(t) \neq 0$.

The analysis of the situation when $x(0-0)=1$ is similar.

**SDBRIDC'** The solutions of the equation
$$Dx(t)=0$$
are the constant functions.

## 2.3 The Logical Gate NOT

The logical gate NOT that computes the complement in the set $\{0,1\}$ is symbolized like in the next figure

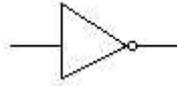

Fig 4

where the gate and the two wires are characterized by delays. It is modeled by one of the next circuits

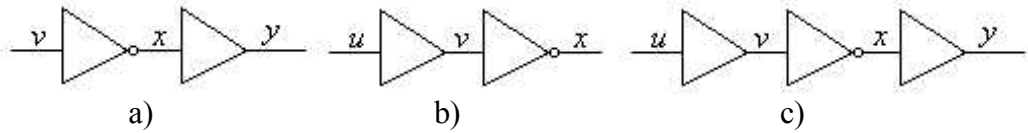

a)          b)          c)

Fig 5

In Fig 5 the logical gate is ideal
$$x(t) = x(0-0) \cdot \chi_{(-\infty,0)}(t) \oplus \overline{v(t)} \cdot \chi_{[0,\infty)}(t) \tag{1}$$
as well as the wires and the delays are localized in the delay circuits. Writing the relations between $u,v$, respectively between $x,y$ follows, like at 2.1. The last step is the elimination (if possible) of the intermediary variables: $x$ at a), $v$ at b), $v$ and $x$ at c). We give some examples.

**SC** Fig 5 c)

The fact that $u$ is of the form
$$u(t) = u(t) \cdot \chi_{(-\infty,t_0)}(t) \oplus u(t_0) \cdot \chi_{[t_0,\infty)}(t) \tag{2}$$
implies that $v$ is of the form
$$v(t) = v(t) \cdot \chi_{(-\infty,t_1)}(t) \oplus u(t_0) \cdot \chi_{[t_1,\infty)}(t) \tag{3}$$
thus, from (1), $x$ is given by
$$x(t) = x(t) \cdot \chi_{(-\infty,t_1)}(t) \oplus \overline{u(t_0)} \cdot \chi_{[t_1,\infty)}(t) \tag{4}$$
and by using SC again for the second delay circuit we get
$$y(t) = y(t) \cdot \chi_{(-\infty,t_2)}(t) \oplus \overline{u(t_0)} \cdot \chi_{[t_2,\infty)}(t) \tag{5}$$
In (2),…,(5) $t_0 \geq 0, t_1 \geq 0, t_2 \geq 0$.

**BDC'** Fig 5 a)
$$\bigcap_{\xi \in [t-d_r,t)} x(\xi) \leq y(t) \leq \bigcup_{\xi \in [t-d_f,t)} x(\xi) \tag{6}$$

$$\bigcap_{\xi \in [t-d_r,t)} (x(0-0) \cdot \chi_{(-\infty,0)}(\xi) \oplus \overline{v(\xi)} \cdot \chi_{[0,\infty)}(\xi)) \leq y(t) \leq$$

$$\leq \bigcup_{\xi \in [t-d_f, t]} (x(0-0) \cdot \chi_{(-\infty,0)}(\xi) \oplus \overline{v(\xi)} \cdot \chi_{[0,\infty)}(\xi)) \qquad \text{(from (1), (6))} \qquad (7)$$

**BDC'** Fig 5 b)

$$\bigcap_{\xi \in [t-d_r, t]} u(\xi) \leq v(t) \leq \bigcup_{\xi \in [t-d_f, t]} u(\xi) \qquad (8)$$

$$\bigcap_{\xi \in [t-d_f, t]} \overline{u(\xi)} \leq \overline{v(t)} \leq \bigcup_{\xi \in [t-d_r, t]} \overline{u(\xi)} \qquad \text{(from (8))} \qquad (9)$$

$$x(0-0) \cdot \chi_{(-\infty,0)}(t) \oplus \bigcap_{\xi \in [t-d_f, t]} \overline{u(\xi)} \cdot \chi_{[0,\infty)}(t) \leq x(t) \leq$$

$$\leq x(0-0) \cdot \chi_{(-\infty,0)}(t) \oplus \bigcup_{\xi \in [t-d_r, t]} \overline{u(\xi)} \cdot \chi_{[0,\infty)}(t) \qquad \text{(from (1), (9))} \qquad (10)$$

**SDBRIDC'** Fig 5a) Some of the next equations are better understood if we take into account the fact that $\forall a \in \{0,1\}, \overline{a} = a \oplus 1$:

$$x(t-0) = x(0-0) \cdot \chi_{(-\infty,0]}(t) \oplus \overline{v(t-0)} \cdot \chi_{(0,\infty)}(t) \qquad \text{(from (1))} \qquad (11)$$

$$v(t-0) = v(0-0) \cdot \chi_{(-\infty,0]}(t) \oplus v(t-0) \cdot \chi_{(0,\infty)}(t) \qquad (12)$$

$$\overline{v(t-0)} = \overline{v(0-0)} \cdot \chi_{(-\infty,0]}(t) \oplus \overline{v(t-0)} \cdot \chi_{(0,\infty)}(t) \qquad \text{(from (12))} \qquad (13)$$

$$x(t-0) = (x(0-0) \oplus \overline{v(0-0)}) \cdot \chi_{(-\infty,0]}(t) \oplus \overline{v(t-0)} \qquad \text{(from (11),(13))} \qquad (14)$$

$$Dx(t) = x(0-0) \cdot \chi_{(-\infty,0]}(t) \oplus x(0-0) \cdot \chi_{(-\infty,0)}(t) \oplus$$

$$\oplus \overline{v(t-0)} \cdot \chi_{(0,\infty)}(t) \oplus \overline{v(t)} \cdot \chi_{[0,\infty)}(t) \qquad \text{(from (1),(11))}$$

$$= x(0-0) \cdot \chi_{\{0\}}(t) \oplus \overline{v(t-0)} \cdot \chi_{(0,\infty)}(t) \oplus \overline{v(0)} \cdot \chi_{\{0\}}(t) \oplus \overline{v(t)} \cdot \chi_{(0,\infty)}(t)$$

$$= (x(0-0) \oplus \overline{v(0)}) \cdot \chi_{\{0\}}(t) \oplus (\overline{v(t-0)} \oplus \overline{v(t)}) \cdot \chi_{(0,\infty)}(t)$$

$$= (x(0-0) \oplus \overline{v(0)}) \cdot \chi_{\{0\}}(t) \oplus Dv(t) \cdot \chi_{(0,\infty)}(t) \qquad (15)$$

$$Dv(t) = (v(0-0) \oplus v(0)) \cdot \chi_{\{0\}}(t) \oplus Dv(t) \cdot \chi_{(0,\infty)}(t) \qquad (16)$$

$$Dx(t) = (x(0-0) \oplus \overline{v(0)} \oplus v(0-0) \oplus v(0)) \cdot \chi_{\{0\}}(t) \oplus Dv(t) \qquad \text{(from (15),(16))}$$

$$= \overline{x(0-0) \oplus v(0-0)} \cdot \chi_{\{0\}}(t) \oplus Dv(t) \qquad (17)$$

$$Dy(t) = (y(t-0) \oplus x(t-0)) \cdot \overline{\bigcup_{\xi \in (t-d,t)} Dx(\xi)} \cdot \chi_{[d,\infty)}(t) \qquad \text{(the hypothesis SDBRIDC')}$$

$$= (y(t-0) \oplus x(0-0) \cdot \chi_{(-\infty,0]}(t) \oplus \overline{v(t-0)} \cdot \chi_{(0,\infty)}(t)) \cdot$$

$$\cdot \overline{\bigcup_{\xi \in (t-d,t)} (\overline{x(0-0) \oplus v(0-0)} \cdot \chi_{\{0\}}(\xi) \oplus Dv(\xi))} \cdot \chi_{[d,\infty)}(t) \qquad \text{(from (11),(17))}$$

$$= (y(t-0) \oplus \overline{v(t-0)}) \cdot \chi_{(0,\infty)}(t) \cdot \qquad (y(0-0) = x(0-0))$$

$$\cdot \overline{\bigcup_{\xi \in (t-d,t)} (\overline{x(0-0) \oplus v(0-0)} \cdot \chi_{\{0\}}(\xi) \oplus Dv(\xi))} \cdot \chi_{[d,\infty)}(t)$$

$$= y(t-0) \oplus \overline{v(t-0)} \cdot \overline{\bigcup_{\xi \in (t-d,t)} Dv(\xi)} \cdot \chi_{[d,\infty)}(t) \qquad (18)$$

**SDBRIDC'** Fig 5b)

$$Dv(t) = (\overline{v(t-0) \oplus u(t-0)}) \cdot \overline{\bigcup_{\xi \in (t-d,t)} Du(\xi)} \cdot \chi_{[d,\infty)}(t) \quad \text{(the hypothesis SDBRIDC')} \quad (19)$$

$$v(t-0) = \overline{x(t-0) \oplus \overline{x(0-0) \oplus v(0-0)} \cdot \chi_{(-\infty,0]}(t)} \quad \text{(from (14))} \quad (20)$$

$$u(t-0) = u(0-0) \cdot \chi_{(-\infty,0]}(t) \oplus u(t-0) \cdot \chi_{(0,\infty)}(t) \quad (21)$$

$$Dx(t) = (\overline{x(t-0) \oplus \overline{x(0-0) \oplus v(0-0)} \cdot \chi_{(-\infty,0]}(t) \oplus u(t-0)}) \cdot \overline{\bigcup_{\xi \in (t-d,t)} Du(\xi)} \cdot \chi_{[d,\infty)}(t)$$

$$\oplus \overline{x(0-0) \oplus v(0-0)} \cdot \chi_{\{0\}}(t) \quad \text{(from (17),(19),(20))}$$

$$= (\overline{x(0-0)} \cdot \chi_{(-\infty,0]}(t) \oplus \overline{x(t-0)} \cdot \chi_{(0,\infty)}(t) \oplus$$

$$\oplus \overline{(x(0-0) \oplus v(0-0))} \cdot \chi_{(-\infty,0]}(t) \oplus u(0-0) \cdot \chi_{(-\infty,0]}(t) \oplus u(t-0) \cdot \chi_{(0,\infty)}(t)) \cdot$$

$$\cdot \overline{\bigcup_{\xi \in (t-d,t)} Du(\xi)} \cdot \chi_{[d,\infty)}(t) \oplus \overline{x(0-0) \oplus v(0-0)} \cdot \chi_{\{0\}}(t) \quad \text{(from (21)}$$

$$= (\overline{x(t-0) \oplus u(t-0)}) \cdot \chi_{(0,\infty)}(t) \cdot \overline{\bigcup_{\xi \in (t-d,t)} Du(\xi)} \cdot \chi_{[d,\infty)}(t) \oplus \overline{x(0-0) \oplus u(0-0)} \cdot \chi_{\{0\}}(t)$$

$$(u(0-0) = v(0-0))$$

$$= \overline{x(t-0) \oplus u(t-0)} \cdot \overline{\bigcup_{\xi \in (t-d,t)} Du(\xi)} \cdot \chi_{[d,\infty)}(t) \oplus \overline{x(0-0) \oplus u(0-0)} \cdot \chi_{\{0\}}(t) \quad (22)$$

A comparison between the forms of (18) and (22) is interesting.

## 2.4 Circuit With Feedback Using a Logical Gate NOT

Let the next circuit, where the logical gate and the wires have delays

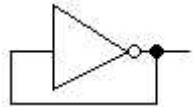

Fig 6

and the circuits to be analyzed are the next ones

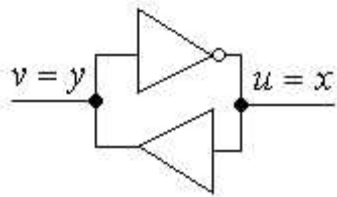

a)

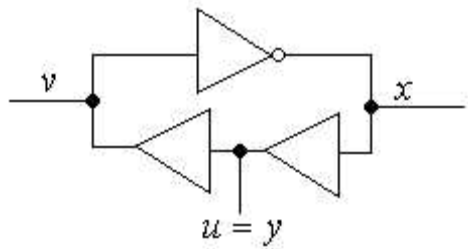

b)

Fig 7

Fig 7 a), respectively Fig 7 b) result from and keep the notations of Fig 5 a), b) (the modeling coincides), respectively of Fig 5 c). In Fig 7 the logical gate is supposed to be ideal

$$x(t) = x(0-0) \cdot \chi_{(-\infty,0)}(t) \oplus \overline{v(t)} \cdot \chi_{[0,\infty)}(t) \tag{1}$$

as well as the wires and the delays are localized in the delay circuits. We shall write the relations between $u, v, x, y$ and we shall try to eliminate three of the four variables.

**SC** Fig 7 a)

We suppose that $x$ is of the form

$$x(t) = x(t) \cdot \chi_{(-\infty,t_0)}(t) \oplus x(t_0) \cdot \chi_{[t_0,\infty)}(t) \tag{2}$$

and from SC this implies that $v$ is of the form

$$v(t) = v(t) \cdot \chi_{(-\infty,t_1)}(t) \oplus v(t_1) \cdot \chi_{[t_1,\infty)}(t) \tag{3}$$

where

$$x(t_0) = v(t_1) \tag{4}$$

and $t_0 \geq 0, t_1 \geq 0$. On the other hand (1), (2) and (3) show that

$$x(t_0) = \overline{v(t_1)} \tag{5}$$

(4) and (5) are contradictory, meaning the falsity of the hypothesis (2). A major difference exists between the facts from 2.2 and the present ones: SC did not bring there information on $x$, that circuit could be stable or unstable, unlike here where (2) is false, the circuit is unstable and instead of (2) we can write

$$x(t) = x(0-0) \cdot \chi_{(-\infty,t_0)}(t) \oplus \overline{x(0-0)} \cdot \chi_{[t_0,t_1)}(t) \oplus$$
$$\oplus x(0-0) \cdot \chi_{[t_1,t_2)}(t) \oplus \overline{x(0-0)} \cdot \chi_{[t_2,t_3)}(t) \oplus ... \tag{6}$$

where $0 \leq t_0 < t_1 < t_2 < ...$ is unbounded.

**BDC'** Fig 7 a)

The system is the next one

$$\bigcap_{\xi \in [t-d_r,t]} x(\xi) \leq v(t) \leq \bigcup_{\xi \in [t-d_f,t]} x(\xi) \tag{7}$$

thus from (1):

$$\bigcap_{\xi \in [t-d_r,t]} (x(0-0) \cdot \chi_{(-\infty,0)}(\xi) \oplus \overline{v(\xi)} \cdot \chi_{[0,\infty)}(\xi)) \leq v(t) \leq$$
$$\leq \bigcup_{\xi \in [t-d_f,t]} (x(0-0) \cdot \chi_{(-\infty,0)}(\xi) \oplus \overline{v(\xi)} \cdot \chi_{[0,\infty)}(\xi)) \tag{8}$$

<u>Case</u> $x(0-0) = 0$

The functions $\bigcap_{\xi \in [t-d_r,t]} \overline{v(\xi)} \cdot \chi_{[0,\infty)}(\xi)$, $v(t)$, $\bigcup_{\xi \in [t-d_f,t]} \overline{v(\xi)} \cdot \chi_{[0,\infty)}(\xi)$ are all null for $t \leq 0$.
At the right of 0, because $\overline{v(0)} = 1$, $\bigcup_{\xi \in [t-d_f,t]} \overline{v(\xi)} \cdot \chi_{[0,\infty)}(\xi)$ becomes 1 and if $v$ continues to remain 0, in the point $t = d_r$, $\bigcap_{\xi \in [t-d_r,t]} \overline{v(\xi)} \cdot \chi_{[0,\infty)}(\xi)$ becomes 1, thus $0 < t_0 \leq d_r$ exists so that $v(t_0) = 1$. At the right of $t_0$, because $\overline{v(t_0)} = 0$, $\bigcap_{\xi \in [t-d_r,t]} \overline{v(\xi)} \cdot \chi_{[0,\infty)}(\xi)$ is 0 and if $v$ continues to be 1, in the point $t_0 + d_f$, $\bigcup_{\xi \in [t-d_f, t)} \overline{v(\xi)} \cdot \chi_{[0,\infty)}(\xi)$ becomes 0, in other words $t_0 < t_1 \leq t_0 + d_f$ exists so that $v(t_1) = 0$ etc. The conclusion is that the unbounded family $t_0, t_1, t_2, ...$ exists so that

$$0 < t_0 \leq d_r \quad (9)$$
$$t_0 < t_1 \leq t_0 + d_f$$
$$t_1 < t_2 \leq t_1 + d_r$$
$$t_2 < t_3 \leq t_2 + d_f$$
$$...$$

and

$$v(t) = \chi_{[t_0, t_1)}(t) \oplus \chi_{[t_2, t_3)}(t) \oplus ... \quad (10)$$

Case $x(0-0) = 1$

In similar conditions with the previous ones, the solutions $v$ of (8) are of the form
$$v(t) = \chi_{(-\infty, t_0)}(t) \oplus \chi_{[t_1, t_2)}(t) \oplus \chi_{[t_3, t_4)}(t) \oplus ... \quad (11)$$

where

$$0 < t_0 \leq d_f \quad (12)$$
$$t_0 < t_1 \leq t_0 + d_r$$
$$t_1 < t_2 \leq t_1 + d_f$$
$$t_2 < t_3 \leq t_2 + d_r$$
$$...$$

Adding AIC to BDC' gives the minimal length of the 0-pulses, respectively of the 1-pulses

in (10): 
$$\delta_r < t_{2k+1} - t_{2k} \quad (13)$$
$$\delta_f < t_{2k+2} - t_{2k+1}$$

in (11): 
$$\delta_f < t_{2k+1} - t_{2k} \quad (14)$$
$$\delta_r < t_{2k+2} - t_{2k+1}$$

for all $k \geq 0$.

RIC is the next one:
$$\overline{v(t-0)} \cdot v(t) \leq \bigcap_{\xi \in [t-\delta_r, t-\delta_r + \mu_r]} x(\xi) \quad (15)$$

$$v(t-0) \cdot \overline{v(t)} \leq \bigcap_{\xi \in [t-\delta_f, t-\delta_f + \mu_f]} \overline{x(\xi)} \quad (16)$$

(10) gives (case $x(0-0) = 0$):
$$\overline{v(t-0)} \cdot v(t) = \chi_{\{t_0, t_2, t_4, ...\}}(t) \quad (17)$$

$$\bigcap_{\xi \in [t-\delta_r, t-\delta_r + \mu_r]} \overline{v(\xi)} \cdot \chi_{[0,\infty)}(\xi) = \bigcap_{\xi \in [t-\delta_r, t-\delta_r + \mu_r]} (\chi_{[0, t_0)}(\xi) \oplus \chi_{[t_1, t_2)}(\xi) \oplus \chi_{[t_3, t_4)}(\xi) \oplus ...) =$$
$$= \chi_{[\delta_r, t_0 + \delta_r - \mu_r)}(t) \oplus \chi_{[t_1 + \delta_r, t_2 + \delta_r - \mu_r)}(t) \oplus \chi_{[t_3 + \delta_r, t_4 + \delta_r - \mu_r)}(t) \oplus ... \quad (18)$$

and inequality (15)
$$\chi_{\{t_0, t_2, t_4, ...\}}(t) \leq \chi_{[\delta_r, t_0 + \delta_r - \mu_r)}(t) \oplus \chi_{[t_1 + \delta_r, t_2 + \delta_r - \mu_r)}(t) \oplus \chi_{[t_3 + \delta_r, t_4 + \delta_r - \mu_r)}(t) \oplus ...$$
is equivalent with

$$\{t_0, t_2, t_4, \ldots\} \subset [\delta_r, t_0 + \delta_r - \mu_r) \vee [t_1 + \delta_r, t_2 + \delta_r - \mu_r) \vee [t_3 + \delta_r, t_4 + \delta_r - \mu_r) \vee \ldots$$

Similarly, from (10) and (16):

$$\{t_1, t_3, t_5, \ldots\} \subset (-\infty, \delta_f - \mu_f) \vee [t_0 + \delta_f, t_1 + \delta_f - \mu_f) \vee [t_2 + \delta_f, t_3 + \delta_f - \mu_f) \vee \ldots$$

Equation (11) (case $x(0-0) = 1$) combined with RIC gives restrictions of the same nature.

**FDC** Fig 7b)

(1) is true together with

$$u(t) = x(t - d_1) \tag{19}$$

$$v(t) = u(t - d_2) \tag{20}$$

where $d_1 \geq 0, d_2 \geq 0$ and by eliminating $u, v$ we obtain

$$x(t) = x(0-0) \cdot \chi_{(-\infty,0)}(t) \oplus \overline{x(t - d_1 - d_2)} \cdot \chi_{[0,\infty)}(t) \tag{21}$$

<u>Case</u> $d_1 + d_2 = 0$

Equation (21) is incompatible.

<u>Case</u> $d_1 + d_2 > 0$

The solution of (21) is

$$x(t) = x(0-0) \cdot \chi_{(-\infty,0)}(t) \oplus \overline{x(0-0)} \cdot \chi_{[0, d_1 + d_2)}(t) \oplus$$
$$\oplus x(0-0) \cdot \chi_{[d_1+d_2, 2(d_1+d_2))}(t) \oplus \overline{x(0-0)} \cdot \chi_{[2(d_1+d_2), 3(d_1+d_2))}(t) \oplus \ldots \tag{22}$$

**DBRIDC** Fig 7 a)

(1) is true together with

$$\overline{v(t-0)} \cdot v(t) = \overline{v(t-0)} \cdot \bigcap_{\xi \in [t-d_r, t-d_r+m_r]} x(\xi) \tag{23}$$

$$v(t-0) \cdot \overline{v(t)} = v(t-0) \cdot \bigcap_{\xi \in [t-d_f, t-d_f+m_f]} \overline{x(\xi)} \tag{24}$$

where $0 \leq m_r \leq d_r, 0 \leq m_f \leq d_f, v(0-0) = x(0-0)$ and by eliminating $x$ we get

$$\overline{v(t-0)} \cdot v(t) = \overline{v(t-0)} \cdot \bigcap_{\xi \in [t-d_r, t-d_r+m_r]} (x(0-0) \cdot \chi_{(-\infty,0)}(\xi) \oplus v(\xi) \cdot \chi_{[0,\infty)}(\xi)) \tag{25}$$

$$v(t-0) \cdot \overline{v(t)} = v(t-0) \cdot \bigcap_{\xi \in [t-d_f, t-d_f+m_f]} (\overline{x(0-0)} \cdot \chi_{(-\infty,0)}(\xi) \oplus v(\xi) \cdot \chi_{[0,\infty)}(\xi)) \tag{26}$$

We suppose that $x(0-0) = 0$.

<u>Case</u> $d_r - m_r > 0, d_f - m_f > 0$

Because in (25) we have $\forall \xi \in [0, d_r), v(\xi) = 0$, the implication is $v(d_r) = 1$. Because in (26) we have $\forall \xi \in [d_r, d_r + d_f), v(\xi) = 1$, the implication is $v(d_r + d_f) = 0$ etc. The solution is

$$v(t) = \chi_{[d_r, d_r + d_f)}(t) \oplus \chi_{[2d_r + d_f, 2d_r + 2d_f)}(t) \oplus \chi_{[3d_r + 2d_f, 3d_r + 3d_f)}(t) \oplus \ldots \tag{27}$$

<u>Case</u> $d_r - m_r = 0$ or $d_f - m_f = 0$

We suppose that $d_r = m_r > 0$ is true. (25) is in this situation

$$\overline{v(t-0)} \cdot v(t) = \overline{v(t-0)} \cdot \bigcap_{\xi \in [t-d_r, t]} \overline{v(\xi)} \cdot \chi_{[0,\infty)}(\xi) = \overline{v(t-0)} \cdot \bigcap_{\xi \in [t-d_r, t)} \overline{v(\xi)} \cdot \chi_{[0,\infty)}(\xi) \cdot \overline{v(t)} \tag{28}$$

For $t < d_r, v(t) = 0$ and at $t = d_r$ we get the contradiction $v(d_r) = \overline{v(d_r)}$. The system is incompatible. The possibilities $d_r = m_r = 0, d_f = m_f > 0, d_f = m_f = 0$ result in incompatible systems too.

The situation when $x(0-0) = 1$ is to be treated similarly.

**SDBRIDC'** Fig 7b)

(1) is true together with

$$Dy(t) = (y(t-0) \oplus x(t-0)) \cdot \overline{\bigcup_{\xi \in (t-d_1, t)} Dx(\xi)} \cdot \chi_{[d_1, \infty)}(t) \tag{29}$$

$$Dv(t) = (v(t-0) \oplus y(t-0)) \cdot \overline{\bigcup_{\xi \in (t-d_2, t)} Dy(\xi)} \cdot \chi_{[d_2, \infty)}(t) \tag{30}$$

where $d_1 > 0, d_2 > 0$. We have $x(0-0) = y(0-0) = v(0-0)$.

We suppose that $x(0-0) = 0$ and (30) used under the form $Dv(0) = 0$ gives $v(0) = 0$.

From (1), $x(0) = 1$. From (29), $y$ becomes 1 at the time instant $d_1$. From (30), $v$ becomes 1 at the time instant $d_1 + d_2$, when in (1) $x$ becomes 0. The conclusion is:

$$x(t) = \chi_{[0, d_1+d_2)}(t) \oplus \chi_{[2d_1+2d_2, 3d_1+3d_2)}(t) \oplus ... \tag{31}$$

$$y(t) = \chi_{[d_1, 2d_1+d_2)}(t) \oplus \chi_{[3d_1+2d_2, 4d_1+3d_2)}(t) \oplus ... \tag{32}$$

$$v(t) = \chi_{[d_1+d_2, 2d_1+2d_2)}(t) \oplus \chi_{[3d_1+3d_2, 4d_1+4d_2)}(t) \oplus ... \tag{33}$$

The situation when $x(0-0) = 1$ is similar.

The solutions are in this case the same like at FDC.

## 2.5 First Conclusions

Some of the facts that were presented in this section have been studied by us some time ago and they have brought a direction of research called pseudo-Boolean differential and integral calculus that is interesting by itself, being related with mathematical analysis ([22] and others). We can define for the functions $\mathbf{R} \to \{0,1\}$ derivatives, integrals, convolution products, distributions etc and many notions and results from the real mathematical analysis have analogues of this type. It is interesting also the study of the equations and of the inequalities written with such functions, in no direct relation with digital electrical engineering.

Another direction of research is the present one, related with the asynchronous circuits. Starting from the known models as well as using the intuition suggested by the literature, modeling is abstracted by defining the delay conditions, shortly the delays, as real time models of the delay circuits. From this moment, the construction of new models is natural. On the other hand, the way that the delay circuit is generalized by the C-element of Muller, the delays are generalized by the 2-delays and generally by the n-delays, that are the models of the C-elements of Muller. An example of this nature is sketched at 11.3.

# 3. Preliminaries

## 3.1 The Boole Algebra with Two Elements

3.1.1 **Definition** The set $\mathbf{B} = \{0,1\}$ is called the *binary Boole algebra*, or the *Boole algebra with two elements*. It is endowed with:

- the order $0 < 1$

- the laws:

    unary: ' $^{-}$ ', the *complement*

    binary: '$\cup$' the *reunion*, '$\cdot$' the *intersection*, '$\oplus$' the *modulo 2 sum*

defined in the next table:

| $^{-}$ | | | $\cup$ | 0 | 1 | | $\cdot$ | 0 | 1 | | $\oplus$ | 0 | 1 |
|---|---|---|---|---|---|---|---|---|---|---|---|---|---|
| | 0 | 1 | 0 | 0 | 1 | | 0 | 0 | 0 | | 0 | 0 | 1 |
| | 1 | 0 | 1 | 1 | 1 | | 1 | 0 | 1 | | 1 | 1 | 0 |

Table 1

We give the usual meaning to the binary relations on $\boldsymbol{B}$: $>, \leq, \geq$.

**3.1.2 Remarks** $(\boldsymbol{B}, ^{-}, \cup, \cdot)$ is a Boole algebra indeed and $(\boldsymbol{B}, \oplus, \cdot)$ is a field, where the inverse of $a$ relative to $\oplus$ is $a$ itself: $a \oplus a = 0$. The relation between the order and the laws of $\boldsymbol{B}$ is expressed by:

$$\forall a, b \in \boldsymbol{B}, a \leq b \Leftrightarrow \overline{a} \geq \overline{b}$$
$$\forall a, b \in \boldsymbol{B}, a \cup b = \max(a, b), a \cdot b = \min(a, b)$$
$$\forall a, b, c \in \boldsymbol{B}, b \leq c \Rightarrow (a \cup b \leq a \cup c), b \leq c \Rightarrow (a \cdot b \leq a \cdot c)$$

**3.1.3 Definition** Let the binary generalized sequence $(x_j)_{j \in J}$. We define

$$\bigcup_{j \in J} x_j = \begin{cases} 1, \text{if } \exists j \in J, x_j = 1 \\ 0, \text{otherwise} \end{cases}, \bigcup_{j \in \emptyset} x_j = 0$$

$$\bigcap_{j \in J} x_j = \begin{cases} 0, \text{if } \exists j \in J, x_j = 0 \\ 1, \text{otherwise} \end{cases}, \bigcap_{j \in \emptyset} x_j = 1$$

**3.1.4 Definition** The functions $f : \boldsymbol{B}^n \to \boldsymbol{B}^m, n, m \geq 1$ and as a special case the functions $f : \boldsymbol{B}^n \to \boldsymbol{B}$ are called *Boolean functions*.

**3.2 Generalities on the $\boldsymbol{R} \to \boldsymbol{B}$ Functions**

**3.2.1 Definition** The next order and laws are induced on $\boldsymbol{B}^{\boldsymbol{R}}$ by those of $\boldsymbol{B}$:

$$f \leq g \Leftrightarrow \forall t, f(t) \leq g(t)$$
$$\forall t, \overline{f}(t) = \overline{f(t)}$$
$$\forall t, (f \cup g)(t) = f(t) \cup g(t)$$
$$\forall t, (f \cdot g)(t) = f(t) \cdot g(t)$$
$$\forall t, (f \oplus g)(t) = f(t) \oplus g(t)$$

where $f, g : \boldsymbol{R} \to \boldsymbol{B}$ and $t \in \boldsymbol{R}$. The fact that we have abusively used the same notations for different orders and laws will cause no misunderstanding.

**3.2.2 Definition** We note with $\chi_A : \boldsymbol{R} \to \boldsymbol{B}$, where $A \subset \boldsymbol{R}$, the *characteristic function* of the set $A$:

$$\chi_A(t) = \begin{cases} 1, t \in A \\ 0, t \notin A \end{cases}$$

**3.2.3 Definition** The *support* of the function $f : \boldsymbol{R} \to \boldsymbol{B}$ is the set $supp\ f \subset \boldsymbol{R}$ defined by

$$supp\ f = \{t \mid t \in \boldsymbol{R}, f(t) = 1\}$$

**3.2.4 Remarks** The next properties are true for all $t \in \mathbf{R}$ and all $f, g : \mathbf{R} \to \mathbf{B}$:
$$\chi_\varnothing(t) = 0, \chi_{\mathbf{R}}(t) = 1$$
$$f(t) = \chi_{supp\ f}(t)$$
$$f \leq g \Leftrightarrow supp\ f \subset supp\ g$$
$$supp\ \overline{f} = \mathbf{R} - supp\ f$$
$$supp\ (f \cup g) = supp\ f \vee supp\ g$$
$$supp\ (f \cdot g) = supp\ f \wedge supp\ g$$
$$supp\ (f \oplus g) = supp\ f \Delta supp\ g$$

In fact we can identify $\mathbf{B}^{\mathbf{R}}$ and $2^{\mathbf{R}}$ from the point of view of the order and of the algebraical properties.

**3.2.5 Definition** We say that $f : \mathbf{R} \to \mathbf{B}$ *has a limit when $t$ tends to infinite* if
$$\exists t', \forall t \geq t', f(t) = f(t')$$
The number $f(t')$ that does not depend on $t'$ is noted with $\lim_{t \to \infty} f(t)$.

## 3.3 Limits and Derivatives. The Continuity and the Differentiability of the $\mathbf{R} \to \mathbf{B}$ Functions

**3.3.1 Definition** Let $f : \mathbf{R} \to \mathbf{B}$. If the function $f^- : \mathbf{R} \to \mathbf{B}$ exists so that
$$\forall t \in \mathbf{R}, \exists \varepsilon > 0, \forall \xi \in (t - \varepsilon, t), f(\xi) = f^-(t)$$
i.e. if for any $t$ and any $\xi < t$ sufficiently close to $t$, $f(\xi)$ depends on $t$ only and not on $\xi$, we say that $f$ *has left limit* or that the *left limit of $f$ exists*. Similarly if $f^+ : \mathbf{R} \to \mathbf{B}$ exists having the property that
$$\forall t \in \mathbf{R}, \exists \varepsilon > 0, \forall \xi \in (t, t + \varepsilon), f(\xi) = f^+(t)$$
is true, we say that $f$ *has right limit* or that the *right limit of $f$ exists*. In the hypothesis that both previous properties are satisfied, we use to say that $f$ is *differentiable*. $f^-(t), f^+(t)$ are sometimes noted with $f(t-0), f(t+0)$ and are called the *left limit,* respectively the *right limit* (*function*) *of $f$*.

**3.3.2 Definition** The functions
$$D_{01}f(t) = \overline{f(t-0)} \cdot f(t), D_{10}f(t) = f(t-0) \cdot \overline{f(t)}$$
$$D^*_{10}f(t) = f(t) \cdot \overline{f(t+0)}, D^*_{01}f(t) = \overline{f(t)} \cdot f(t+0)$$
are called the *left* and the *right semi-derivatives of $f$* and the functions
$$Df(t) = f(t-0) \oplus f(t) = \overline{f(t-0)} \cdot f(t) \cup f(t-0) \cdot \overline{f(t)}$$
$$D^*f(t) = f(t+0) \oplus f(t) = \overline{f(t+0)} \cdot f(t) \cup f(t+0) \cdot \overline{f(t)}$$
are called the *left*, respectively the *right derivative of $f$*.

**3.3.3 Definition** $f$ is *left continuous*, respectively *right continuous*, if
$$\forall t, f(t) = f(t-0)$$
$$\forall t, f(t) = f(t+0)$$

**3.3.4 Remark** We can prove [22] that the only differentiable $\mathbf{R} \to \mathbf{B}$ functions that are both left continuous and right continuous (on an interval) are the two constant functions (on that interval). If the differentiable function $f$ is right continuous, then it is constant (on an interval) iff $Df(t) = 0$ (on that interval) and a dual property holds for the differentiable left continuous functions.

3.3.5 **Examples** a) $$f(t) = \chi_{[0,1]}(t) \oplus \chi_{\{2\}}(t)$$
is a differentiable function, that is neither left, nor right continuous. More precisely
$$f(t-0) = \chi_{(0,1]}(t), f(t+0) = \chi_{[0,1)}(t)$$
$$Df(t) = \chi_{\{0\}}(t) \oplus \chi_{\{2\}}(t), D*f(t) = \chi_{\{1\}}(t) \oplus \chi_{\{2\}}(t)$$

b) The function
$$f(t) = \bigcup_{n \geq 1} \chi_{(\frac{1}{2n+1}, \frac{1}{2n}]}(t)$$
is left continuous, as it has the property that $f(t-0)$ exists and it is equal with $f(t)$. $f(t+0)$ does not exist however, because for $t = 0$, we have
$$\forall \varepsilon > 0, \exists \xi, \xi' \in (0, \varepsilon), f(\xi) \neq f(\xi')$$

## 3.4 The Properties of the Limits and of the Derivatives

3.4.1 **Theorem** If $f, g$ have left limit, respectively right limit, then the next properties hold

a) $D\overline{f} = \overline{Df}$
b) $D(f \oplus g) = Df \oplus Dg$
c) $D(f \cdot g) = f \cdot Dg \oplus g \cdot Df \oplus Df \cdot Dg$

respectively the dual properties.

**Proof** a) $D\overline{f}(t) = \overline{f}(t-0) \oplus \overline{f}(t) = \overline{f(t-0)} \oplus \overline{f(t)} = 1 \oplus f(t-0) \oplus 1 \oplus f(t) =$
$= f(t-0) \oplus f(t) = Df(t)$

3.4.2 **Theorem** a) If $f$ has left limit, then $f^-$ has left limit and
$$(f^-)^- = f^-$$

b) If $f$ is differentiable, then $f^-$ has right limit and
$$(f^-)^+ = f^+$$

the dual statements of a), b) being also true.

**Proof** a) Let $t$ arbitrary and fixed, $t' < t$ and $f^-$ with the property
$$\forall \xi \in (t',t), f(\xi) = f^-(t)$$
We fix the numbers $\xi, \omega$ arbitrary with $t' < \xi < \omega < t$. We have $f(\xi) = f(\omega)$ and, as $f(\xi)$ depends only on $\omega$ -not on $\xi$ - we get $f(\xi) = f^-(\omega)$. On the other hand $f^-(\omega) = f^-(t)$ thus $f^-(\omega)$ is independent on $\omega$ -it depends only on $t$ -and we get $f^-(\omega) = (f^-)^-(t)$. We have obtained $(f^-)^-(t) = f^-(\omega) = f^-(t)$.

3.4.3 **Corollary** a) For $f$ with left limit, $Df$ has left limit and
$$DDf = Df$$

b) If $f$ is differentiable, then
$$D*Df = Df$$

**Proof** a) $DDf(t) = D(f(t-0) \oplus f(t)) = f(t-0) \oplus f(t-0) \oplus f(t-0) \oplus f(t) = Df(t)$
b) $D*Df(t) = D*(f(t-0) \oplus f(t)) = f(t+0) \oplus f(t+0) \oplus f(t-0) \oplus f(t) = Df(t)$

3.4.4 **Remark** The way that the left, respectively the right limits are iterated shows that we do not need to work with derivatives of the second or higher order.

3.4.5 **Notation** We note with $\tau^d : R \to R$ the translation with $d \in R$:
$$\tau^d(t) = t - d$$

3.4.6 **Theorem** Let $d \in R$ arbitrary. $f$ has left limit iff $f \circ \tau^d$ has left limit. Similar properties hold for the left continuity, the differentiability etc.
**Proof** Obvious.

3.4.7 **Remark** The compatibility between the Boolean laws and the translation is given by
$$\overline{f} \circ \tau^d = \overline{f \circ \tau^d}$$
$$(f \cup g) \circ \tau^d = f \circ \tau^d \cup g \circ \tau^d$$
etc and the compatibility between the limits, the derivatives and respectively the translation is expressed by the equations
$$f^- \circ \tau^d = (f \circ \tau^d)^-, \; f^+ \circ \tau^d = (f \circ \tau^d)^+$$
$$Df \circ \tau^d = D(f \circ \tau^d), \; D*f \circ \tau^d = D*(f \circ \tau^d)$$

**3.5 Conventions Concerning the Drawings of the Graphics of the $R \to B$ Functions**

3.5.1 **Conventions** In order to make easier the understanding of the $R \to B$ functions, we make the next *conventions* concerning the drawing of their graphics:
    a) the two values 0,1 are not written on the vertical axis. They are supposed to be known, the only necessary convention is that the low value be associated with 0 and the high value be associated with 1
    b) we draw vertical lines in these points where the function switches (the discontinuity points), even if the vertical lines do not belong to the graphic
    c) we put bullets on the vertical lines that are drawn like at b), underlining this way the points that actually belong to the graphic (the value of the function in the switching point)
    d) we avoid writing values on the time axis, i.e. the horizontal one, whenever this causes no misunderstanding.

3.5.2 **Example** We give the example of the next figure, where we have drawn the graphics of the functions from 3.3.5 a).

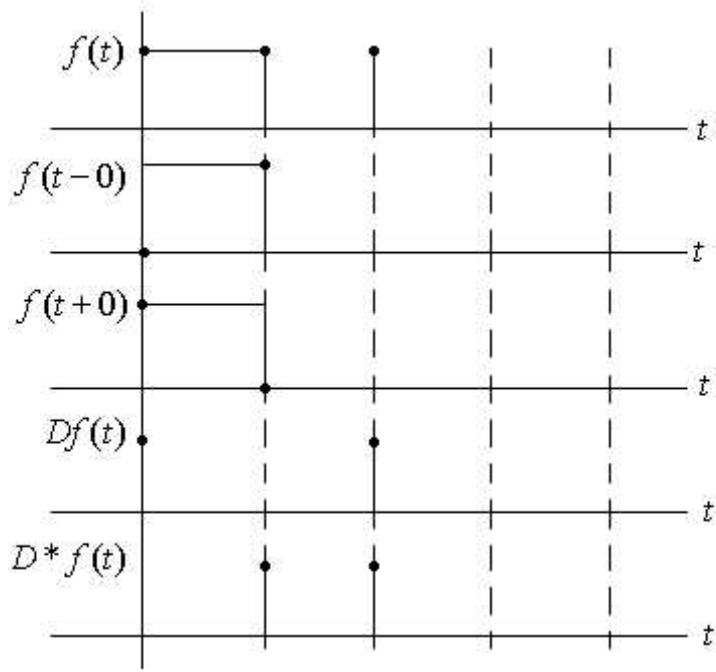

Fig 8

## 4. Signals

### 4.1 The Definition of the Signals

**4.1.1 Definition** A function $f : \mathbf{R} \to \mathbf{B}$ is called (*electrical*) *signal*[2], if
    i) it is differentiable
    ii) it is right continuous
    iii) $supp\ Df \subset [0, \infty)$

The set of the signals is noted with $S$ and the set of the non-empty subsets of $S$ is noted $P*(S)$.

**4.1.2 Remark** We give an interpretation to the previous definition. The next property, that can be inferred from i)
$$\forall t', t'' \in \mathbf{R}, t' < t'' \Rightarrow (t', t'') \wedge supp\ Df \text{ is a finite set}$$
is the *finite variability*, a signal switches finitely many times in any bounded time interval and it is an inertia request. The right continuity of $f$ is a property of *causality*, or *non-anticipation* and it is related to the fact that the present depends on the past and maybe on the present itself, but not on the future. The request iii) is related to the initial time that is 0 -but this condition is also one of non-anticipation. The anticipative systems having an evolution with the time axis reversed and where the present depends on the future and maybe on the present itself, are modeled by differentiable left continuous functions $f$ with
$$supp\ D*f \subset (-\infty, 0]$$
    We have just indicated how the dual notion to that of signal is to be defined.

**4.1.3 Theorem** The next conditions are equivalent for the function $f : \mathbf{R} \to \mathbf{B}$:

---

[2] These functions, or similar functions with the same role in modeling, have also been called in literature (see [1], [14], [16]) *Boolean signals*, *piecewise constant signals*, *piecewise continuous signals*, *non-zeno signals* and respectively *finite variability signals*.

a) $f$ is signal

b) the unbounded family $0 \leq t_0 < t_1 < t_2 < ...$ exists so that
$$f(t) = f(-1) \cdot \chi_{(-\infty, t_0)}(t) \oplus f(t_0) \cdot \chi_{[t_0, t_1)}(t) \oplus f(t_1) \cdot \chi_{[t_1, t_2)}(t) \oplus ...$$

**Sketch of proof** $a) \Rightarrow b)$  4.1.1 i) implies the existence of an upper and lower unbounded family $... < t_{-1} < t_0 < t_1 < ...$ where $t_0$ can be taken $\geq 0$ without loosing the generality, satisfying the property that $\forall t$,

$$f(t) = ... \oplus f(t_{-1}) \cdot \chi_{\{t_{-1}\}}(t) \oplus f(\frac{t_{-1}+t_0}{2}) \cdot \chi_{(t_{-1}, t_0)}(t) \oplus$$

$$\oplus f(t_0) \cdot \chi_{\{t_0\}}(t) \oplus f(\frac{t_0+t_1}{2}) \cdot \chi_{(t_0, t_1)}(t) \oplus f(t_1) \cdot \chi_{\{t_1\}}(t) \oplus ...$$

4.1.1 ii), 4.1.1 iii) and the previous equation imply b).

$b) \Rightarrow a)$ It is shown that $\forall t$, three possibilities exist:

i)      $t < t_0$; then $f(t-0) = f(t+0) = f(t)$

ii)     $t \geq t_0$ and $\exists k, t = t_k$; then $f(t-0) = \begin{cases} f(t_{k-1}), & k \geq 1 \\ f(-1), & k = 0 \end{cases}$, $f(t+0) = f(t_k)$

iii)    $t \geq t_0$ and $\exists k, t \in (t_k, t_{k+1})$; then $f(t-0) = f(t+0) = f(t_k)$

**4.1.4 Theorem** Let $f \in S$ an arbitrary signal and $d \in \mathbf{R}$. The following statements are true:

a) $f \circ \tau^d$ is differentiable and right continuous

b) $f \circ \tau^d \in S \Leftrightarrow supp\, Df \circ \tau^d \subset [0, \infty)$

c) if $d \geq 0$, then $f \circ \tau^d \in S$.

**Proof** a) results from 4.1.3 b), because $f \circ \tau^d$ has the same form like $f$ except the request $0 \leq t_0$. b) results from $Df \circ \tau^d = D(f \circ \tau^d)$ (see Remark 3.4.7) and from a). For c), we take in consideration the fact that
$$supp\, D(f \circ \tau^d) = \{t \mid Df(t-d) = 1\} = \{t+d \mid Df(t) = 1\} = d + supp\, Df$$
and b) also, implication $\Leftarrow$:
$$d \geq 0 \text{ and } supp\, Df \subset [0, \infty) \Rightarrow d + supp\, Df \subset [0, \infty) \Rightarrow$$
$$\Rightarrow supp\, D(f \circ \tau^d) \subset [0, \infty) \Rightarrow f \circ \tau^d \in S$$

**4.1.5 Remarks** $(S, \leq)$ is an ordered set, with $\leq$ induced from $\mathbf{B}^\mathbf{R}$ and the constant functions 0, 1 are the null and respectively the universal element of $S$. $(S, \overline{\phantom{x}}, \cup, \cdot)$ is a Boolean algebra and $(S, \oplus, \cdot)$ is a commutative ring.

In $S$, the next equivalence holds, see Remark 3.3.4:

a) $f$ is constant (on certain intervals)

b) $Df = 0$ (on certain intervals).

**4.1.6 Definition** Let $f \in S$. By 1-*pulse*, respectively 0-*pulse* of the signal $f$ we mean the existence of the numbers $t' < t''$ with the property that
$$\forall \xi \in [t', t''), f(\xi) = 1 \text{ and } f(t'-0) = f(t'') = 0$$
$$\forall \xi \in [t', t''), f(\xi) = 0 \text{ and } f(t'-0) = f(t'') = 1$$
is true. In this case we say that $f$ has a 1-pulse, respectively a 0-pulse of *length* $t'' - t'$.

## 4.2 Useful Lemmas

**4.2.1 Theorem** Let $f$ some differentiable function and the numbers $0 \leq m \leq d$. The functions
$$\Phi(t) = \bigcap_{\xi \in [t-d, t-d+m]} f(\xi)$$
$$\Psi(t) = \bigcup_{\xi \in [t-d, t-d+m]} f(\xi)$$
are differentiable and they satisfy

$$\Phi(t-0) = f(t-d-0) \cdot \bigcap_{\xi \in [t-d, t-d+m)} f(\xi) \tag{1}$$

$$\Phi(t+0) = \bigcap_{\xi \in (t-d, t-d+m]} f(\xi) \cdot f(t-d+m+0) \tag{2}$$

$$\Psi(t-0) = f(t-d-0) \cup \bigcup_{\xi \in [t-d, t-d+m)} f(\xi) \tag{3}$$

$$\Psi(t+0) = \bigcup_{\xi \in (t-d, t-d+m]} f(\xi) \cup f(t-d+m+0) \tag{4}$$

**Proof** If $m = 0$, then $\Phi(t) = \Psi(t) = f(t-d)$ is differentiable and we use Definition 3.1.3 ($\bigcap_{\xi \in \emptyset} f(\xi) = 1$, $\bigcup_{\xi \in \emptyset} f(\xi) = 0$).

We suppose now that $m > 0$. Let $t$ arbitrary and fixed. The left limit of $f$ in $t-d$ shows the existence of $\varepsilon_1 > 0$ with
$$\forall \xi \in (t-d-\varepsilon_1, t-d), f(\xi) = f(t-d-0)$$
and the left limit of $f$ in $t-d+m$ shows the existence of $\varepsilon_2 > 0$ so that
$$\forall \xi \in (t-d+m-\varepsilon_2, t-d+m), f(\xi) = f(t-d+m-0)$$
For any $0 < \varepsilon < \min(\varepsilon_1, \varepsilon_2, m)$ we infer
$$\Phi(t-\varepsilon) = \bigcap_{\xi \in [t-d-\varepsilon, t-d+m-\varepsilon]} f(\xi) = \bigcap_{\xi \in [t-d-\varepsilon, t-d)} f(\xi) \cdot \bigcap_{\xi \in [t-d, t-d+m-\varepsilon]} f(\xi) =$$
$$= f(t-d-0) \cdot \bigcap_{\xi \in [t-d, t-d+m-\varepsilon]} f(\xi) =$$
$$= f(t-d-0) \cdot \bigcap_{\xi \in [t-d, t-d+m-\varepsilon]} f(\xi) \cdot f(t-d+m-0) =$$
$$= f(t-d-0) \cdot \bigcap_{\xi \in [t-d, t-d+m-\varepsilon]} f(\xi) \cdot \bigcap_{\xi \in (t-d+m-\varepsilon, t-d+m)} f(\xi) =$$
$$= f(t-d-0) \cdot \bigcap_{\xi \in [t-d, t-d+m)} f(\xi)$$
Because the value of $\Phi(t-\varepsilon)$ does not depend on $\varepsilon$, we get $\Phi(t-\varepsilon) = \Phi(t-0)$ and because $t$ was arbitrary, (1) is proved.

The right limit of $f$ in $t-d$ shows the existence of $\varepsilon_3 > 0$ so that
$$\forall \xi \in (t-d, t-d+\varepsilon_3), f(\xi) = f(t-d+0)$$
and on the other hand the right limit of $f$ in $t-d+m$ shows the existence of $\varepsilon_4 > 0$ with
$$\forall \xi \in (t-d+m, t-d+m+\varepsilon_4), f(\xi) = f(t-d+m+0)$$
We take some $0 < \varepsilon' < \min(\varepsilon_3, \varepsilon_4, m)$ for which we have

$$\Phi(t+\varepsilon') = \bigcap_{\xi\in[t-d+\varepsilon',t-d+m+\varepsilon']} f(\xi) = \bigcap_{\xi\in[t-d+\varepsilon',t-d+m]} f(\xi) \cdot \bigcap_{\xi\in(t-d+m,t-d+m+\varepsilon']} f(\xi) =$$

$$= \bigcap_{\xi\in[t-d+\varepsilon',t-d+m]} f(\xi) \cdot f(t-d+m+0) =$$

$$= f(t-d+0) \cdot \bigcap_{\xi\in[t-d+\varepsilon',t-d+m]} f(\xi) \cdot f(t-d+m+0) =$$

$$= \bigcap_{\xi\in(t-d,t-d+\varepsilon')} f(\xi) \cdot \bigcap_{\xi\in[t-d+\varepsilon',t-d+m]} f(\xi) \cdot f(t-d+m+0) =$$

$$= \bigcap_{\xi\in(t-d,t-d+m]} f(\xi) \cdot f(t-d+m+0)$$

The fact that $\Phi(t+\varepsilon')$ does not depend on $\varepsilon'$ shows that $\Phi(t+\varepsilon') = \Phi(t+0)$ and because $t$ was arbitrary, (2) is true. $\Phi$ is differentiable.

The proof for $\Psi$ is similar.

**4.2.2 Theorem** In the previous conditions and with the previous notations, we have:
$$\overline{\Phi(t-0)} \cdot \Phi(t) = \overline{f(t-d-0)} \cdot \bigcap_{\xi\in[t-d,t-d+m]} f(\xi)$$

$$\Phi(t-0) \cdot \overline{\Phi(t)} = f(t-d-0) \cdot \bigcap_{\xi\in[t-d,t-d+m)} f(\xi) \cdot \overline{f(t-d+m)}$$

$$\overline{\Psi(t-0)} \cdot \Psi(t) = \overline{f(t-d-0)} \cdot \bigcup_{\xi\in[t-d,t-d+m)} f(\xi) \cdot f(t-d+m)$$

$$\Psi(t-0) \cdot \overline{\Psi(t)} = f(t-d-0) \cdot \overline{\bigcup_{\xi\in[t-d,t-d+m]} f(\xi)}$$

**Proof**
$$\overline{\Phi(t-0)} \cdot \Phi(t) = \overline{f(t-d-0) \cdot \bigcap_{\xi\in[t-d,t-d+m)} f(\xi)} \cdot \bigcap_{\xi\in[t-d,t-d+m]} f(\xi) =$$

$$= (\overline{f(t-d-0)} \cup \overline{\bigcap_{\xi\in[t-d,t-d+m)} f(\xi)}) \cdot \bigcap_{\xi\in[t-d,t-d+m]} f(\xi) = \overline{f(t-d-0)} \cdot \bigcap_{\xi\in[t-d,t-d+m]} f(\xi)$$

$$\overline{\Psi(t-0)} \cdot \Psi(t) = \overline{f(t-d-0) \cup \bigcup_{\xi\in[t-d,t-d+m)} f(\xi)} \cdot \bigcup_{\xi\in[t-d,t-d+m]} f(\xi) =$$

$$= \overline{f(t-d-0)} \cdot \overline{\bigcup_{\xi\in[t-d,t-d+m)} f(\xi)} \cdot (\bigcup_{\xi\in[t-d,t-d+m)} f(\xi) \cup f(t-d+m)) =$$

$$= \overline{f(t-d-0)} \cdot \overline{\bigcup_{\xi\in[t-d,t-d+m)} f(\xi)} \cdot f(t-d+m)$$

**4.2.3 Theorem** If $f$ is signal, then $\Phi, \Psi$ are signals.

**Proof** If $m=0$ and $\Phi(t)=\Psi(t)=f(t-d)$, then the differentiable function $f(t-d)$ is signal, from Theorem 4.1.4 c). We consider from this moment that $m>0$.

We know from Theorem 4.2.1 that $\Phi$ is differentiable and we must show that it satisfies 4.1.1 ii) and 4.1.1 iii). The right continuity of $f$ in $t-d$ shows that $\varepsilon_1 > 0$ exists with
$$\forall \xi \in [t-d, t-d+\varepsilon_1), f(\xi) = f(t-d)$$
and the right continuity of $f$ in $t-d+m$ shows the existence of $\varepsilon_2 > 0$ so that
$$\forall \xi \in (t-d+m, t-d+m+\varepsilon_2), f(\xi) = f(t-d+m)$$

Let $0 < \varepsilon < \min(\varepsilon_1, \varepsilon_2, m)$ and we conclude

$$\Phi(t+\varepsilon) = \bigcap_{\xi \in [t-d+\varepsilon, t-d+m+\varepsilon]} f(\xi) = \bigcap_{\xi \in [t-d+\varepsilon, t-d+m]} f(\xi) \cdot \bigcap_{\xi \in (t-d+m, t-d+m+\varepsilon]} f(\xi) =$$

$$= \bigcap_{\xi \in [t-d+\varepsilon, t-d+m]} f(\xi) \cdot f(t-d+m) = \bigcap_{\xi \in [t-d+\varepsilon, t-d+m]} f(\xi) =$$

$$= f(t-d) \cdot \bigcap_{\xi \in [t-d+\varepsilon, t-d+m]} f(\xi) =$$

$$= \bigcap_{\xi \in [t-d, t-d+\varepsilon)} f(\xi) \cdot \bigcap_{\xi \in [t-d+\varepsilon, t-d+m]} f(\xi) = \bigcap_{\xi \in [t-d, t-d+m]} f(\xi) = \Phi(t)$$

Thus $\Phi(t+\varepsilon) = \Phi(t+0) = \Phi(t)$. $\Phi$ is right continuous.

Moreover, the property 4.1.1 iii) is fulfilled since from 4.2.2:

$$D\Phi(t) = \overline{\Phi(t-0)} \cdot \Phi(t) \cup \Phi(t-0) \cdot \overline{\Phi(t)} =$$

$$= \overline{f(t-d-0)} \cdot \bigcap_{\xi \in [t-d, t-d+m]} f(\xi) \cup f(t-d-0) \cdot \bigcap_{\xi \in [t-d, t-d+m)} f(\xi) \cdot \overline{f(t-d+m)} \le$$

$$\le \overline{f(t-d-0)} \cdot f(t-d) \cup f(t-d+m-0) \cdot \overline{f(t-d+m)} \le$$

$$\le \overline{f(t-d-0)} \cdot f(t-d) \cup f(t-d-0) \cdot \overline{f(t-d)} \cup$$

$$\cup \overline{f(t-d+m-0)} \cdot f(t-d+m) \cup f(t-d+m-0) \cdot \overline{f(t-d+m)} =$$

$$= D(f \circ \tau^d)(t) \cup D(f \circ \tau^{d-m})(t)$$

and, because $f \circ \tau^d, f \circ \tau^{d-m}$ are signals

$$supp\ D(f \circ \tau^d) \subset [0, \infty), supp\ D(f \circ \tau^{d-m}) \subset [0, \infty)$$

we have

$$supp\ D\Phi \subset supp\ D(f \circ \tau^d) \vee supp\ D(f \circ \tau^{d-m}) \subset [0, \infty)$$

$\Phi$ is signal.

The proof for $\Psi$ is dual.

## 5. An Overview of the Delays: Informal Definitions

5.1 We present now some of the intuitive knowledge that has generated the efforts represented by the present work.

At least two things are understood by the word *delay*: a *real non-negative number d* see 5.2, 5.3 and a *logical condition* see 5.4 and the following. These two occur usually together in definitions, since a complete separation is very difficult.

5.2 **Informal definitions** As real non-negative number, the word *delay* is a short form for one of the following i), ii).

i) *propagation delay*, or *transport delay* [10] representing the '*time interval between a transition in an input to the gate and a corresponding output transition. If the output transition changes from 0 to 1, the delay is rising, otherwise falling*'. The same notion is called in [12], [13] *transmission delay for transitions*.

ii) *inertial delay* representing [10] the '*minimum amount of time during which an input signal must persist to affect a change at the output*". In [12], [13] the same notion is called *threshold for cancellation* and in [14] *latency delay*.

5.3 **Convention** The distinct numbers 5.2 i) and 5.2 ii) are generally taken to be equal [14] when the last exists, i.e. in the presence of inertia. We quote the next opinion:[12], [13]: '*A common form of implementation of the inertial delay model is the one in which the transmission delay d for transitions is the same as the threshold for cancellation. In other* words, when a transition appears at the input, the transition will appear at the output after *d* unless a second transition occurs within that period'.

**5.4 Classification of delays** The logical condition called *delay condition* defines a model and idioms like '*fixed delay*' will often be used as a short form for '*fixed delay condition*' or '*fixed delay model*' etc. The 'delays' to follow are all logical conditions defined informally.

Thus, from the *timing properties* point of view, we have
  a) *unbounded delays*
  b) *bounded delays*
  c) *fixed delays*
and from the *memory properties* point of view, the delays are:
  i) *pure delays*, i.e. delays without memory
  ii) *inertial delays*, i.e. delays with memory.

**5.5 Informal definition** The *unbounded delays* are defined
  a) [7]: '*a delay may take on any finite value*'
  b) [11]: '*no bound on the magnitude is known a priori, except that it is positive and finite*'.

**5.6 Remark** The unbounded delay model is evaluated in [5] to be '*robust to delay variations*', but '*unrealistically conservative*'.

**5.7 Informal definition** The *bounded delays* are defined in the next manner:
  a) [7]: '*a delay may have any value in a given time interval*'
  b) [11]: a delay is bounded '*if an upper and lower bound on its magnitude are known before the synthesis or analysis of the circuit is begun*'
  c) [5] '*every component is assumed to have an uncertain delay, that lies between given upper and lower bounds. The delay bounds take into account potential delay variations due to statistical fluctuations in the fabrication process, variations in ambient temperature, power supply etc*'.
  d) [10]: '*In practice, manufactured circuits of the same design may have different gate delays due to manufacturing fluctuations in delay related parameters such as capacitance, resistivity and transistors sizes. To be practical, we need to provide an analysis for not just a manufactured instance of a design but the entire family of manufactured circuits of the same design. To model manufacturing uncertainties, we assume the gate delays to be variable within closed intervals. Therefore a complete delay analysis determines the delays of circuits with variable gate delays…*'

**5.8 Remarks** The bounded delay model is considered to be the most realistic one from the three: unbounded, bounded and fixed delays.

To be remarked in this context the necessary supplementary conditions (of speed-independence, or delay insensitiveness for example) of invariance relative to the variation of the delays, required in the synthesis procedure of the circuits.

On the other hand, non-conflicting differences occur in the approaches when defining and using the unbounded and the bounded delays: from having no lower bounds or upper bounds, the poorest case of the delay model, to having four such bounds $d_{r,\min} \leq d_{r,\max}$, $d_{f,\min} \leq d_{f,\max}$, the richest case of bounded delay that makes use of the distinction between the rising and the falling delays. The more detailed the model is, the more difficult is its handling and the more realistic are its results.

**5.9 Informal definition** The *fixed delays* are a special case of bounded delays, when '*a delay is assumed to have a fixed value*', [7] and the lower bounds of the delays are equal with the upper bounds of the delays making the delay be fixed, known.

**5.10 Remark** The fixed delay model is considered to be very unrealistic, in the sense that small variations of the delays due to variations in the ambient temperature, power supply,… cause great, unacceptable differences between the model and the modeled circuit. [5]: '*Since it is almost impossible to obtain a precise delay of a component in a chip, this is not a realistic model for timing verification purpose*'.

**5.11 Informal definition** The *pure delays*, or *ideal delays* are defined like this.

a) [11]: a delay is considered to be pure '*if it transmits each event on its input to its output, i.e. it corresponds to a pure translation in time of the input waveform*'.

b) [5]: '*a pure delay simply shifts a waveform in time without altering its shape*'.

The same idea is found in [10], where the pure delay timed Boolean functions are defined.

**5.12 Remark** [13] refers to the pure delays, by considering that '*This model is unrealistic in the sense that practical gates will not transit a pulse caused by two transitions very close together whereas the model guarantees that every transition will be at the output irrespective of the proximity of the successive pulses*'.

**5.13 Informal definition** The *inertial delays* (or *latency delays*) have generated the most controversies, see also [5], [10]. The next opinions are generally accepted.

a) [12], [13] The inertial delays '*model the fact that the practical circuits will not respond to two transitions which are very close together. The inertial delay model is one in which input transitions are replicated at the output after some period of time unless two transitions occur at the input within some defined period, in which case neither transition is transmitted*'.

b) [11]: '*pulses shorter than or equal to the delay magnitude are not transmitted*', see also Convention 5.3.

c) [7] '*an inertial delay has a threshold period d. Pulses of duration less than d are filtered out*', compatible with Convention 5.3 again.

**5.14 Alternative definition** In [1], [14] the authors show intuitively, like above, what inertia is and then two variants of fixed, respectively bounded inertial delays are mentioned. We reproduce only the second variant from [1], called there *non-deterministic inertial delay*, for making the exposure as simple as possible and for the same reason we have changed the language and the notations. $x^0 \in \boldsymbol{B}$ and the real numbers $0 \leq d_{min} \leq d_{max}$ are given and the requests are

i) $\forall t \in [0, d_{min}), x(t) = x^0$ initialization

ii) $\forall t \geq d_{min}$, $Dx(t) = 1 \Rightarrow \exists t' \in supp\ Du \wedge [t - d_{max}, t - d_{min}]$ such that $x(t) = u(t')$ and $(t', t) \wedge supp\ Du = \varnothing$

iii) $\forall t \in supp\ Du, (t, t + d_{max}] \wedge supp\ Du \neq \varnothing$ or $[t + d_{min}, t + d_{max}] \wedge supp\ Dx \neq \varnothing$

**5.15 Remark** In [1] we can find the next observation relative to definition 5.14: '*one could assume that changes should persist for at least $l_1$ time units but propagated after $l_2, l_2 > l_1$ time*', in other words one could abandon for the sake of accuracy the 'common form of implementation of the inertial delay model' from Convention 5.3.

**5.16 Alternative definition** We refer now to the approach from [4], where two variants of fixed, respectively of bounded inertial delays are given also, from which we reproduce the second one under the form:

i) $Dx(t) = 1 \Rightarrow \forall \xi \in [t - d_{min}, t), u(\xi) = x(t)$

ii) $\forall a \in \boldsymbol{B}, ((\forall \xi \in [t, t + d_{max}), u(\xi) = a) \Rightarrow$

$$\Rightarrow (\exists \tau \in [t, t+d_{\max}), \forall \xi \in [\tau, t+d_{\max}), x(\xi) = a))$$

**5.17 Remark** We make brief comments and a comparison between 5.14 and 5.16:

- terminological differences with our work occur, that the reader is asked to pay not very much attention at his first reading of this text.

- in the second definition, initialization is missing. If we start by definition from null initial conditions, then initialization is not necessary and if we reason for any possible initial value, then initialization is missing too. The possibility that initialization is missing at 5.14 also is given by the value $d_{\min} = 0$.

- 5.14 ii) and 5.16 i) essentially express the same idea, the first condition being stronger.

- 5.14 iii) is a negligent way of expressing the idea 5.16 ii). This negligence is symptomatic in the sense that in the non-formalized theories that we refer to, it produces no effects (we quote in this context with amusement one of our mentors, professor A. C. Albu from Timisoara that is specialized in the Foundations of Mathematics who called our attention once that '*all the research is made in a non-formalized manner*').

## 6. Delays

### 6.1 Stability. Rising and Falling Transmission Delays for Transitions

**6.1.1 Definition** Let $u, x$ two signals, called *input* (or *control*) and respectively *state* (or *output*). The next property

$$\forall a \in \boldsymbol{B}, (\exists t_1, \forall t \geq t_1, u(t) = a) \Rightarrow (\exists t_2, \forall t \geq t_2, x(t) = a)$$

is called the *stability condition* (SC). We say that the couple $(u, x)$ satisfies SC.

We also call stability condition the function $Sol_{SC} : S \rightarrow P^*(S)$ defined by:

$$Sol_{SC}(u) = \{x \mid (u, x) \text{ satisfies SC}\}$$

**6.1.2 Remark** SC states the next cause-effect relation between $u$ and $x$: if $\lim_{t \rightarrow \infty} u(t)$ does not exist, then $Sol_{SC}(u) = S$ and if $\lim_{t \rightarrow \infty} u(t)$ exists then, whichever it might be, $\lim_{t \rightarrow \infty} x(t)$ exists and $\lim_{t \rightarrow \infty} x(t) = \lim_{t \rightarrow \infty} u(t)$. On the other hand, the next 'non-anticipative' statement

$$\forall a \in \boldsymbol{B}, (\exists t_1, \forall t \geq t_1, u(t) = a) \Rightarrow (\exists t_2 \geq t_1, \forall t \geq t_2, x(t) = a)$$

is equivalent with SC.

**6.1.3 Definition** We suppose that $(u, x)$ satisfies SC, that $\lim_{t \rightarrow \infty} u(t)$ exists and that *supp Du* $\neq \emptyset$, *supp Dx* $\neq \emptyset$ [3]. We note

$$t_1 = \max \text{ supp } Du, t_2 = \max \text{ supp } Dx$$

The *transmission delay for transitions* (of $x$ relative to $u$) is the number $d \geq 0$ given by

$$d = \max(0, t_2 - t_1)$$

If

$$\overline{u(t_1 - 0)} \cdot u(t_1) = \overline{x(t_2 - 0)} \cdot x(t_2) = 1$$

then $d$ is called *rising* and if

$$u(t_1 - 0) \cdot \overline{u(t_1)} = x(t_2 - 0) \cdot \overline{x(t_2)} = 1$$

---
[3] *supp Dx, supp Dy* are finite, non-empty in this case.

then it is called *falling*. If *supp Du*, respectively *supp Dx* is empty, then $t_1$, respectively $t_2$ is by definition 0 and if $\lim_{t \to \infty} u(t)$ does not exist, then $d$ is not defined.

## 6.2 Delays

**6.2.1 Definition** A *delay condition* (DC) or shortly a *delay* is a function $i : S \to P^*(S)$ with
$$\forall u, i(u) \subset Sol_{SC}(u)$$

**6.2.1 Remark** The delays are the models of the delay circuits, i.e. of the circuits that compute the identity $1_B$. In practice, we usually work with systems of equations or inequalities in $u, x$ and for each $u$, $i(u)$ represents the set of solutions of these systems. Definition 6.2.1 asks that such solutions always exist and that the systems be stable.

**6.2.3 Examples** of DC's. a) $i(u) = \{u\}$ is usually noted with $I$. More general, the equation $i(u) = \{u \circ \tau^d\}$ defines a DC when $d \geq 0$ (see Theorem 4.1.4 c)) noted with $I_d$.

b) $i(u) = \{x \mid \exists d \geq 0, x(t) = u(t) \cdot \chi_{[d,\infty)}(t)\}$ and
$$i(u) = \{x \mid \exists d \geq 0, x(t) = \chi_{(-\infty,d)}(t) \oplus u(t) \cdot \chi_{[d,\infty)}(t)\}$$
are DC's.

c) $i(u) = Sol_{SC}(u)$ is a DC called the *unbounded delay*.

**6.2.4 Theorem** Let $U \subset S$, the DC's $i, j$ and the arbitrary function $\varphi : S \to P^*(S)$.

a) If $\forall u, i(u) \wedge U \neq \emptyset$, then the next equation defines a DC:
$$(i \wedge U)(u) = i(u) \wedge U$$

b) If $i, j$ satisfy $\forall u, i(u) \wedge j(u) \neq \emptyset$, then $i \wedge j$ is a DC defined by:
$$(i \wedge j)(u) = i(u) \wedge j(u)$$

c) Items a), b) are generalized by: if $\forall u, i(u) \wedge \varphi(u) \neq \emptyset$, then $i \wedge \varphi$ is a DC
$$(i \wedge \varphi)(u) = i(u) \wedge \varphi(u)$$

g) The function $i \vee j$ that is defined in the next manner is a DC:
$$(i \vee j)(u) = i(u) \vee j(u)$$

**Proof** c) The fact that $i \wedge \varphi$ takes values in $P^*(S)$ is assured by the hypothesis $\forall u, i(u) \wedge \varphi(u) \neq \emptyset$. Furthermore, for all $u$ we have
$$(i \wedge \varphi)(u) = i(u) \wedge \varphi(u) \subset i(u) \subset Sol_{SC}(u)$$

## 6.3 Determinism

**6.3.1 Definition** Let the DC $i$. If $\forall u, i(u)$ has a single element, then it is called *deterministic* and otherwise it is called *non-deterministic*.

**6.3.2 Remark** By interpreting $i$ as the set of the solutions of a system, its determinism indicates that the solution is unique for all $u$ and this allows identifying a deterministic DC with a function $i : S \to S$. The meaning of the non-deterministic delays consists in the fact that in an electrical circuit to one input $u$ there correspond several possible outputs $x \in i(u)$ depending on the variations in ambient temperature, power supply, on the technology etc.

**6.3.3 Examples** At 6.2.3 $I, I_d$ are deterministic and the other DC's are non-deterministic.

Let $U \subset S$ and the DC's $i, j$. At 6.2.4 a) and respectively at 6.2.4 b), if $i$ is deterministic then $i \wedge U(= i)$ is deterministic and respectively $i \wedge j(= i)$ is deterministic.

**6.3.4 Theorem** For $0 \leq m \leq d$, the next functions are deterministic DC's
$$x(t) = \bigcap_{\xi \in [t-d, t-d+m]} u(\xi)$$
$$x(t) = \bigcup_{\xi \in [t-d, t-d+m]} u(\xi)$$

**Proof** They are signals from Theorem 4.2.3. If $\exists d', \forall t \geq d', u(t) = 0$ and this is equivalent with $u(t) \leq \chi_{(-\infty, d')}(t)$ then
$$x(t) = \bigcap_{\xi \in [t-d, t-d+m]} u(\xi) \leq \bigcap_{\xi \in [t-d, t-d+m]} \chi_{(-\infty, d')}(\xi) = \chi_{(-\infty, d+d'-m)}(t)$$
i.e. $\forall t \geq d + d' - m, x(t) = 0$. Similarly, if $\exists d', \forall t \geq d', u(t) = 1$ and this is equivalent with $u(t) \geq \chi_{[d', \infty)}(t)$, then
$$x(t) = \bigcap_{\xi \in [t-d, t-d+m]} u(\xi) \geq \bigcap_{\xi \in [t-d, t-d+m]} \chi_{[d', \infty)}(\xi) = \chi_{[d+d', \infty)}(t)$$
in other words $\forall t \geq d + d', x(t) = 1$. We have shown that $x \in Sol_{SC}(u)$.

The proof for $x(t) = \bigcup_{\xi \in [t-d, t-d+m]} u(\xi)$ is dual.

## 6.4 Order

**6.4.1 Definition** For the DC's $i, j$ we define $i \subset j$ by
$$\forall u, i(u) \subset j(u)$$

**6.4.2 Remarks** The inclusion $\subset$ defines an order –that is not total- in the set of the DC's. $Sol_{SC}$ is the unit of this ordered set: any DC $i$ satisfies $i \subset Sol_{SC}$.

We interpret the inclusion $i \subset j$ by the fact that the first system contains more restrictive conditions than the second and modeling in the first case is more precise than in the second case. In particular, a deterministic DC $i$ contains the maximal information and the DC $Sol_{SC}$ contains the minimal information about the modeled delay circuit.

**6.4.3 Theorem** Any DC $j$ includes a deterministic DC $i$.
**Proof** For any $u$, the axiom of choice allows choosing from the set $j(u)$ a point $x$ and we define $i(u) = \{x\}$. $i(u)$ is non-empty and satisfies $i(u) \subset j(u) \subset Sol_{SC}(u)$, in other words $i$ is a deterministic DC.

**6.4.4 Corollary** In the inclusion $i \subset j$ of DC's, if $j$ is deterministic, then $i = j$.
**Proof** In the previous proof, the only possibility of choosing $i$ is
$$\forall u, i(u) = j(u)$$

**6.4.5 Examples** Let $U \subset S$ and the DC's $i, j$. If $\forall u, i(u) \wedge U \neq \emptyset$, then $i \wedge U \subset i$ and if $\forall u, i(u) \wedge j(u) \neq \emptyset$, then $i \wedge j \subset i \subset i \vee j$.

## 6.5 Time Invariance

**6.5.1 Definition** The DC $i$ is called *time invariant* if
$$\forall u, \forall x, \forall d \in \mathbf{R}, (u \circ \tau^d \in S \text{ and } x \in i(u)) \Rightarrow (x \circ \tau^d \in S \text{ and } x \circ \tau^d \in i(u \circ \tau^d))$$
and if the previous property is not true, then $i$ is called *time variable*.

**6.5.2 Remarks** We mention also a weaker version of time invariance, that we shall not use:

$$\forall u, \forall x, \forall d \in \mathbf{R}, (u \circ \tau^d \in S \text{ and } x \in i(u) \text{ and } x \circ \tau^d \in S) \Rightarrow (x \circ \tau^d \in i(u \circ \tau^d))$$

Let us suppose now that the signals $f : \mathbf{R} \to \mathbf{B}$ would have been defined more generally by any of the next equivalent conditions, see Theorem 4.1.3:

a) $f$ is differentiable, right continuous and $\exists t_0, \text{supp } Df \subset [t_0, \infty)$

b) the unbounded family $t_0 < t_1 < t_2 < \ldots$ exists so that

$$f(t) = f(t_0 - 0) \cdot \chi_{(-\infty, t_0)}(t) \oplus f(t_0) \cdot \chi_{[t_0, t_1)}(t) \oplus f(t_1) \cdot \chi_{[t_1, t_2)}(t) \oplus \ldots$$

i.e. we relax the condition $\text{supp } Df \subset [0, \infty)$ at a) and respectively we omit the condition $0 \leq t_0$ at b). We keep the conditions of lower boundness of $\text{supp } Df$ and respectively of $(t_n)$, that both mean the existence of *some* initial time instant $t_0$ and let us note with $\widetilde{S}$ the set of these signals. Obviously $S \subset \widetilde{S}$. In such conditions, a time invariant DC $\widetilde{i}$ is a function $\widetilde{i} : \widetilde{S} \to P^*(\widetilde{S})$ ($P^*(\widetilde{S}) = \{A \mid A \subset \widetilde{S}, A \neq \emptyset\}$) satisfying

i) $\forall u \in \widetilde{S}, \forall x \in \widetilde{i}(u), \forall a \in \mathbf{B}, (\exists t_1, \forall t \geq t_1, u(t) = a) \Rightarrow (\exists t_2, \forall t \geq t_2, x(t) = a)$

ii) $\forall u \in S, \forall x \in S, \forall d \in \mathbf{R},$

$$(u \circ \tau^d \in S \text{ and } x \in \widetilde{i}(u)) \Rightarrow (x \circ \tau^d \in S \text{ and } x \circ \tau^d \in \widetilde{i}(u \circ \tau^d))$$

i.e. ii) reproduces 6.5.1 written with $S$ not with $\widetilde{S}$. We note with $i = \widetilde{i}_{|S}$ the restriction of $\widetilde{i}$ at $S$. We have the next

**6.5.3 Theorem** a) $\forall u \in S, \widetilde{i}(u) \subset S$

b) $i$ is a time invariant DC

**Proof** a) We suppose against all reason the existence of some $u \in S$ and $x \in \widetilde{i}(u)$ so that $x \notin S$, thus $\text{supp } Dx \neq \emptyset$ and $t_0 = \min \text{supp } Dx$ satisfies $t_0 < 0$. Then for any $d \in (t_0, 0)$ we have that $u \circ \tau^{-d} \in S$ and $x \in \widetilde{i}(u)$ are both true, but $x \circ \tau^{-d} \in S$ is false because, see also the proof of Theorem 4.1.4 c)

$$\min \text{supp } D(x \circ \tau^{-d}) = -d + \min \text{supp } Dx = -d + t_0 < 0$$

This is in contradiction with the fact that $\widetilde{i}$ satisfies the time invariance condition 6.5.2 ii).

b) We take into account a). The three properties that must be fulfilled, i.e. the fact that $\forall u \in S, i(u) \neq \emptyset$, that $\forall u \in S, i(u) \subset Sol_{SC}(u)$ and the condition of time invariance 6.5.1 are satisfied by $i$ because they are satisfied by $\widetilde{i}$.

**6.5.4 Remark** If we try to replace 6.5.2 ii) with 6.5.1 written with $\widetilde{S}$ instead of $S$, then this time invariance condition becomes, see also Theorem 6.5.6 for something similar:

$$\forall u \in \widetilde{S}, \forall x \in \widetilde{S}, \forall d \in \mathbf{R}, x \in \widetilde{i}(u) \Rightarrow x \circ \tau^d \in \widetilde{i}(u \circ \tau^d)$$

The property is reasonable and it constitutes an alternative definition of time invariance for the DC's $\widetilde{i} : \widetilde{S} \to P^*(\widetilde{S})$, but it is too weak to produce the validity of the statement 6.5.3 a).

On the other hand, all the important delay conditions that we are interested in are time invariant. Theorem 6.5.3 allows choosing for them the initial time instant be 0 and this was already anticipated by Definition 4.1.1. We shall always use from this moment $S$ in our work, not $\widetilde{S}$.

**6.5.5 Examples** a) We show that $I_d$ is time invariant, where $d \geq 0$: for $d' \in \mathbf{R}$ arbitrary, $u \circ \tau^{d'} \in S$ and $x \in I_d(u)$, i.e. $x = u \circ \tau^d$ imply

$$x \circ \tau^{d'} = (u \circ \tau^d) \circ \tau^{d'} = u \circ \tau^{d+d'} = (u \circ \tau^{d'}) \circ \tau^d \in S$$

from Theorem 4.1.4 c), resulting the fact that $x \circ \tau^{d'} \in I_d(u \circ \tau^{d'})$ too.

b) We show that $Sol_{SC}$ is time variable and we give the next counterexample. For $u = 1, d = -1$ and $x = \chi_{[0,\infty)}$, the prerequisites of 6.5.1 is fulfilled under the form $1 = 1 \circ \tau^{-1} \in S$ and $\chi_{[0,\infty)} \in Sol_{SC}(1)$, but the conclusion is false, since $\chi_{[0,\infty)} \circ \tau^{-1} = \chi_{[-1,\infty)} \notin S$.

c) Let the time invariant DC's $i, j$. Then $i \vee j$ is time invariant and if $\forall u, i(u) \wedge j(u) \neq \emptyset$, then $i \wedge j$ is time invariant too.

**6.5.6 Theorem** Let $i$ a time invariant DC. We have the next equivalence:
$$\forall u, \forall x, \forall d \geq 0, x \in i(u) \Leftrightarrow x \circ \tau^d \in i(u \circ \tau^d)$$

**Proof** $\Rightarrow$ $u \circ \tau^d \in S$ and $x \in i(u)$ are both true, taking into account Theorem 4.1.4 c). We apply the time invariance of $i$.

$\Leftarrow$ $(u \circ \tau^d) \circ \tau^{-d} \in S$ and $x \circ \tau^d \in i(u \circ \tau^d)$ are true. By using 6.5.1 we get $(x \circ \tau^d) \circ \tau^{-d} \in i((u \circ \tau^d) \circ \tau^{-d})$.

### 6.6 Constancy

**6.6.1 Definition** A DC $i$ is called *constant*, if $d_r \geq 0, d_f \geq 0$ exist so that for any $u$ and any $x \in i(u)$ we have

$$\overline{x(t-0)} \cdot x(t) \leq \overline{u(t - d_r)} \tag{1}$$
$$x(t-0) \cdot \overline{x(t)} \leq \overline{u(t - d_f)} \tag{2}$$

If the previous property is not satisfied then $i$ is called *non-constant*.

**6.6.2 Examples** a) $I_d$ is constant, $d \geq 0$ because from $x(t) = u(t-d)$ we infer
$$\overline{x(t-0)} \cdot x(t) = \overline{u(t-d-0)} \cdot u(t-d) \leq \overline{u(t-d)}$$
$$x(t-0) \cdot \overline{x(t)} = u(t-d-0) \cdot \overline{u(t-d)} \leq \overline{u(t-d)}$$

b) Let $U \subset S$ and the DE's $i, j$ from which $i$ is constant. If defined, $i \wedge U$ and $i \wedge j$ are constant and in general a DC included in a constant DC is constant. If $j$ is constant, then the property $\forall u, i(u) \wedge j(u) = \emptyset$ makes $i \vee j$ be not constant in general due to the possibility $d_r^i \neq d_r^j$, respectively $d_f^i \neq d_f^j$, but if $\exists u, i(u) \wedge j(u) \neq \emptyset$, then $d_r, d_f$ uniquely exist so that $\forall x \in i(u) \wedge j(u)$ 6.6.1 (1), 6.6.1 (2) are fulfilled and in such circumstances $i \vee j$ is constant.

**6.6.3 Theorem** The deterministic DC's (see Theorem 6.3.4) defined by the next equations
$$x(t) = \bigcap_{\xi \in [t-d, t-d+m]} u(\xi) \tag{1}$$
$$x(t) = \bigcup_{\xi \in [t-d, t-d+m]} u(\xi) \tag{2}$$

where $0 \leq m \leq d$ are time invariant and constant.

**Proof** We give the proof for (1) and the proof for (2) is similar.

The time invariance Let $d' \in R$ arbitrary so that $u \circ \tau^{d'} \in S$. Then
$$(x \circ \tau^{d'})(t) = x(t-d') = \bigcap_{\xi \in [t-d-d', t-d-d'+m]} u(\xi) = \bigcap_{\xi+d' \in [t-d, t-d+m]} u(\xi) =$$
$$= \bigcap_{\xi \in [t-d, t-d+m]} u(\xi - d') = \bigcap_{\xi \in [t-d, t-d+m]} (u \circ \tau^{d'})(\xi) \qquad (3)$$
shows that $x \circ \tau^{d'} \in S$ and $x \circ \tau^{d'} \in i(u \circ \tau^{d'})$, i.e. $i$ defined by (1) is time invariant.

The constancy From 4.2.2 we get
$$\overline{x(t-0)} \cdot x(t) = \overline{u(t-d-0)} \cdot \bigcap_{\xi \in [t-d, t-d+m]} u(\xi) \le u(t-d)$$
$$x(t-0) \cdot \overline{x(t)} = u(t-d-0) \cdot \overline{\bigcap_{\xi \in [t-d, t-d+m]} u(\xi)} \cdot \overline{u(t-d+m)} \le \overline{u(t-d+m)}$$

**6.6.4 Remark** Constancy means that $x$ is allowed to switch only if $u$ has anticipated that possibility $d_r$, respectively $d_f$ time units before. It does not imply the uniqueness of $d_r, d_f$ and at 6.6.3 (1) we have an example when $d_r$ is not unique.

**6.7 Rising-Falling Symmetry**

**6.7.1 Definition** The DC $i$ is called (*rising-falling*) *symmetrical* if one of the next equivalent properties is true:
   a) $\forall u, i(\overline{u}) = \{\overline{x} \mid x \in i(u)\}$
   b) $\forall u, \forall x, x \in i(u) \Leftrightarrow \overline{x} \in i(\overline{u})$
and respectively (*rising-falling*) *asymmetrical* otherwise.

**6.7.2 Examples** a) $I_d$ with $d \ge 0$ arbitrary is symmetrical because
$$\overline{x} \in I_d(\overline{u}) \Leftrightarrow \overline{x(t)} = \overline{u(t-d)} \Leftrightarrow x(t) = u(t-d) \Leftrightarrow x \in I_d(u)$$
is true for all $u$ and all $x$.

b) $Sol_{SC}$ is symmetrical also and let for this an arbitrary $u$. If $\lim_{t \to \infty} u(t)$ does not exist, then $\lim_{t \to \infty} \overline{u(t)}$ does not exist and
$$Sol_{SC}(\overline{u}) = Sol_{SC}(u) = S$$
from where we infer the desired property. Let us suppose now that $\lim_{t \to \infty} u(t)$ exists, thus $\lim_{t \to \infty} \overline{u(t)}$ exists and we note with $c, \overline{c} \in B$ these limits. We take an $x \in Sol_{SC}(u)$ and we have the existence of $d \ge 0$ with $x(t) = x(t) \cdot \chi_{(-\infty,d)}(t) \oplus c \cdot \chi_{[d,\infty)}(t)$. Because $\overline{x(t)} \cdot \chi_{(-\infty,d)}(t) \oplus \overline{c} \cdot \chi_{[d,\infty)}(t) \in Sol_{SC}(\overline{u})$, we infer $\overline{x} \in Sol_{SC}(\overline{u})$ resulting that $x \in Sol_{SC}(u) \Rightarrow \overline{x} \in Sol_{SC}(\overline{u})$. The inverse implication is similarly proved.

c) The DC's defined at 6.2.3 b) are asymmetrical. We suppose for the first of them that $x \in i(u)$, i.e. $d \ge 0$ exists so that $x(t) = u(t) \cdot \chi_{[d,\infty)}(t)$. Because $\overline{x(t)} = \chi_{(-\infty,d)}(t) \oplus \overline{u(t)} \cdot \chi_{[d,\infty)}(t) \notin i(\overline{u})$, we infer that the symmetry is not fulfilled.

d) Let $U \subset S$ with $x \in U \Rightarrow \overline{x} \in U$ and $i, j$ symmetrical DC's. Then $i \vee j$ is symmetrical and, when defined, the DC's $i \wedge U$ and $i \wedge j$ are symmetrical too.

## 6.8 Serial Connection

**6.8.1 Definition** For the DC's $i, j$ we note with $k = i \circ j$ the function $k: S \to P^*(S)$ defined in the next way
$$k(u) = \{y \mid \exists x, x \in j(u) \text{ and } y \in i(x)\}$$
$k$ is called the *serial connection* of the DC's $i$ and $j$.

**6.8.2 Theorem** (*the compatibility between the serial connection and consistency, stability, determinism, time invariance and symmetry*)
    a) $k$ is a DC.
    b) If $i, j$ are deterministic, then $k$ is deterministic.
    c) If $i, j$ are time invariant, then $k$ is time invariant.
    d) If $i, j$ are symmetrical, then $k$ is symmetrical.

**Proof** a) Let $u$ arbitrary for which $j(u) \neq \emptyset$ shows that $x \in j(u)$ exists and $i(x) \neq \emptyset$ shows that $y \in i(x)$ exists, thus $y \in k(u)$ exists. On the other hand, if $\lim_{t \to \infty} u(t)$ exists, then $\lim_{t \to \infty} x(t)$ exists and $\lim_{t \to \infty} y(t)$ exists also and the three limits are equal.

    b) The determinism of $k$ is equivalent with the fact that the composition of the $S \to S$ functions (see Remark 6.3.2) is a $S \to S$ function.

    c) We must show that for any $u, y \in S, d \in \mathbf{R}$ we have the implication
$$(u \circ \tau^d \in S \text{ and } y \in k(u)) \Rightarrow (y \circ \tau^d \in S \text{ and } y \circ \tau^d \in k(u \circ \tau^d))$$
true. If $y \in k(u)$, let $x \in S$ whose existence is guaranteed by the definition of $k$ so that $x \in j(u)$ and $y \in i(x)$. By applying 6.5.1 to $j$, because $u \circ \tau^d \in S$, we have that $x \circ \tau^d \in S$ and $x \circ \tau^d \in j(u \circ \tau^d)$. We apply again 6.5.1 to $i$ and we obtain $y \circ \tau^d \in S$ and $y \circ \tau^d \in i(x \circ \tau^d)$. Thus $y \circ \tau^d \in k(u \circ \tau^d)$.

    d) $\overline{y} \in k(\overline{u}) \Leftrightarrow \exists \overline{x}, \overline{x} \in j(\overline{u}) \text{ and } \overline{y} \in i(\overline{x}) \Leftrightarrow \exists x, x \in j(u) \text{ and } y \in i(x) \Leftrightarrow y \in k(u)$

**6.8.3 Counterexample** showing that the serial connection $k$ of the constant DC's $i, j$ is not necessarily constant and we consider for this the DC defined by
$$i(u) = Sol_{SC}(u) \wedge \{x \mid \overline{x(t-0)} \cdot x(t) \leq u(t - d_r) \text{ and } x(t-0) \cdot \overline{x(t)} \leq \overline{u(t - d_f)}\}$$
where $d_r \geq 0, d_f \geq 0$. $i$ is obviously constant.

    Let $u = \chi_{[\tau, \infty)}$ for which $1 \in i(u)$ and $\chi_{[d, \infty)} \in i(1)$ take place for any $\tau \geq 0$ and any $d \geq 0$. In order that $k = i \circ i$ be constant, $d_r'' \geq 0$ must exist so that for any $u = \chi_{[\tau, \infty)}$, the signal $y = \chi_{[d, \infty)}$ should satisfy
$$\chi_{\{d\}}(t) = \overline{y(t-0)} \cdot y(t) \leq u(t - d_r'') = \chi_{[d_r'' + \tau, \infty)}(t) \quad (1)$$

Let such a $d_r''$ fixed; by choosing $d$ and $\tau$ so that $d < d_r'' + \tau$ be true, (1) is false. $k$ is not constant.

**6.8.4 Remark** The set of the DC's is a non-commutative semi-group relative to the serial connection having the unit $I$:
$$i \circ I = I \circ i = i$$
Three of its important sub-semi-groups are represented by the deterministic, the time invariant and respectively the symmetrical DC's.

**6.8.5 Theorem** (*the compatibility between the serial connection and the order*) Let the DC's $i, j, k$. The next implications are true:
$$i \subset j \Rightarrow i \circ k \subset j \circ k$$
$$j \subset k \Rightarrow i \circ j \subset i \circ k$$

**Proof** Let $u$ and $y \in (i \circ k)(u)$, meaning that some $x$ exists with $x \in k(u)$ and $y \in i(x)$; because $y \in j(x)$, we obtain $y \in (j \circ k)(u)$.

On the other hand, if we suppose that $y \in (i \circ j)(u)$, then $x$ exists so that $x \in j(u)$ and $y \in i(x)$. We get $x \in k(u)$ implying $y \in (i \circ k)(u)$.

**6.8.6 Theorem** (*the compatibility between the serial connection and* $\wedge, \vee$) Let $U \subset S$ and the DC's $i, j, k$.

a) If $\forall u, i(u) \wedge U \neq \emptyset$, then $\forall u, (i \circ j)(u) \wedge U \neq \emptyset$ and
$$(i \wedge U) \circ j = (i \circ j) \wedge U$$
If $\forall u, j(u) \wedge U \neq \emptyset$, then we have
$$i \circ (j \wedge U) \subset i \circ j$$

b) If $\forall u, i(u) \wedge j(u) \neq \emptyset$, then $\forall u, (i \circ k)(u) \wedge (j \circ k)(u) \neq \emptyset$ and
$$(i \wedge j) \circ k \subset (i \circ k) \wedge (j \circ k)$$
If $\forall u, j(u) \wedge k(u) \neq \emptyset$, then $\forall u, (i \circ j)(u) \wedge (i \circ k)(u) \neq \emptyset$ and
$$i \circ (j \wedge k) \subset (i \circ j) \wedge (i \circ k)$$

c) We have
$$(i \vee j) \circ k = (i \circ k) \vee (j \circ k)$$
$$i \circ (j \vee k) = (i \circ j) \vee (i \circ k)$$

**Proof** a) $\forall u, ((i \wedge U) \circ j)(u) = \{y \mid \exists x, y \in i(x) \text{ and } y \in U \text{ and } x \in j(u)\} = ((i \circ j) \wedge U)(u)$

$\forall u, (i \circ (j \wedge U))(u) = \{y \mid \exists x, y \in i(x) \text{ and } x \in j(u) \text{ and } x \in U\} \subset$
$\subset \{y \mid \exists x, y \in i(x) \text{ and } x \in j(u)\} = (i \circ j)(u)$

b) $\forall u, ((i \wedge j) \circ k)(u) = \{y \mid \exists x, y \in i(x) \text{ and } y \in j(x) \text{ and } x \in k(u)\} \subset$
$\subset \{y \mid \exists x, \exists x', y \in i(x) \text{ and } x \in k(u) \text{ and } y \in j(x') \text{ and } x' \in k(u)\} = ((i \circ k) \wedge (j \circ k))(u)$

$\forall u, (i \circ (j \wedge k))(u) = \{y \mid \exists x, y \in i(x) \text{ and } x \in j(u) \text{ and } x \in k(u)\} \subset$
$\subset \{y \mid \exists x, \exists x', y \in i(x) \text{ and } x \in j(u) \text{ and } y \in i(x') \text{ and } x' \in k(u)\} = ((i \circ j) \wedge (i \circ k))(u)$

c) $\forall u, ((i \vee j) \circ k)(u) = \{y \mid \exists x, (y \in i(x) \text{ or } y \in j(x)) \text{ and } x \in k(u)\} =$
$= \{y \mid \exists x, (y \in i(x) \text{ and } x \in k(u)) \text{ or } (y \in j(x) \text{ and } x \in k(u))\} = ((i \circ k) \vee (j \circ k))(u)$

$\forall u, (i \circ (j \vee k))(u) = \{y \mid \exists x, y \in i(x) \text{ and } (x \in j(u) \text{ or } x \in k(u))\} =$
$= \{y \mid \exists x, (y \in i(x) \text{ and } x \in j(u)) \text{ or } (y \in i(x) \text{ and } x \in k(u))\} = ((i \circ j) \vee (i \circ k))(u)$

# 7. Bounded Delays

## 7.1 The Consistency Condition

**7.1.1 Theorem** Let $0 \leq m_r \leq d_r, 0 \leq m_f \leq d_f$ be given. The following statements are equivalent:

a) $$\forall u \in S, \bigcap_{\xi \in [t-d_r, t-d_r+m_r]} u(\xi) \leq \bigcup_{\xi \in [t-d_f, t-d_f+m_f]} u(\xi)$$

b) $$\forall u \in S, \bigcap_{\xi \in [t-d_r, t-d_r+m_r]} u(\xi) \cdot \bigcap_{\xi \in [t-d_f, t-d_f+m_f]} \overline{u(\xi)} = 0$$

    c) $\quad\quad\quad\quad\quad\quad d_r - m_r \leq d_f \text{ and } d_f - m_f \leq d_r$

    d) $\quad\quad\quad \forall u \in S, \exists x \in S \quad \bigcap_{\xi \in [t-d_r, t-d_r+m_r]} u(\xi) \leq x(t) \leq \bigcup_{\xi \in [t-d_f, t-d_f+m_f]} u(\xi)$

**Proof** $a) \Leftrightarrow b)$: $\forall u, \bigcap_{\xi \in [t-d_r, t-d_r+m_r]} u(\xi) \leq \bigcup_{\xi \in [t-d_f, t-d_f+m_f]} u(\xi)$

$\Leftrightarrow \forall u, \overline{\bigcap_{\xi \in [t-d_r, t-d_r+m_r]} u(\xi)} \cup \bigcup_{\xi \in [t-d_f, t-d_f+m_f]} u(\xi) = 1$

$\Leftrightarrow \forall u, \overline{\bigcap_{\xi \in [t-d_r, t-d_r+m_r]} u(\xi)} \cdot \overline{\bigcap_{\xi \in [t-d_f, t-d_f+m_f]} \overline{u(\xi)}} = 1$

$\Leftrightarrow \forall u, \bigcap_{\xi \in [t-d_r, t-d_r+m_r]} u(\xi) \cdot \bigcap_{\xi \in [t-d_f, t-d_f+m_f]} \overline{u(\xi)} = 0$

$b) \Leftrightarrow c)$: $\forall u, \bigcap_{\xi \in [t-d_r, t-d_r+m_r]} u(\xi) \cdot \bigcap_{\xi \in [t-d_f, t-d_f+m_f]} \overline{u(\xi)} = 0$

$\Leftrightarrow \forall t, [t-d_r, t-d_r+m_r] \wedge [t-d_f, t-d_f+m_f] \neq \emptyset$

($\Rightarrow$ if $\exists t, [t-d_r, t-d_r+m_r] \wedge [t-d_f, t-d_f+m_f] = \emptyset$, then $u$ exists so that $\forall \xi \in [t-d_r, t-d_r+m_r], u(\xi) = 1$ and $\forall \xi \in [t-d_f, t-d_f+m_f], u(\xi) = 0$ contradiction with the hypothesis; $\Leftarrow$ if $\forall t, [t-d_r, t-d_r+m_r] \wedge [t-d_f, t-d_f+m_f] \neq \emptyset$, then $\forall u, \forall t$, $\forall \xi \in [t-d_r, t-d_r+m_r] \wedge [t-d_f, t-d_f+m_f]$, $u(\xi) = 1$ and $u(\xi) = 0$ cannot be both true and the conclusion results)

$\Leftrightarrow \forall t, \neg(t - d_r + m_r < t - d_f \text{ or } t - d_f + m_f < t - d_r)$

$\Leftrightarrow \forall t, t - d_r + m_r \geq t - d_f \text{ and } t - d_f + m_f \geq t - d_r$

$\Leftrightarrow d_r - m_r \leq d_f \text{ and } d_f - m_f \leq d_r$

$a) \Leftrightarrow d)$ is obvious if we take into account the fact that $\bigcap_{\xi \in [t-d_r, t-d_r+m_r]} u(\xi)$, $\bigcup_{\xi \in [t-d_f, t-d_f+m_f]} u(\xi)$ are signals, see Theorem 4.2.3.

**7.1.2 Definition** Any of the properties 7.1.1 a),…,7.1.1 d) is called the *consistency condition* (of the bounded delay condition) (CC$_{BDC}$).

**7.1.3 Remark** Let us see some special cases of satisfaction of CC$_{BDC}$. If $d_r = d_f = d$ and $m_r = m_f = m$, CC$_{BDC}$ is fulfilled under the form $d \geq d - m$. CC$_{BDC}$ is also fulfilled if $m_r = d_r$ and $m_f = d_f$. If $m_r = m_f = 0$, then the satisfaction of CC$_{BDC}$ is equivalent with $d_r = d_f$.

## 7.2 Bounded Delays

**7.2.1 Theorem** The next system

$$\bigcap_{\xi \in [t-d_r, t-d_r+m_r]} u(\xi) \leq x(t) \leq \bigcup_{\xi \in [t-d_f, t-d_f+m_f]} u(\xi) \tag{1}$$

where $u, x \in S$ and $0 \leq m_r \leq d_r$, $0 \leq m_f \leq d_f$ defines a DC if and only if CC$_{BDC}$ is satisfied.

**Proof** If $CC_{BDC}$ is true, then (1) has solutions $x$ for any $u$ and all the solutions satisfy $x \in Sol_{SC}(u)$, see Theorem 6.3.4. If $CC_{BDC}$ is not fulfilled, then some $u$ exists so that (1) has no solutions.

**7.2.2 Definition** We suppose that $CC_{BDC}$ is true: $d_r - m_r \leq d_f, d_f - m_f \leq d_r$. The system 7.2.1 (1) is called the *bounded delay condition* (BDC); $u, x$ are the *input* (or the *control*), respectively the *state* (or the *output*); $m_r, m_f$ are the (*rising, falling*) *memories* (or *thresholds for cancellation*) and $d_r, d_f$ are the (*rising, falling*) *upper bounds of the* (*transmission*) *delays* (*for transitions*). The differences $d_f - m_f$, respectively $d_r - m_r$ are called the (*rising, falling*) *lower bounds of the* (*transmission*) *delays* (*for transitions*). We say that the tuple $(u, m_r, d_r, m_f, d_f, x)$ satisfies BDC.

We also call bounded delay condition the function $Sol_{BDC}^{m_r, d_r, m_f, d_f} : S \to P^*(S)$ defined by

$$Sol_{BDC}^{m_r, d_r, m_f, d_f}(u) = \{x \mid (u, m_r, d_r, m_f, d_f, x) \text{ satsifies BDC}\}$$

**7.2.3 Theorem** a) The next system

$$\bigcap_{\xi \in [t-d_{r,\max}, t-d_{f,\min}]} u(\xi) \leq x(t) \leq \bigcup_{\xi \in [t-d_{f,\max}, t-d_{r,\min}]} u(\xi) \tag{1}$$

where $u, x \in S$, $0 \leq d_{r,\min} \leq d_{r,\max}$, $0 \leq d_{f,\min} \leq d_{f,\max}$ defines a DC if and only if

$$d_{r,\min} \leq d_{f,\max} \tag{2}$$

$$d_{f,\min} \leq d_{r,\max} \tag{3}$$

b) When (2), (3) are fulfilled, the system (1) and BDC from 7.2.1 (1) are equivalent, in the sense that by a suitable choice of the parameters $m_r, d_r, m_f, d_f, d_{r,\min}, d_{r,\max}, d_{f,\min}, d_{f,\max}$ their solutions coincide for all $u$.

**Proof** a) <u>If</u> (2), (3) show that the sets $[t - d_{r,\max}, t - d_{f,\min}], [t - d_{f,\max}, t - d_{r,\min}]$ are non-empty and the inequalities $0 \leq d_{r,\min} \leq d_{r,\max}$, $0 \leq d_{f,\min} \leq d_{f,\max}$ show that the meet $[t - d_{r,\max}, t - d_{f,\min}] \wedge [t - d_{f,\max}, t - d_{r,\min}]$ is non-empty because

$$t - d_{f,\max} \leq t - d_{f,\min} \text{ and } t - d_{r,\max} \leq t - d_{r,\min}$$

thus (1) has a solution for any $u$, similarly with the proof of 7.1.1. It is shown that for any $u$ and any solution $x$ of (1) we have $x \in Sol_{SC}(u)$.

<u>Only if</u> If $d_{r,\min} > d_{f,\max}$ then $[t - d_{f,\max}, t - d_{r,\min}] = \emptyset$ shows that $\forall t, x(t) = 0$ in (1) implying the existence of some $u$ so that (1) has no solutions. Similarly for $d_{f,\min} > d_{r,\max}$.

b) We define

$$d_r = d_{r,\max} \tag{4}$$

$$d_f = d_{f,\max} \tag{5}$$

$$m_r = d_{r,\max} - d_{f,\min} \tag{6}$$

$$m_f = d_{f,\max} - d_{r,\min} \tag{7}$$

**7.2.4 Theorem** (*the property of compatibility between the initial values of the input and of the*

*output for BDC*) If $x \in Sol_{BDC}^{m_r,d_r,m_f,d_f}(u)$, then $x(0-0) = u(0-0)$.

**Proof** $u(0-0) = 0$ is equivalent with $u(t) \leq \chi_{[0,\infty)}(t)$. We have in this case

$$x(t) \leq \bigcup_{\xi \in [t-d_f, t-d_f+m_f]} u(\xi) \leq \bigcup_{\xi \in [t-d_f, t-d_f+m_f]} \chi_{[0,\infty)}(\xi) = \chi_{[d_f-m_f,\infty)}(t) \leq \chi_{[0,\infty)}(t)$$

i.e. $x(0-0) = 0$.

$u(0-0) = 1$ is equivalent with $u(t) \geq \chi_{(-\infty,0)}(t)$ and we have

$$x(t) \geq \bigcap_{\xi \in [t-d_r, t-d_r+m_r]} u(\xi) \geq \bigcap_{\xi \in [t-d_r, t-d_r+m_r]} \chi_{(-\infty,0)}(\xi) = \chi_{(-\infty, d_r-m_r)}(t) \geq \chi_{(-\infty,0)}(t)$$

i.e. $x(0-0) = 1$.

**7.2.5 Theorem** $Sol_{BDC}^{m_r,d_r,m_f,d_f}(c) = \{c\}, \forall c \in \mathbf{B}$.

**Proof** If $\forall t, u(t) = c$, then $\forall t$, $\bigcap_{\xi \in [t-d_r, t-d_r+m_r]} u(\xi) = \bigcup_{\xi \in [t-d_f, t-d_f+m_f]} u(\xi) = c$.

**7.2.6 Remark** We analyze BDC in the hypothesis that $m_r > 0$ in the following manner.

a) $x(0-0) = 0$ and $u(t) = \chi_{[0,\tau)}(t)$, with $0 < \tau \leq m_r$, i.e. we apply at the input a 'short', 'insufficiently persistent' 1-pulse. The situation is the one from the next figure, where we remark that $\bigcap_{\xi \in [t-d_r, t-d_r+m_r]} u(\xi)$ is identically null.

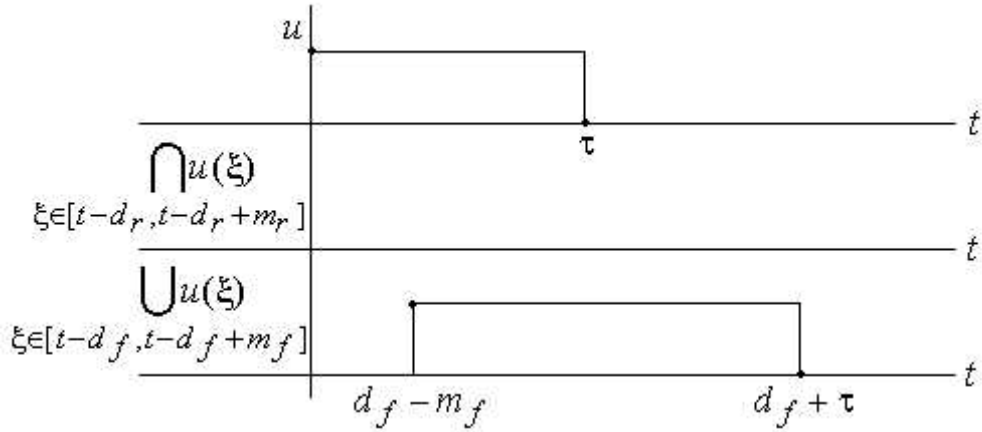

Fig 9

$$x(t) = x(t) \cdot \chi_{[d_f-m_f, d_f+\tau)}(t)$$

$t \in (-\infty, d_f - m_f)$; $x(t) = 0$ necessarily, the pulse from the input did not alter the output, that keeps the initial value

$t \in [d_f - m_f, d_f + \tau)$; $x(t) = 0$ or $x(t) = 1$, 0 and 1 are both possible values of the output, that could be altered by the pulse from the input

$t \in [d_f + \tau, \infty)$; $x(t) = 0$ necessarily, the value 1 from the input cannot alter the output any longer

We observe that the 1-pulses of length shorter than or equal to $m_r$ (the dual situation is true also) at the input are not necessarily transmitted to the output, thus BDC includes the special case when such pulses are surely not transmitted. The last property strengthens the inertial character of BDC.

b) $x(0-0) = 0$ and $u(t) = \chi_{[0,\tau)}(t)$, with $\tau > m_r$, in other words a *'long'*, *'persistent'* 1-pulse is applied at the input. This corresponds to the drawing in Fig 10, where $\bigcap_{\xi \in [t-d_r, t-d_r+m_r]} u(\xi)$ is not null:

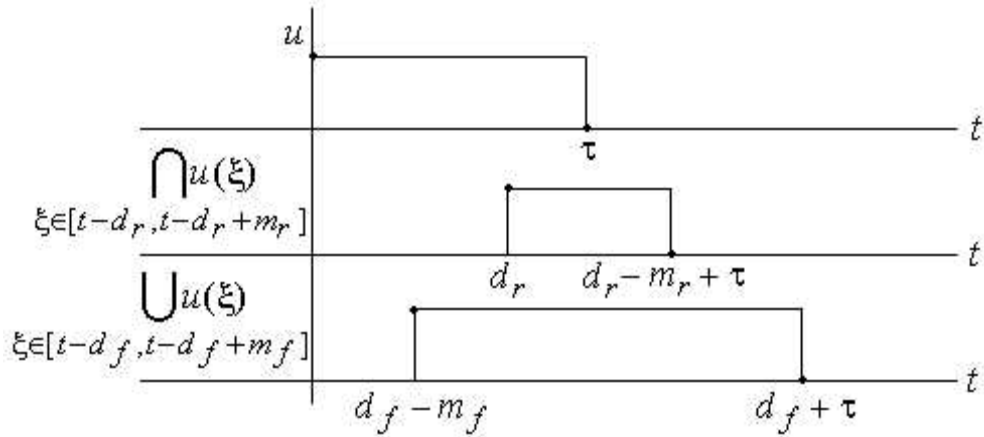

Fig 10

$$x(t) = x(t) \cdot \chi_{[d_f-m_f, d_r)}(t) \oplus \chi_{[d_r, d_r-m_r+\tau)}(t) \oplus x(t) \cdot \chi_{[d_r-m_r+\tau, d_f+\tau)}(t)$$

$t \in (-\infty, d_f - m_f)$; $x(t) = 0$ necessarily, the input pulse did not alter the output

$t \in [d_f - m_f, d_r)$; $x(t) = 0$ or $x(t) = 1$, 0 and 1 are possible values of the output, that could be altered by the input pulse

$t \in [d_r, d_r - m_r + \tau)$; $x(t) = 1$ necessarily, the persistent input pulse has propagated to the output

$t \in [d_r - m_r + \tau, d_f + \tau)$; $x(t) = 0$ or $x(t) = 1$, 0 and 1 are possible values of $x$, that could still be influenced by the input pulse

$t \in [d_f + \tau, \infty)$; $x(t) = 0$ necessarily, the state of balance=stability when $x$ depends only on the value $\lim_{t \to \infty} u(t) = 0$ and not on the input pulse.

We conclude that if $u$ is 1 for strictly more than $m_r$ time units, then $x$ becomes 1 with a delay of $d \in [d_f - m_f, d_r]$ time units and, in a dual manner, that if $u$ is 0 for strictly more than $m_f$ time units then $x$ becomes 0 with a delay of $d \in [d_r - m_r, d_f]$ time units. Fig 10 best shows why $d_f - m_f = d_{r,\min}$ and $d_r = d_{r,\max}$, see 7.2.3 (4),…, 7.2.3 (7), were previously called the rising lower bound and upper bound of the delays, together with the dual taxonomy.

This analysis has shown, by the use of the words *possible* and *necessary* that was made on purpose, the existence of interference not only with temporal logic –this was obvious from the very beginning- but also with modal logic. The idea is natural to all these authors that consider temporal logic be a *'modal logic aiming to express time dependent processes'*, together with Mihaela Malita and Mircea Malita ('*The Foundations of Artificial Intelligence, 1 Propositional Logics*', ed. Tehnica, Bucuresti, 1987, in Romanian).

Two interpretations of BDC exist, that we shall call the positive and the negative interpretation.

The *positive interpretation* of BDC is the following: it is natural that the value of the output be arbitrary when the input is not sufficiently persistent (in the previous case a) for $t \in [d_f - m_f, d_f + \tau)$ and in the previous case b), for $t \in [d_f - m_f, d_r) \vee [d_r - m_r + \tau, d_f + \tau))$. This fact could also be modeled by the replacement of the set $\{0,1\}$ with $\{0, \frac{1}{2}, 1\}$ -some authors do so- but this has algebraical disadvantages, because the set $\{0, \frac{1}{2}, 1\}$ is poorer algebraically than $\{0,1\}$. Non-determinism is another way (ours!) of solving this problem.

The *negative interpretation* of BDC: it is not natural that at some time moment $t_1$ we should have $x(t_1 - 0) \cdot \overline{x(t_1)} = 1$, while $\forall t \leq t_1, u(t-0) \cdot \overline{u(t)} = 0$ in the sense that $x$ switches in a manner that is not compatible with the (more or less) (left) local behavior of $u$.

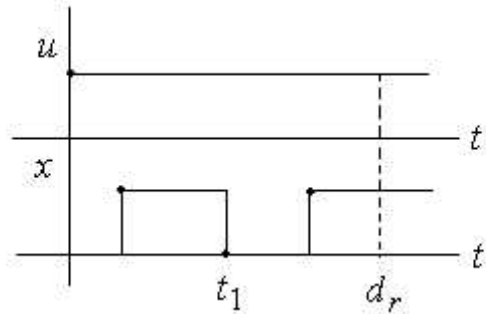

Fig 11

In other words, pulses on the output may exist that do not reproduce the pulses on the input.

### 7.3 The Properties of the Bounded Delays

**7.3.1 Theorem** (*The representation of the solutions of BDC*). Let $u, x \in S$ and the numbers $0 \leq m_r \leq d_r, 0 \leq m_f \leq d_f$ so that CC$_{BDC}$ is fulfilled. The following statements are equivalent:

a) $x \in Sol_{BDC}^{m_r, d_r, m_f, d_f}(u)$

b) $\exists y \in S$ so that $x$ satisfies

$$x(t) = \bigcap_{\xi \in [t-d_r, t-d_r+m_r]} u(\xi) \cup y(t) \cdot \bigcup_{\xi \in [t-d_f, t-d_f+m_f]} u(\xi)$$

**Proof** We can take $y(t) = x(t)$.

**7.3.2 Remark** Let $0 \leq m_r \leq d_r, 0 \leq m_f \leq d_f$ and $0 \leq m'_r \leq d'_r, 0 \leq m'_f \leq d'_f$ so that CC$_{BDC}$ is fulfilled twice: $d_r \geq d_f - m_f, d_f \geq d_r - m_r$, respectively $d'_r \geq d'_f - m'_f, d'_f \geq d'_r - m'_r$. The next statements are equivalent:

a) $\forall u, Sol_{BDC}^{m_r,d_r,m_f,d_f}(u) \wedge Sol_{BDC}^{m_r',d_r',m_f',d_f'}(u) \neq \emptyset$

b) $d_r' \geq d_f - m_f, d_f' \geq d_r - m_r, d_r \geq d_f' - m_f', d_f \geq d_r' - m_r'$

and if one of them is satisfied, then $Sol_{BDC}^{m_r,d_r,m_f,d_f} \wedge Sol_{BDC}^{m_r',d_r',m_f',d_f'}$ exists under the form

$$\bigcap_{\xi \in [t-d_r, t-d_r+m_r]} u(\xi) \cup \bigcap_{\xi \in [t-d_r', t-d_r'+m_r']} u(\xi) \leq x(t) \leq \bigcup_{\xi \in [t-d_f, t-d_f+m_f]} u(\xi) \cdot \bigcup_{\xi \in [t-d_f', t-d_f'+m_f']} u(\xi)$$

but it is not a BDC. On the other hand $Sol_{BDC}^{m_r,d_r,m_f,d_f} \vee Sol_{BDC}^{m_r',d_r',m_f',d_f'}$ exists without other consistency requirements than those represented by the validity of the two $CC_{BDC}$'s and it is not a BDC either:

$$\bigcap_{\xi \in [t-d_r, t-d_r+m_r]} u(\xi) \cdot \bigcap_{\xi \in [t-d_r', t-d_r'+m_r']} u(\xi) \leq x(t) \leq \bigcup_{\xi \in [t-d_f, t-d_f+m_f]} u(\xi) \cup \bigcup_{\xi \in [t-d_f', t-d_f'+m_f']} u(\xi)$$

**7.3.3 Theorem** Let $0 \leq m_r \leq d_r, 0 \leq m_f \leq d_f$ so that $CC_{BDC}$ is true. The next statements are equivalent:

a) $Sol_{BDC}^{m_r,d_r,m_f,d_f}$ is deterministic

b) the upper bounds and the lower bounds of the delays coincide
$$d_r = d_f - m_f, d_f = d_r - m_r$$

c) the memories are null
$$m_r = m_f = 0$$

d) the bounded delay degenerates in a translation
$$\exists d \geq 0, Sol_{BDC}^{m_r,d_r,m_f,d_f} = I_d$$

**Proof** $a) \Rightarrow b)$ The hypothesis states that $Sol_{BDC}^{m_r,d_r,m_f,d_f}$ has exactly one element, i.e.

$$\forall u, \bigcap_{\xi \in [t-d_r, t-d_r+m_r]} u(\xi) = \bigcup_{\xi \in [t-d_f, t-d_f+m_f]} u(\xi) \tag{1}$$

We give $u$ the values $\chi_{[0,\infty)}, \chi_{(-\infty,0)}$ and we obtain

$$\bigcap_{\xi \in [t-d_r, t-d_r+m_r]} \chi_{[0,\infty)}(\xi) = \chi_{[d_r,\infty)}(t) \tag{2}$$

$$\bigcup_{\xi \in [t-d_f, t-d_f+m_f]} \chi_{[0,\infty)}(\xi) = \chi_{[d_f-m_f,\infty)}(t) \tag{3}$$

$$\bigcap_{\xi \in [t-d_r, t-d_r+m_r]} \chi_{(-\infty,0)}(\xi) = \chi_{(-\infty, d_r-m_r)}(t) \tag{4}$$

$$\bigcup_{\xi \in [t-d_f, t-d_f+m_f]} \chi_{(-\infty,0)}(\xi) = \chi_{(-\infty, d_f)}(t) \tag{5}$$

(1) implies the equality of the functions from (2) and (3) and of the functions from (4) and (5) and this indicates the satisfaction of b).

$b) \Rightarrow c)$ We add the two relations term by term and we get $m_r + m_f = 0$. c) is true.

$c) \Rightarrow d)$ From the hypothesis $c)$ and from $CC_{BDC}$ we infer $d_r \geq d_f, d_f \geq d_r$ i.e. $d_r = d_f = d$ and 7.2.1 (1) becomes

$$u(t-d) \le x(t) \le u(t-d)$$

in other words $Sol_{BDC}^{0,d,0,d}(u) = \{u(t-d)\}$.

$d) \Rightarrow a)$ Obvious, since $I_d$ is deterministic.

**7.3.4 Theorem** Let $0 \le m_r \le d_r, 0 \le m_f \le d_f$ and $0 \le m_r' \le d_r', 0 \le m_f' \le d_f'$ so that $CC_{BDC}$ is fulfilled for each of them. The next statements are equivalent

a) $$Sol_{BDC}^{m_r,d_r,m_f,d_f} \subset Sol_{BDC}^{m_r',d_r',m_f',d_f'}$$

b) $\forall u,$
$$\bigcap_{\xi \in [t-d_r',t-d_r'+m_r']} u(\xi) \le \bigcap_{\xi \in [t-d_r,t-d_r+m_r]} u(\xi) \le \bigcup_{\xi \in [t-d_f,t-d_f+m_f]} u(\xi) \le \bigcup_{\xi \in [t-d_f',t-d_f'+m_f']} u(\xi)$$

c) $$[t-d_r', t-d_r' + m_r'] \supset [t-d_r, t-d_r + m_r]$$
$$[t-d_f, t-d_f + m_f] \subset [t-d_f', t-d_f' + m_f']$$

d) $$d_r' - m_r' \le d_r - m_r \le d_f \le d_f'$$
$$d_f' - m_f' \le d_f - m_f \le d_r \le d_r'$$

**Proof** Obvious.

**7.3.5 Theorem** $Sol_{BDC}^{m_r,d_r,m_f,d_f}$ is time invariant.

**Proof** We choose $u, x \in S$ and $d \in \mathbf{R}$ arbitrary so that $u \circ \tau^d \in S$ and $x \in Sol_{BDC}^{m_r,d_r,m_f,d_f}(u)$ be true. From the supposition that $u(0-0) = 0$, i.e. $u \circ \tau^d(t) \le \chi_{[0,\infty)}(t)$ and from

$$(x \circ \tau^d)(t) \le \bigcup_{\xi \in [t-d_f,t-d_f+m_f]} (u \circ \tau^d)(\xi) \le \bigcup_{\xi \in [t-d_f,t-d_f+m_f]} \chi_{[0,\infty)}(\xi) = \chi_{[d_f-m_f,\infty)}(t) \le \chi_{[0,\infty)}(t)$$

we infer $supp\, D_{01}(x \circ \tau^d) \subset [0,\infty)$; and if $u(0-0) = 1$, i.e. $u \circ \tau^d(t) \ge \chi_{(-\infty,0)}(t)$, then

$$(x \circ \tau^d)(t) \ge \bigcap_{\xi \in [t-d_r,t-d_r+m_r]} (u \circ \tau^d)(\xi) \ge \bigcap_{\xi \in [t-d_r,t-d_r+m_r]} \chi_{(-\infty,0)}(\xi) = \chi_{(-\infty,d_r-m_r)}(t) \ge \chi_{(-\infty,0)}(t)$$

from where $supp\, D_{10}(x \circ \tau^d) \subset [0,\infty)$. Because

$$supp\, D(x \circ \tau^d) = supp\, D_{01}(x \circ \tau^d) \vee supp\, D_{10}(x \circ \tau^d) \subset [0,\infty)$$

we conclude that $x \circ \tau^d \in S$. The inequalities

$$\bigcap_{\xi \in [t-d_r,t-d_r+m_r]} (u \circ \tau^d)(\xi) \le (x \circ \tau^d)(t) \le \bigcup_{\xi \in [t-d_f,t-d_f+m_f]} (u \circ \tau^d)(\xi)$$

that we have already made use of show that $x \circ \tau^d \in Sol_{BDC}^{m_r,d_r,m_f,d_f}(u \circ \tau^d)$.

**7.3.6 Theorem** $Sol_{BDC}^{m_r,d_r,m_f,d_f}$ is symmetrical if and only if $d_r = d_f$ and $m_r = m_f$.

**Proof** <u>If</u> We note with $d$ the common value of $d_r = d_f$ and with $m$ the common value of $m_r = m_f$. We get

$$\overline{x} \in Sol_{BDC}^{m,d,m,d}(\overline{u}) \Leftrightarrow \bigcap_{\xi \in [t-d,t-d+m]} \overline{u(\xi)} \le \overline{x(t)} \le \bigcup_{\xi \in [t-d,t-d+m]} \overline{u(\xi)}$$

$$\Leftrightarrow \overline{\bigcup_{\xi\in[t-d,t-d+m]}u(\xi)} \leq \overline{x(t)} \leq \overline{\bigcap_{\xi\in[t-d,t-d+m]}u(\xi)}$$

$$\Leftrightarrow \bigcap_{\xi\in[t-d,t-d+m]}u(\xi) \leq x(t) \leq \bigcup_{\xi\in[t-d,t-d+m]}u(\xi) \Leftrightarrow x \in Sol_{BDC}^{m,d,m,d}(u)$$

<u>Only if</u> For $u = \chi_{[0,\tau)}$ with $\tau > m_r, \tau > m_f$, we have

$$\bigcap_{\xi\in[t-d_r,t-d_r+m_r]}\overline{u(\xi)} = \chi_{(-\infty,d_r-m_r)\vee[\tau+d_r,\infty)}(t)$$

$$\bigcup_{\xi\in[t-d_f,t-d_f+m_f]}\overline{u(\xi)} = \chi_{(-\infty,d_f)\vee[\tau+d_f-m_f,\infty)}(t)$$

$$\bigcap_{\xi\in[t-d_r,t-d_r+m_r]}u(\xi) = \chi_{[d_r,\tau+d_r-m_r)}(t)$$

$$\bigcup_{\xi\in[t-d_f,t-d_f+m_f]}u(\xi) = \chi_{[d_f-m_f,\tau+d_f)}(t)$$

The systems of inequalities that are inferred from $\overline{x} \in Sol_{BDC}^{m_r,d_r,m_f,d_f}(\overline{\chi_{[0,\tau)}})$:

$$\chi_{(-\infty,d_r-m_r)\vee[\tau+d_r,\infty)}(t) \leq \overline{x(t)} \leq \chi_{(-\infty,d_f)\vee[\tau+d_f-m_f,\infty)}(t) \text{ i.e.}$$

$$\chi_{[d_f,\tau+d_f-m_f)}(t) \leq x(t) \leq \chi_{[d_r-m_r,\tau+d_r)}(t)$$

and respectively from $x \in Sol_{BDC}^{m_r,d_r,m_f,d_f}(\chi_{[0,\tau)})$:

$$\chi_{[d_r,\tau+d_r-m_r)}(t) \leq x(t) \leq \chi_{[d_f-m_f,\tau+d_f)}(t)$$

are equivalent –i.e. they have the same solutions- only if $d_r = d_f$ and $m_r = m_f$.

**7.3.7 Theorem** Let the numbers $0 \leq m_r \leq d_r, 0 \leq m_f \leq d_f$ and $0 \leq m'_r \leq d'_r, 0 \leq m'_f \leq d'_f$ so that $d_r \geq d_f - m_f, d_f \geq d_r - m_r$ and $d'_r \geq d'_f - m'_f, d'_f \geq d'_r - m'_r$ are true. Then $Sol_{BDC}^{m_r+m'_r,d_r+d'_r,m_f+m'_f,d_f+d'_f}$ is a BDC and we have

$$Sol_{BDC}^{m'_r,d'_r,m'_f,d'_f} \circ Sol_{BDC}^{m_r,d_r,m_f,d_f} = Sol_{BDC}^{m_r+m'_r,d_r+d'_r,m_f+m'_f,d_f+d'_f}$$

**Proof** We observe that $d_f + d'_f \geq d_r + d'_r - m_r - m'_r$, $d_r + d'_r \geq d_f + d'_f - m_f - m'_f$, thus $CC_{BDC}$ is fulfilled again and $Sol_{BDC}^{m_r+m'_r,d_r+d'_r,m_f+m'_f,d_f+d'_f}$ makes sense.

We prove

$$Sol_{BDC}^{m'_r,d'_r,m'_f,d'_f} \circ Sol_{BDC}^{m_r,d_r,m_f,d_f} \subset Sol_{BDC}^{m_r+m'_r,d_r+d'_r,m_f+m'_f,d_f+d'_f}$$

Let $u$ arbitrary and $y \in Sol_{BDC}^{m'_r,d'_r,m'_f,d'_f} \circ Sol_{BDC}^{m_r,d_r,m_f,d_f}(u)$, i.e. $x \in Sol_{BDC}^{m_r,d_r,m_f,d_f}(u)$ exists so that $y \in Sol_{BDC}^{m'_r,d'_r,m'_f,d'_f}(x)$. We infer $u(0-0) = x(0-0) = y(0-0)$, i.e. the compatibility with the initial conditions is satisfied. Moreover:

$$\bigcap_{\omega\in[\xi-d_r,\xi-d_r+m_r]} u(\omega) \leq x(\xi) \leq \bigcup_{\omega\in[\xi-d_f,\xi-d_f+m_f]} u(\omega) \tag{1}$$

$$\bigcap_{\xi\in[t-d'_r,t-d'_r+m'_r]} x(\xi) \leq y(t) \leq \bigcup_{\xi\in[t-d'_f,t-d'_f+m'_f]} x(\xi) \tag{2}$$

from where

$$\bigcap_{\xi\in[t-d_r-d'_r,t-d_r-d'_r+m_r+m'_r]} u(\xi) = \bigcap_{\xi\in[t-d'_r,t-d'_r+m'_r]} \bigcap_{\omega\in[\xi-d_r,\xi-d_r+m_r]} u(\omega) \leq$$

$$\leq \bigcap_{\xi\in[t-d'_r,t-d'_r+m'_r]} x(\xi) \leq y(t) \leq \bigcup_{\xi\in[t-d'_f,t-d'_f+m'_f]} x(\xi) \leq$$

$$\leq \bigcup_{\xi\in[t-d'_f,t-d'_f+m'_f]} \bigcup_{\omega\in[\xi-d_f,\xi-d_f+m_f]} u(\omega) = \bigcup_{\xi\in[t-d_f-d'_f,t-d_f-d'_f+m_f+m'_f]} u(\xi)$$

thus $y \in Sol_{BDC}^{m_r+m'_r,d_r+d'_r,m_f+m'_f,d_f+d'_f}(u)$.

We prove

$$Sol_{BDC}^{m_r+m'_r,d_r+d'_r,m_f+m'_f,d_f+d'_f} \subset Sol_{BDC}^{m'_r,d'_r,m'_f,d'_f} \circ Sol_{BDC}^{m_r,d_r,m_f,d_f}$$

and we must show that for any $y$ with

$$\bigcap_{\xi\in[t-d_r-d'_r,t-d_r-d'_r+m_r+m'_r]} u(\xi) \leq y(t) \leq \bigcup_{\xi\in[t-d_f-d'_f,t-d_f-d'_f+m_f+m'_f]} u(\xi) \tag{3}$$

some $x$ exists so that (1), (2) be fulfilled. The property is true for $t < 0$ and let against all reason $t_0 \geq 0$ be the first time instant when the property is false. We suppose that $y(t_0) = 0$. From (3)

$$\bigcap_{\xi\in[t_0-d'_r,t_0-d'_r+m'_r]} \bigcap_{\omega\in[\xi-d_r,\xi-d_r+m_r]} u(\omega) = 0$$

i.e.

$$\exists \xi_0 \in [t_0 - d'_r, t_0 - d'_r + m'_r], \bigcap_{\omega\in[\xi_0-d_r,\xi_0-d_r+m_r]} u(\omega) = 0$$

(1) shows that we can choose $x(\xi_0) = 0$ and (2) is true, because

$$\bigcap_{\xi\in[t_0-d'_r,t_0-d'_r+m'_r]} x(\xi) = x(\xi_0) = 0$$

in contradiction with the supposition that was made on the existence of $t_0$. The same result is obtained if we suppose $y(t_0) = 1$.

**7.3.8 Corollary** The set of the BDC's is, relative to the serial connection, a commutative semi-group where the unit is $I$.

**7.3.9 Remarks** If $i$ is an arbitrary DC, then by its serial connection with a BDC we obtain an arbitrary DC (not a BDC).

We combine now the order of the BDC's from 7.3.4 with Theorem 6.8.5 of compatibility between the serial connection and the order and with Theorem 7.3.7 related with the serial connection of the BDC's in the next manner. Let the BDC's $Sol_{BDC}^{m_r,d_r,m_f,d_f}$,

$Sol_{BDC}^{m'_r,d'_r,m'_f,d'_f}$ and $Sol_{BDC}^{m''_r,d''_r,m''_f,d''_f}$, where $0 \le m_r \le d_r$, $0 \le m_f \le d_f$, $0 \le m'_r \le d'_r$, $0 \le m'_f \le d'_f$, $0 \le m''_r \le d''_r$, $0 \le m''_f \le d''_f$ and $CC_{BDC}$ is fulfilled three times. The implication

$$Sol_{BDC}^{m'_r,d'_r,m'_f,d'_f} \subset Sol_{BDC}^{m''_r,d''_r,m''_f,d''_f} \Rightarrow$$

$$\Rightarrow Sol_{BDC}^{m'_r,d'_r,m'_f,d'_f} \circ Sol_{BDC}^{m_r,d_r,m_f,d_f} \subset Sol_{BDC}^{m''_r,d''_r,m''_f,d''_f} \circ Sol_{BDC}^{m_r,d_r,m_f,d_f}$$

means that

$$d''_r - m''_r \le d'_r - m'_r \le d'_f \le d''_f$$
$$d''_f - m''_f \le d'_f - m'_f \le d'_r \le d''_r$$

implies

$$d_r + d''_r - m_r - m''_r \le d_r + d'_r - m_r - m'_r \le d_f + d'_f \le d_f + d''_f$$
$$d_f + d''_f - m_f - m''_f \le d_f + d'_f - m_f - m'_f \le d_r + d'_r \le d_r + d''_r$$

The other situation from 6.8.5 is similar

$$Sol_{BDC}^{m'_r,d'_r,m'_f,d'_f} \subset Sol_{BDC}^{m''_r,d''_r,m''_f,d''_f} \Rightarrow$$

$$\Rightarrow Sol_{BDC}^{m_r,d_r,m_f,d_f} \circ Sol_{BDC}^{m'_r,d'_r,m'_f,d'_f} \subset Sol_{BDC}^{m_r,d_r,m_f,d_f} \circ Sol_{BDC}^{m''_r,d''_r,m''_f,d''_f}$$

**7.3.10 Theorem** If $x \in Sol_{BDC}^{m_r,d_r,m_f,d_f}(u)$, the next inequalities are fulfilled:

$$u(t-d_r-0) \cdot \bigcap_{\xi \in [t-d_r, t-d_r+m_r)} u(\xi) \le x(t-0) \le u(t-d_f-0) \cup \bigcup_{\xi \in [t-d_f, t-d_f+m_f)} u(\xi)$$

**Proof** We take the left limit in BDC and use 4.2.1.

### 7.4 Fixed and Inertial Delays

**7.4.1 Corollary** of Theorem 7.3.3. The deterministic BDC's are given by the equation

$$x(t) = u(t-d) \qquad (1)$$

where $d \ge 0$; the non-deterministic BDC's consist in the system 7.2.1 (1), where $m_r + m_f > 0$.

**7.4.2 Definition** For $u, x \in S$ and $d \ge 0$, the relation 7.4.1 (1) is called the *fixed delay condition* (FDC). The delay defined by this equation is also called *pure*, *ideal* or *non-inertial*.

We also call fixed delay condition the function $I_d$.

A delay different from FDC is called *inertial*.

**7.4.3 Remark** A special case of inclusion at 7.3.4 consists in the situation when the left BDC is deterministic. Let $d \in [d_r - m_r, d_r] \wedge [d_f - m_f, d_f]$; then the statements

$$t-d_r \le t-d \le t-d_r+m_r, t-d_f \le t-d \le t-d_f+m_f$$

$$\bigcap_{\xi \in [t-d_r, t-d_r+m_r]} u(\xi) \le u(t-d) \le \bigcup_{\xi \in [t-d_f, t-d_f+m_f]} u(\xi)$$

make us conclude that $I_d \subset Sol_{BDC}^{m_r,d_r,m_f,d_f}$.

**7.4.4 Corollary** FDC is deterministic (Example 6.3.3, Theorem 7.3.3), time invariant (Example 6.5.5, Theorem 7.3.5), constant (Example 6.6.2) and symmetrical (Example 6.7.2, Theorem 7.3.6). The serial connection of the FDC's coincides with the composition of the translations: for $d \geq 0, d' \geq 0$ we have

$$I_d \circ I_{d'} = I_{d'} \circ I_d = I_{d+d'}$$

**7.4.5 Remark** At Definition 7.4.2 inertia was defined to be the property of the DC's of being not ideal. In particular the non-deterministic DC's, for example the non-trivial BDC's where $m_r + m_f > 0$ are inertial.

# 8. Absolute Inertial Delays

## 8.1 Absolute Inertia

**8.1.1 Theorem** Let the numbers $\delta_r \geq 0, \delta_f \geq 0$. When $x \in S$, the next conditions are equivalent:

a) $\quad \overline{x(t-0)} \cdot x(t) \leq \bigcap_{\xi \in [t, t+\delta_r]} x(\xi)$

$\quad x(t-0) \cdot \overline{x(t)} \leq \bigcap_{\xi \in [t, t+\delta_f]} \overline{x(\xi)}$

b) $\quad \overline{x(t-0)} \cdot x(t) \leq \overline{x(t-0)} \cdot \bigcap_{\xi \in [t, t+\delta_r]} x(\xi)$

$\quad x(t-0) \cdot \overline{x(t)} \leq x(t-0) \cdot \bigcap_{\xi \in [t, t+\delta_f]} \overline{x(\xi)}$

c) $\quad \overline{x(t-0)} \cdot x(t) = \overline{x(t-0)} \cdot \bigcap_{\xi \in [t, t+\delta_r]} x(\xi)$

$\quad x(t-0) \cdot \overline{x(t)} = x(t-0) \cdot \bigcap_{\xi \in [t, t+\delta_f]} \overline{x(\xi)}$

d) $\quad \overline{x(t-0)} \cdot x(t) \leq \overline{x(t-\delta_f-0)} \cdot \bigcap_{\xi \in [t-\delta_f, t)} x(\xi)$

$\quad x(t-0) \cdot \overline{x(t)} \leq x(t-\delta_r-0) \cdot \bigcap_{\xi \in [t-\delta_r, t)} \overline{x(\xi)}$

e) $\quad \forall d < d', \overline{x(d-0)} \cdot x(d) = 1$ and $x(d'-0) \cdot \overline{x(d')} = 1 \Rightarrow d'-d > \delta_r$

$\quad \forall d < d', x(d-0) \cdot \overline{x(d)} = 1$ and $\overline{x(d'-0)} \cdot x(d') = 1 \Rightarrow d'-d > \delta_f$

**Proof** $a) \Rightarrow b)$ Both terms of the first inequality from a) are multiplied with $\overline{x(t-0)}$ and the first inequality from b) results.

$b) \Rightarrow a) \quad \overline{x(t-0)} \cdot x(t) \leq \overline{x(t-0)} \cdot \bigcap_{\xi \in [t, t+\delta_r]} x(\xi) \leq \bigcap_{\xi \in [t, t+\delta_r]} x(\xi)$

$a) \Rightarrow c)$ In the system

$$\overline{x(t-0)} \cdot x(t) \leq \bigcap_{\xi \in [t, t+\delta_r]} x(\xi) \leq x(t)$$

we multiply all the terms with $\overline{x(t-0)}$.

$c) \Rightarrow a)$ $\qquad \overline{x(t-0)} \cdot x(t) = \overline{x(t-0)} \cdot \bigcap_{\xi \in [t, t+\delta_r]} x(\xi) \leq \bigcap_{\xi \in [t, t+\delta_r]} x(\xi)$

$a) \Rightarrow e)$ Let $d < d'$ arbitrary so that
$$\overline{x(d-0)} \cdot x(d) = 1 \text{ and } x(d'-0) \cdot \overline{x(d')} = 1$$

a) states that $\bigcap_{\xi \in [d, d+\delta_r]} u(\xi) = 1$ and $\bigcap_{\xi \in [d', d'+\delta_f]} \overline{u(\xi)} = 1$, thus $[d, d+\delta_r] \wedge [d', d'+\delta_f] = \emptyset$ from where $d + \delta_r < d'$ and the conclusion is $d' - d > \delta_r$.

$e) \Rightarrow a)$ We suppose that a) is not true, i.e. $d$ exists with
$$\overline{x(d-0)} \cdot x(d) = 1 \text{ and } \bigcap_{\xi \in [d, d+\delta_r]} x(\xi) = 0$$

meaning the existence of $d' \in (d, d+\delta_r]$ where $x$ switches from 1 to 0
$$x(d'-0) \cdot \overline{x(d')} = 1$$

We have $d' - d \leq \delta_r$, contradiction with e).

$d) \Rightarrow e)$ Let us take two numbers $d < d'$ so that
$$\overline{x(d-0)} \cdot x(d) = 1 \text{ and } x(d'-0) \cdot \overline{x(d')} = 1$$

resulting
$$\overline{x(d-\delta_f - 0)} \cdot \bigcap_{\xi \in [d-\delta_f, d)} \overline{x(\xi)} = 1 \text{ and } x(d'-\delta_r - 0) \cdot \bigcap_{\xi \in [d'-\delta_r, d')} x(\xi) = 1$$

In other words $\varepsilon_1 > 0$ and $\varepsilon_2 > 0$ exist with the property
$$\forall \xi \in [d - \delta_f - \varepsilon_1, d), x(\xi) = 0$$
$$\forall \xi \in [d' - \delta_r - \varepsilon_2, d'), x(\xi) = 1$$
$$[d - \delta_f - \varepsilon_1, d) \wedge [d' - \delta_r - \varepsilon_2, d') = \emptyset$$

The last empty intersection gives the conclusion that
$$d \leq d' - \delta_r - \varepsilon_2$$

is true, i.e. $d' - d \geq \delta_r + \varepsilon_2 > \delta_r$.

$e) \Rightarrow d)$ If d) is not true then $d'$ exists so that
$$\overline{x(d'-0)} \cdot x(d') = 1 \text{ and } \overline{x(d' - \delta_f - 0)} \cdot \bigcap_{\xi \in [d'-\delta_f, d')} \overline{x(\xi)} = 0$$

This means that for any $\varepsilon > 0$ some $d \in [d' - \delta_f - \varepsilon, d')$ exists with
$$x(d-0) \cdot \overline{x(d)} = 1$$

The inequality $d' - \delta_f - \varepsilon \leq d$ that is true for all $\varepsilon > 0$ gives $d' - d \leq \delta_f$, thus e) is not true.

**8.1.2 Definition** Any of the properties 8.1.1 a),…, 8.1.1 e) is called the *absolute inertial condition* (AIC). $\delta_r, \delta_f$ are called the (*rising, falling*) *inertial parameters*. If AIC is fulfilled, we say that the tuple $(\delta_r, \delta_f, x)$ satisfies AIC. When $\delta_r = \delta_f = 0$, AIC is called *trivial* and if $\delta_r > 0$ or $\delta_f > 0$, then AIC is called *non-trivial*.

We also call absolute inertial condition the set $Sol_{AIC}^{\delta_r, \delta_f} \subset S$ defined by
$$Sol_{AIC}^{\delta_r, \delta_f} = \{x \mid (\delta_r, \delta_f, x) \text{ satisfies AIC}\}$$

**8.1.3 Remarks** For all $t<0$ and all $x$, AIC is trivially fulfilled since $\overline{x(t-0)}\cdot x(t) = x(t-0)\cdot \overline{x(t)} = 0$. This property represents the compatibility between RIC and the initial value of $x$.

The interpretation of AIC results from Fig 12. We observe how the switch $\overline{x(t_2-0)}\cdot x(t_2) = 1$ in the 8.1.1 a) version assures that $x$ will remain 1 for a time interval of length $t_3-t_2 > \delta_r$ and in the 8.1.1 d) version assures that $x$ has remained 0 for a time interval of length $t_2-t_1 > \delta_f$. When $t$ runs in $\mathbf{R}$, the two conditions are equivalent.

To be remarked the way that any of 8.1.1 a),…, 8.1.1 e) degenerates in the trivial situation $\delta_r = \delta_f = 0$: $Sol_{AIC}^{0,0} = S$. To be remarked also the intermediary situations when one of $\delta_r > 0, \delta_f = 0$, respectively $\delta_r = 0, \delta_f > 0$ is true and the inclusions $Sol_{AIC}^{\delta_r,\delta_f} \subset S$ are strict. In fact we have

$$Sol_{AIC}^{\delta_r,\delta_f} \subset Sol_{AIC}^{\delta_r',\delta_f'} \Leftrightarrow \delta_r \geq \delta_r' \text{ and } \delta_f \geq \delta_f'$$

$$Sol_{AIC}^{\delta_r,\delta_f} \wedge Sol_{AIC}^{\delta_r',\delta_f'} = Sol_{AIC}^{\max(\delta_r,\delta_r'),\max(\delta_f,\delta_f')}$$

$$Sol_{AIC}^{\delta_r,\delta_f} \vee Sol_{AIC}^{\delta_r',\delta_f'} = Sol_{AIC}^{\min(\delta_r,\delta_r'),\min(\delta_f,\delta_f')}$$

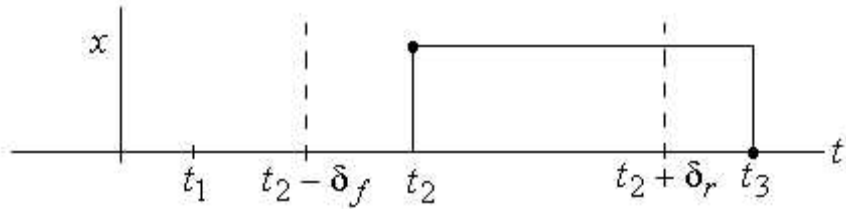

Fig 12

The set $Sol_{AIC}^{\delta_r,\delta_f}$ is not closed under the Boolean laws if $\delta_r > 0$ or $\delta_f > 0$, for example $\chi_{[0,2)}, \chi_{[1,3)} \in Sol_{AIC}^{1,0}$ and $\chi_{[1,2)} = \chi_{[0,2)} \cdot \chi_{[1,3)} \notin Sol_{AIC}^{1,0}$.

**8.2 Absolute Inertial Delays**

**8.2.1 Definition** Let the numbers $\delta_r \geq 0$, $\delta_f \geq 0$. If the DC $i$ satisfies $\forall u, i(u) \subset Sol_{AIC}^{\delta_r,\delta_f}$, then it is called *absolute inertial delay condition* (AIDC). We also say that $i$ *satisfies AIC* (represented by $Sol_{AIC}^{\delta_r,\delta_f}$).

**8.2.2 Remark** Intuitively, we say that AIDC expresses a cause-effect relationship between an input and an inertial state, so that for any form of $u$, the variations of the delayed signal $x$ cannot be faster than the satisfaction of AIC allows.

**8.2.3 Examples** A DC $i$ with the property $\forall u, i(u) \wedge Sol_{AIC}^{\delta_r, \delta_f} \neq \emptyset$ defines the AIDC $i \wedge Sol_{AIC}^{\delta_r, \delta_f}$. For example $Sol_{SC} \wedge Sol_{AIC}^{\delta_r, \delta_f}$ is AIDC for any $\delta_r \geq 0, \delta_f \geq 0$ and $I_d \wedge Sol_{AIC}^{\delta_r, \delta_f}$, $d \geq 0$ is AIDC for $\delta_r = \delta_f = 0$ only.

On the other hand, the next functions

$$i(u) = \begin{cases} 1, \text{if } \exists \lim_{t \to \infty} u(t) \text{ and } \lim_{t \to \infty} u(t) = 1 \\ 0, \text{otherwise} \end{cases}$$

$$i(u) = \begin{cases} 0, \text{if } \exists \lim_{t \to \infty} u(t) \text{ and } \lim_{t \to \infty} u(t) = 0 \\ 1, \text{otherwise} \end{cases}$$

$$i(u) = \begin{cases} \{1\} \vee \{\chi_{[\tau, \infty)} \mid \tau \geq 0\}, \text{if } \exists \lim_{t \to \infty} u(t) \text{ and } \lim_{t \to \infty} u(t) = 1 \\ \{0\} \vee \{\chi_{(-\infty, \tau)} \mid \tau \geq 0\}, \text{otherwise} \end{cases}$$

$$i(u) = \begin{cases} \{0\} \vee \{\chi_{(-\infty, \tau)} \mid \tau \geq 0\}, \text{if } \exists \lim_{t \to \infty} u(t) \text{ and } \lim_{t \to \infty} u(t) = 0 \\ \{1\} \vee \{\chi_{[\tau, \infty)} \mid \tau \geq 0\}, \text{otherwise} \end{cases}$$

are DC's satisfying AIC for all $\delta_r \geq 0, \delta_f \geq 0$.

**8.2.4 Theorem** Let $0 \leq m \leq d$. The deterministic DC (see Theorem 6.3.4) $x(t) = \bigcap_{\xi \in [t-d, t-d+m]} u(\xi)$ satisfies

$$x(t-0) \cdot \overline{x(t)} \leq \bigcap_{\xi \in [t, t+m]} \overline{x(\xi)}$$

and its dual $x(t) = \bigcup_{\xi \in [t-d, t-d+m]} u(\xi)$ satisfies

$$\overline{x(t-0)} \cdot x(t) \leq \bigcap_{\xi \in [t, t+m]} x(\xi)$$

**Proof** We show the first of these two properties. The hypothesis states
$$d'' > d' \text{ and } x(d'-0) \cdot \overline{x(d')} = 1 \text{ and } \overline{x(d''-0)} \cdot x(d'') = 1$$
from where we infer (see 4.2.2)
$$u(d'-d-0) \cdot \bigcap_{\xi \in [d'-d, d'-d+m)} u(\xi) \cdot \overline{u(d'-d+m)} = 1$$
$$\overline{u(d''-d-0)} \cdot \bigcap_{\xi \in [d''-d, d''-d+m]} u(\xi) = 1$$

i.e. $d'-d+m = d''-d-\varepsilon$ for some $\varepsilon > 0$ and eventually $d''-d' > m$.

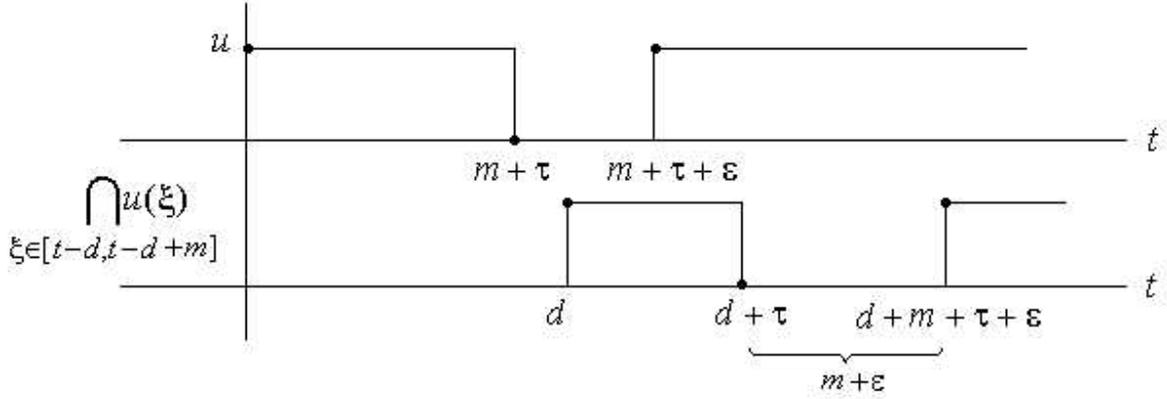

Fig 13

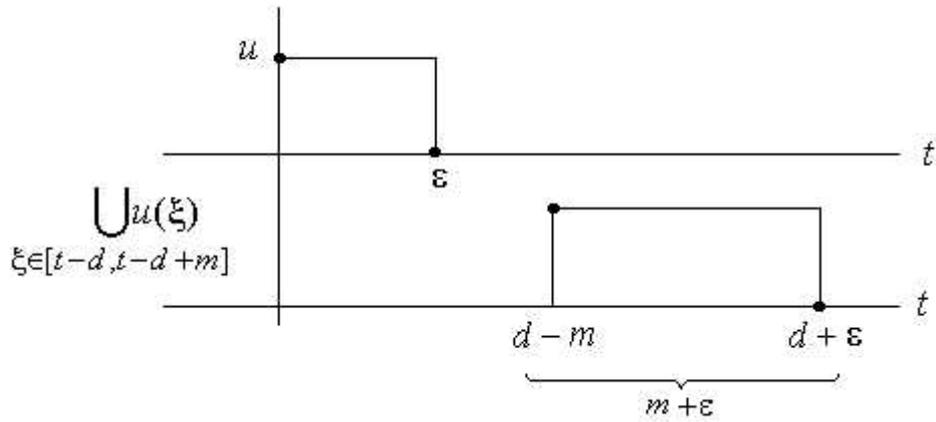

Fig 14

**8.2.5 Theorem** Let $0 \leq m_r \leq d_r, 0 \leq m_f \leq d_f$ so that $d_f \geq d_r - m_r, d_r \geq d_f - m_f$ is true. The system

$$\overline{x(t-0)} \cdot x(t) = \overline{x(t-0)} \cdot \bigcap_{\xi \in [t-d_r, t-d_r+m_r]} u(\xi) \quad (1)$$

$$x(t-0) \cdot \overline{x(t)} = x(t-0) \cdot \bigcap_{\xi \in [t-d_f, t-d_f+m_f]} \overline{u(\xi)} \quad (2)$$

satisfies the next properties:

a) (*the property of compatibility of (1), (2) with the initial conditions*) $x(0-0) = u(0-0)$ and for any $t < 0$ we have

$$\overline{x(t-0)} \cdot x(t) = \overline{x(t-0)} \cdot \bigcap_{\xi \in [t-d_r, t-d_r+m_r]} u(\xi) = 0 \quad (3)$$

$$x(t-0) \cdot \overline{x(t)} = x(t-0) \cdot \bigcap_{\xi \in [t-d_f, t-d_f+m_f]} \overline{u(\xi)} = 0 \quad (4)$$

b) (1), (2) have a unique solution $x \in S$.

c) (*the property of compatibility of the system with the final conditions*) (1), (2) define a deterministic DC $i$: $x \in Sol_{SC}(u)$ and if $\exists \lim\limits_{t \to \infty} u(t)$, then $t_1$ exists so that $\forall t \geq t_1$ we have (3), (4) fulfilled.

d) $x \in Sol_{AIC}^{d_f - d_r + m_r, d_r - d_f + m_f}$

e) $x \in Sol_{BDC}^{m_r, d_r, m_f, d_f}$

f) (1), (2) are time invariant

g) (1), (2) are constant

h) (1), (2) are symmetrical iff $d_r = d_f$ and $m_r = m_f$.

**Proof** a) If $x(0-0) \neq u(0-0)$, then we suppose $x(0-0) = 0, u(0-0) = 1$. This implies that for any $t < 0 \leq d_r - m_r$ we have

$$0 = \overline{x(t-0)} \cdot x(t) \neq \overline{x(t-0)} \cdot \bigcap_{\xi \in [t-d_r, t-d_r + m_r]} u(\xi) = 1$$

contradiction and similarly for $x(0-0) = 1, u(0-0) = 0$. Thus $x(t-0) = u(t-0) = c$ for any $t < 0$, where $c \in B$ is some constant and we get $\overline{x(t-0)} \cdot x(t) = x(t-0) \cdot \overline{x(t)} = c \cdot \overline{c} = 0$. At the same time, $t < 0$ implies that $\bigcap\limits_{\xi \in [t-d_r, t-d_r + m_r]} u(\xi)$ is equal with $c$ and $\bigcap\limits_{\xi \in [t=d_f, t-d_f + m_f]} \overline{u(\xi)}$ is equal with $\overline{c}$ so that

$$\overline{x(t-0)} \cdot \bigcap_{\xi \in [t-d_r, t-d_r + m_r]} u(\xi) = \overline{c} \cdot c = 0$$

$$x(t-0) \cdot \bigcap_{\xi \in [t-d_f, t-d_f + m_f]} \overline{u(\xi)} = c \cdot \overline{c} = 0$$

and the truth of a) results.

b) The solution is unique for $t < 0$ and is given by $x(t) = u(0-0)$. The non-uniqueness means the existence of a time instant $t_1 \geq 0$ with the property that $\forall t < t_1$ the solution is unique and on the other hand $x(t_1) = 0, x(t_1) = 1$ satisfy both (1), (2). We suppose $x(t_1 - 0) = 0$ and (1) becomes

$$x(t_1) = \bigcap_{\xi \in [t_1 - d_r, t_1 - d_r + m_r]} u(\xi)$$

from where either $x(t_1) = 0$, or $x(t_1) = 1$ is true but not both. Similarly for $x(t_1 - 0) = 1$. As $t_1$ was arbitrary, the solution is unique.

c) We suppose without loosing the generality that $\exists t_1 \geq 0, \forall t \geq t_1, u(t) = 1$ and we have two possibilities:

c.i) $x(t_1 + d_r - 0) = 0$. From (1) we have $x(t_1 + d_r) = 1$ and $\forall t > t_1 + d_r$, $x(t) = 1$ and (3), (4) are fulfilled

c.ii) $x(t_1 + d_r - 0) = 1$. Then $\forall t \geq t_1 + d_r$, $x(t) = 1$ and (3), (4) are fulfilled too.

d) The hypothesis is

$$d' > d \text{ and } \overline{x(d-0)} \cdot x(d) = 1 \text{ and } x(d'-0) \cdot \overline{x(d')} = 1$$

resulting

$$\bigcap_{\xi \in [d-d_r, d-d_r + m_r]} u(\xi) \cdot \bigcap_{\xi \in [d'-d_f, d'-d_f + m_f]} \overline{u(\xi)} = 1$$

$$\Rightarrow [d-d_r, d-d_r+m_r] \wedge [d'-d_f, d'-d_f+m_f] = \varnothing$$
$$\Rightarrow d-d_r+m_r < d'-d_f \text{ or } d'-d_f+m_f < d-d_r$$
$$\Rightarrow d'-d > d_f - d_r + m_r$$

(the inequality $d'-d < d_f - m_f - d_r$ represents a contradiction, because the left term is strictly positive and from $d_f - m_f \leq d_r$, we get that the right term is non-positive).

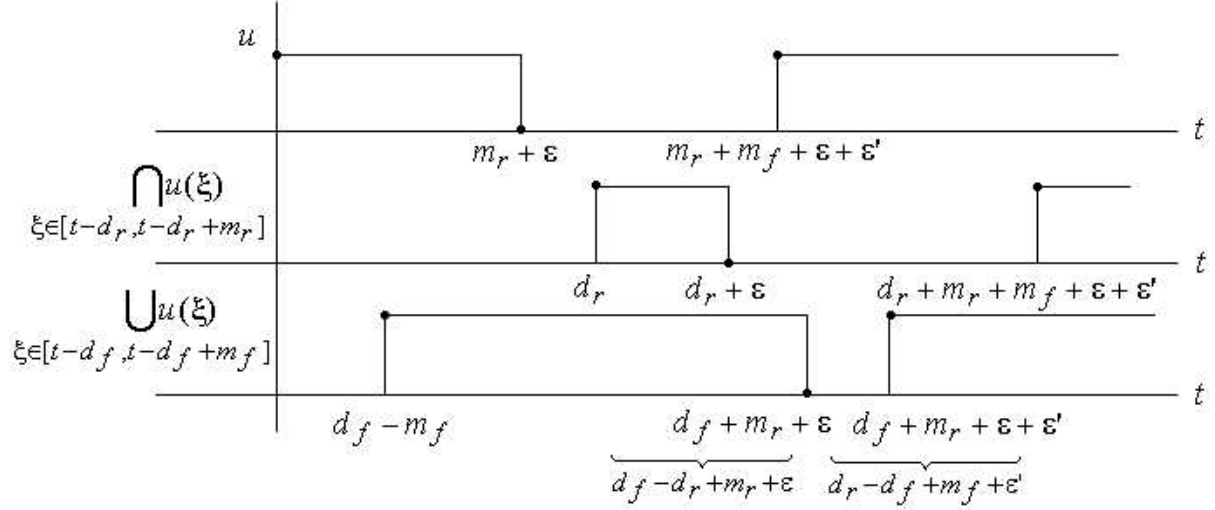

Fig 15

e) We suppose against all reason that $t, u$ exist so that $\bigcap_{\xi \in [t-d_r, t-d_r+m_r]} u(\xi) = 1$ and $x(t) = 0$; we get from (1) $\overline{x(t-0)} = 0$ and $x(t-0) = 1$ and on the other hand, because $\bigcap_{\xi \in [t-d_f, t-d_f+m_f]} \overline{u(\xi)} = 0$ (from $CC_{BDC}$) equation (2) gives $\overline{x(t)} = 0$, thus $x(t) = 1$, contradiction. In other words $\bigcap_{\xi \in [t-d_r, t-d_r+m_r]} u(\xi) \leq x(t)$. The inequality $x(t) \leq \bigcup_{\xi \in [t-d_f, t-d_f+m_f]} u(\xi)$ is similarly proved.

f) We show that $i$ is time invariant and the hypothesis is $u \circ \tau^d \in S$ and $x = i(u)$, where $d \in \mathbf{R}$ is arbitrary. Because $i(u \circ \tau^d) \in S$ and $x \circ \tau^d = i(u \circ \tau^d)$ (as resulted by the replacement in (1), (2) of $u, x$ with $x \circ \tau^d, u \circ \tau^d$), the fulfillment of this property follows.

g) The inequalities
$$\overline{x(t-0)} \cdot x(t) = \overline{x(t-0)} \cdot \bigcap_{\xi \in [t-d_r, t-d_r+m_r]} u(\xi) \leq u(t-d_r)$$
$$x(t-0) \cdot \overline{x(t)} = x(t-0) \cdot \bigcap_{\xi \in [t-d_f, t-d_f+m_f]} \overline{u(\xi)} \leq \overline{u(t-d_f)}$$

show that $i$ is constant.

h) $\overline{x} \in i(\overline{u}) \Leftrightarrow x \in i(u)$ is equivalent, from the form of (1), (2) with the fact that

$$\bigcap_{\xi \in [t-d_r, t-d_r+m_r]} u(\xi) = \bigcap_{\xi \in [t-d_f, t-d_f+m_f]} u(\xi)$$

and with the fact that $d_r = d_f, m_r = m_f$.

**8.2.6 Remark** If in 8.2.5 we put $m_r = m_f = 0$, then $CC_{BDC}$ implies $d_r = d_f = d$ and the system

$$\overline{x(t-0)} \cdot x(t) = \overline{x(t-0)} \cdot u(t-d)$$
$$x(t-0) \cdot \overline{x(t)} = x(t-0) \cdot \overline{u(t-d)}$$

is equivalent with FDC (it has the same solution).

**8.2.7 Theorem** a) $Sol_{SC} \wedge Sol_{AIC}^{\delta_r, \delta_f}$ is time variable.

b) $Sol_{SC} \wedge Sol_{AIC}^{\delta_r, \delta_f}$ is symmetrical iff $\delta_r = \delta_f$.

**Proof** a) See Example 6.5.5 b) where the time variability of $Sol_{SC}$ is shown.

b) <u>If</u> Like at 6.7.2 b), by replacing $Sol_{SC}$ with $Sol_{SC} \wedge Sol_{AIC}^{\delta, \delta}$.
<u>Only if</u> We suppose against all reason that $\delta_r < \delta_f$; by taking a $\delta \in (\delta_r, \delta_f)$ we have for some $u$ (with $\lim_{t \to \infty} u(t) = 1$ false) that $\chi_{[0,\delta)} \in Sol_{AIC}^{\delta_r, \delta_f}$ and $\overline{\chi_{[0,\delta)}} \notin Sol_{AIC}^{\delta_r, \delta_f}$, contradiction with the symmetry request:

$$\forall u, Sol_{SC}(\overline{u}) \wedge Sol_{AIC}^{\delta_r, \delta_f} = \{\overline{x} \mid x \in Sol_{SC}(u) \wedge Sol_{AIC}^{\delta_r, \delta_f}\}$$

Similarly for $\delta_f < \delta_r$.

**8.2.8 Theorem** Being given the non-negative numbers $\delta_r, \delta_f, \delta_r', \delta_f'$ and the DC's $i, j$, the next properties are true:

a) $(i \wedge Sol_{AIC}^{\delta_r', \delta_f'}) \circ (j \wedge Sol_{AIC}^{\delta_r, \delta_f}) = (i \circ (j \wedge Sol_{AIC}^{\delta_r, \delta_f})) \wedge Sol_{AIC}^{\delta_r', \delta_f'} \subset (i \circ j) \wedge Sol_{AIC}^{\delta_r', \delta_f'}$

b) If $\forall u, i(u) \subset Sol_{AIC}^{\delta_r, \delta_f}, j(u) \subset Sol_{AIC}^{\delta_r, \delta_f}$ then $\forall u, (i \circ j)(u) \subset Sol_{AIC}^{\delta_r', \delta_f'}$.

**Proof** a) From 6.8.6 a).

b) $\forall u, (i \circ j)(u) = ((i \wedge Sol_{AIC}^{\delta_r', \delta_f'}) \circ (j \wedge Sol_{AIC}^{\delta_r, \delta_f}))(u) \stackrel{a)}{\subset} ((i \circ j) \wedge Sol_{AIC}^{\delta_r', \delta_f'})(u) \subset (i \circ j)(u)$

thus $i \circ j = (i \circ j) \wedge Sol_{AIC}^{\delta_r', \delta_f'}$. The conclusion follows.

## 8.3 The Consistency Condition

**8.3.1 Theorem** The numbers $0 \leq m_r \leq d_r$, $0 \leq m_f \leq d_f$ are given so that $d_f \geq d_r - m_r, d_r \geq d_f - m_f$ is true and let also $\delta_r \geq 0, \delta_f \geq 0$. The next statements are equivalent:

a) $\delta_r + \delta_f \leq m_r + m_f$

b) For any $u \in S$, the system of inequalities

$$\bigcap_{\xi \in [t-d_r, t-d_r+m_r]} u(\xi) \leq x(t) \leq \bigcup_{\xi \in [t-d_f, t-d_f+m_f]} x(\xi)$$

$$\overline{x(t-0)} \cdot x(t) \leq \bigcap_{\xi \in [t, t+\delta_r]} x(\xi)$$

$$x(t-0) \cdot \overline{x(t)} \leq \bigcap_{\xi \in [t, t+\delta_f]} \overline{x(\xi)}$$

has a solution $x \in S$.

**Proof** $a) \Rightarrow b)$ The next possibilities exist:

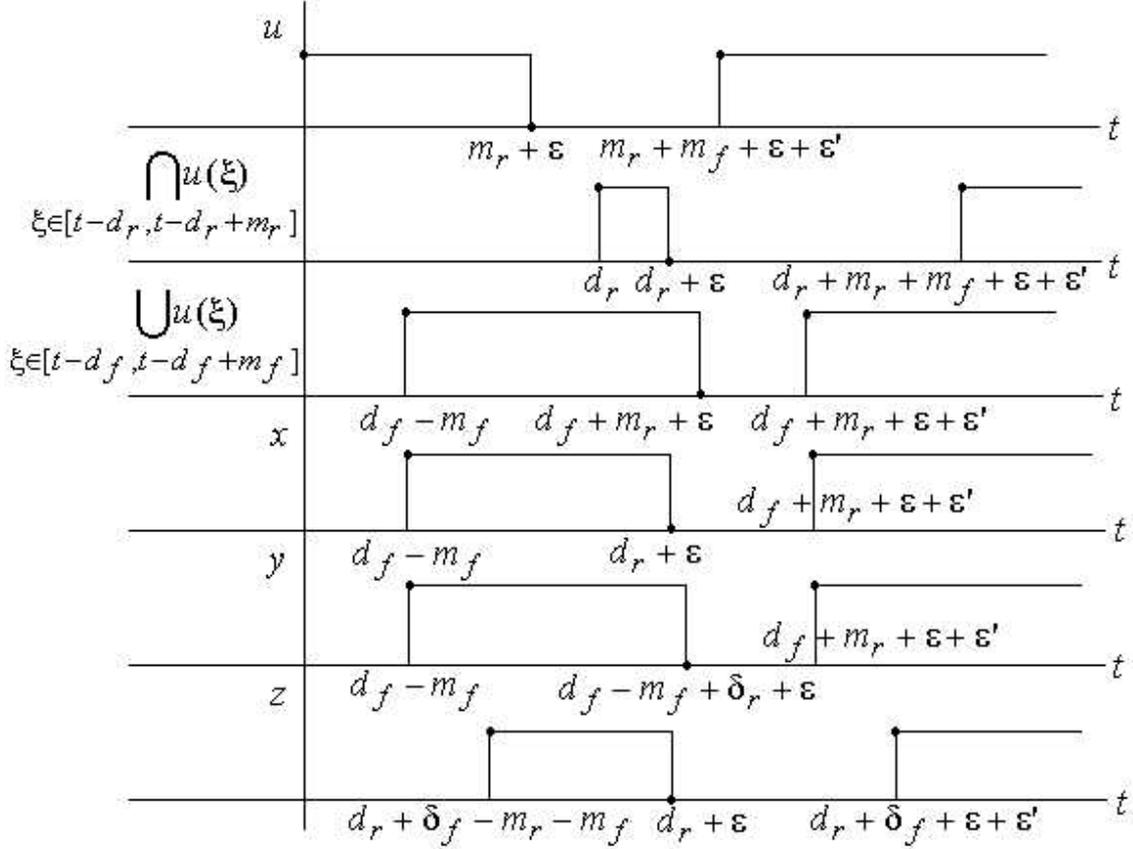

Fig 16

i) $d_r - d_f + m_f \geq \delta_r, d_f - d_r + m_r \geq \delta_f$

ii) $d_r - d_f + m_f < \delta_r, d_f - d_r + m_r \geq \delta_f$

iii) $d_r - d_f + m_f \geq \delta_r, d_f - d_r + m_r < \delta_f$

because the fourth possibility, i.e. $d_r - d_f + m_f < \delta_r, d_f - d_r + m_r < \delta_f$ gives the contradiction $\delta_r + \delta_f > m_r + m_f$. It is sufficient without restricting the generality to consider the input

$$u(t) = \chi_{[0, m_r + \varepsilon) \vee [m_r + m_f + \varepsilon + \varepsilon', \infty)}(t)$$

that produces a 1-pulse followed by a 0-pulse at the output, where $\varepsilon, \varepsilon' > 0$ are arbitrary, for which the existing solutions of the system in the three cases i), ii), iii) have been noted in Fig 16 with $x, y, z$. For $d_r - d_f + m_f \geq \delta_r$, $d_f - d_r + m_r \geq \delta_f$ (case i)) and

$$x(t) = \chi_{[d_f - m_f, d_r + \varepsilon) \vee [d_f + m_r + \varepsilon + \varepsilon', \infty)}(t)$$

it is obvious that BDC and AIC:
$$d_r + \varepsilon - d_f + m_f > \delta_r, d_f + m_r + \varepsilon + \varepsilon' - d_r - \varepsilon > \delta_f$$
are fulfilled. In the case of
$$y(t) = \chi_{[d_f - m_f, d_f - m_f + \delta_r + \varepsilon) \vee [d_f + m_r + \varepsilon + \varepsilon', \infty)}(t)$$
if $d_r - d_f + m_f < \delta_r$ (consequence of case ii)) we have BDC and AIC fulfilled under the form
$$d_r + \varepsilon < d_f - m_f + \delta_r + \varepsilon \leq d_f + m_r + \varepsilon$$
$$d_f - m_f + \delta_r + \varepsilon - d_f + m_f > \delta_r, d_f + m_r + \varepsilon + \varepsilon' - d_f + m_f - \delta_r - \varepsilon > \delta_f$$
and similarly for
$$z(t) = \chi_{[d_r + \delta_f - m_r - m_f, d_r + \varepsilon) \vee [d_r + \delta_f + \varepsilon + \varepsilon', \infty)}(t)$$
if $d_f - d_r + m_r < \delta_f$ (consequence of case iii)) where BDC and AIC are fulfilled again:
$$d_f - m_f < d_r + \delta_f - m_r - m_f \leq d_r$$
$$d_r + \varepsilon - d_r - \delta_f + m_r + m_f > \delta_r, d_r + \delta_f + \varepsilon + \varepsilon' - d_r - \varepsilon > \delta_f$$
$b) \Rightarrow a)$ We consider the input
$$u(t) = \chi_{[0, m_r + \varepsilon) \vee [m_r + m_f + 2\varepsilon, 2m_r + m_f + 3\varepsilon) \vee [2(m_r + m_f) + 4\varepsilon, 3m_r + 2m_f + 5\varepsilon) \vee \ldots}(t)$$
where $\varepsilon > 0$ is arbitrary

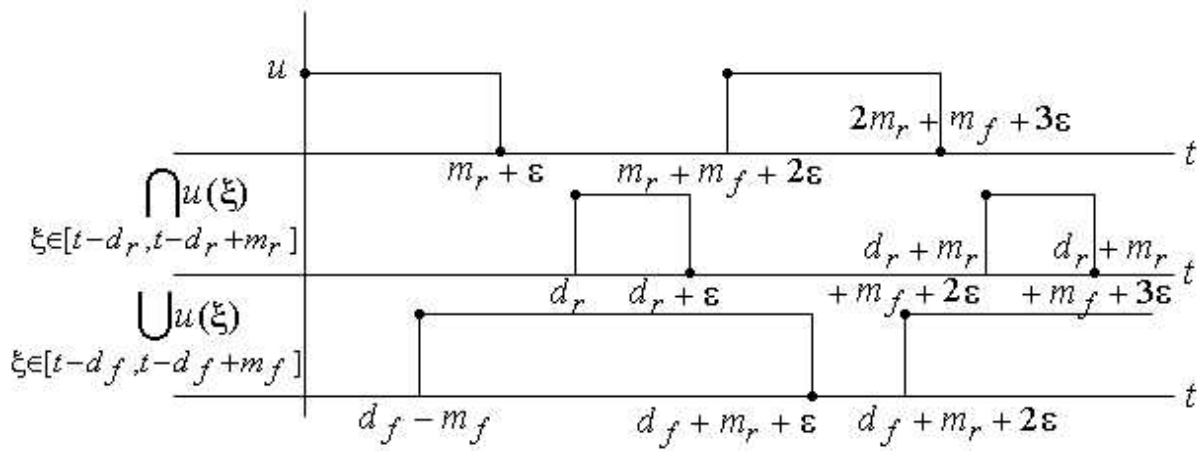

Fig 17

The hypothesis states the existence of a state $x$ of the form
$$x(t) = \chi_{[t_1, t_2) \vee [t_3, t_4) \vee \ldots}(t)$$
where
$$\delta_r < t_2 - t_1$$
$$\delta_f < t_3 - t_2$$
$$\delta_r < t_4 - t_3$$
$$\ldots$$
and $t_1 < t_2 < t_3 < t_4 < \ldots$ are arbitrary numbers so that
$$d_f - m_f \leq t_1 \leq d_r$$
$$d_r + \varepsilon \leq t_2 \leq d_f + m_r + \varepsilon$$

$$d_f + m_r + 2\varepsilon \leq t_3 \leq d_r + m_r + m_f + 2\varepsilon$$
$$d_r + m_r + m_f + 3\varepsilon \leq t_4 \leq d_f + 2m_r + m_f + 3\varepsilon$$
$$d_f + 2m_r + m_f + 4\varepsilon \leq t_5 \leq d_r + 2m_r + 2m_f + 4\varepsilon$$
$$\ldots$$

We infer
$$\delta_r + \delta_f < t_3 - t_1 \leq d_r - d_f - m_r + 2(m_r + m_f) + 2\varepsilon \leq 2(m_r + m_f) + 2\varepsilon$$
$$2(\delta_r + \delta_f) < t_5 - t_1 \leq d_r - d_f - m_r + 3(m_r + m_f) + 4\varepsilon \leq 3(m_r + m_f) + 4\varepsilon$$
$$3(\delta_r + \delta_f) < t_7 - t_1 \leq d_r - d_f - m_r + 4(m_r + m_f) + 6\varepsilon \leq 4(m_r + m_f) + 6\varepsilon$$
$$\ldots$$

and thus
$$\forall n \geq 1, \forall \varepsilon > 0, \delta_r + \delta_f < \frac{n+1}{n}(m_r + m_f) + 2\varepsilon$$

**8.3.2 Definition** Given the numbers $0 \leq m_r \leq d_r, 0 \leq m_f \leq d_f$ and $\delta_r \geq 0, \delta_f \geq 0$, the system of inequalities
$$d_r \geq d_f - m_f, d_f \geq d_r - m_r$$
$$\delta_r + \delta_f \leq m_r + m_f$$
is called the *consistency condition* (of the bounded absolute inertial delay condition) (CC$_{BAIDC}$).

**8.3.3 Remark** (*Special cases of* CC$_{BAIDC}$) If $m_r = m_f = 0$ then CC$_{BAIDC}$ becomes $d_r = d_f$ and $\delta_r = \delta_f = 0$ and the system 8.3.1 b) is equivalent with FDC. For $m_r = d_r, m_f = d_f$ CC$_{BAIDC}$ becomes $\delta_r + \delta_f \leq d_r + d_f$; and in the symmetrical case $m_r = m_f = m$, $d_r = d_f = d, \delta_r = \delta_f = \delta$, CC$_{BAIDC}$ means $\delta \leq m$.

## 8.4 Bounded Absolute Inertial Delays

**8.4.1 Definition** Let $u, x \in S$ and the numbers $0 \leq m_r \leq d_r, 0 \leq m_f \leq d_f, 0 \leq \delta_r, 0 \leq \delta_f$ so that $d_f \geq d_r - m_r, d_r \geq d_f - m_f$ and $\delta_r + \delta_f \leq m_r + m_f$ are true. The system of inequalities

$$\bigcap_{\xi \in [t-d_r, t-d_r+m_r]} u(\xi) \leq x(t) \leq \bigcup_{\xi \in [t-d_f, t-d_f+m_f]} x(\xi) \quad (1)$$

$$\overline{x(t-0)} \cdot x(t) \leq \bigcap_{\xi \in [t, t+\delta_r]} x(\xi) \quad (2)$$

$$x(t-0) \cdot \overline{x(t)} \leq \bigcap_{\xi \in [t, t+\delta_f]} \overline{x(\xi)} \quad (3)$$

is called the *bounded absolute inertial delay condition* (BAIDC). When BAIDC is fulfilled, we say that the tuple $(u, m_r, d_r, m_f, d_f, \delta_r, \delta_f, x)$ satisfies BAIDC.

We also call bounded absolute inertial delay condition the DC $Sol_{BDC}^{m_r, d_r, m_f, d_f} \wedge Sol_{AIC}^{\delta_r, \delta_f}$.

**8.4.2 Remark** BAIDC is a DC indeed, from Theorem 8.3.1. Its interpretation is obvious: $x$ is a solution of BAIDC if it switches from 0 to 1 with a delay $d \in [d_f - m_f, d_r]$ and then it keeps the value 1 for strictly more than $\delta_r$ time units and dually if it switches from 1 to 0 with a delay $d \in [d_r - m_r, d_f]$ and, after the switch, a new switch may happen after strictly more than $\delta_f$ time units.

BAIDC keeps the negative interpretation of BDC, see 7.2.6.

**8.4.3 Theorem** We suppose that $0 \leq m_r \leq d_r$, $0 \leq m_f \leq d_f$, $0 \leq m'_r \leq d'_r$, $0 \leq m'_f \leq d'_f$ and $\delta_r \geq 0$, $\delta_f \geq 0$, $\delta'_r \geq 0$, $\delta'_f \geq 0$ are given and they satisfy CC$_{\text{BAIDC}}$ twice. In such conditions $Sol_{BDC}^{m_r,d_r,m_f,d_f} \wedge Sol_{AIC}^{\delta_r,\delta_f}$, $Sol_{BDC}^{m'_r,d'_r,m'_f,d'_f} \wedge Sol_{AIC}^{\delta'_r,\delta'_f}$,

$Sol_{BDC}^{m_r+m'_r,d_r+d'_r,m_f+m'_f,d_f+d'_f} \wedge Sol_{AIC}^{\delta'_r,\delta'_f}$ are BAIDC's and the next property of serial connection holds:

$$(Sol_{BDC}^{m'_r,d'_r,m'_f,d'_f} \wedge Sol_{AIC}^{\delta'_r,\delta'_f}) \circ (Sol_{BDC}^{m_r,d_r,m_f,d_f} \wedge Sol_{AIC}^{\delta_r,\delta_f}) \subset$$

$$\subset Sol_{BDC}^{m_r+m'_r,d_r+d'_r,m_f+m'_f,d_f+d'_f} \wedge Sol_{AIC}^{\delta'_r,\delta'_f}$$

**Proof** We observe that $\delta'_r + \delta'_f \leq m'_r + m'_f$ implies $\delta'_r + \delta'_f \leq m_r + m'_r + m_f + m'_f$ thus, taking into account Theorems 7.3.7 and 8.2.8, the result follows.

# 9. Relative Inertial Delays

## 9.1 Relative Inertia

**9.1.1 Theorem** Let the real numbers $0 \leq \mu_r \leq \delta_r, 0 \leq \mu_f \leq \delta_f$. The next statements are equivalent

a) $\overline{x(t-0)} \cdot x(t) \leq \bigcap_{\xi \in [t-\delta_r, t-\delta_r+\mu_r]} u(\xi)$

$x(t-0) \cdot \overline{x(t)} \leq \bigcap_{\xi \in [t-\delta_f, t-\delta_f+\mu_f]} \overline{u(\xi)}$

b) $\overline{x(t-0)} \cdot x(t) \leq \overline{x(t-0)} \cdot \bigcap_{\xi \in [t-\delta_r, t-\delta_r+\mu_r]} u(\xi)$

$x(t-0) \cdot \overline{x(t)} \leq x(t-0) \cdot \bigcap_{\xi \in [t-\delta_f, t-\delta_f+\mu_f]} \overline{u(\xi)}$

where $u, x \in S$.

**Proof** Similar with the $a) \Leftrightarrow b)$ part of the proof from 8.1.1.

**9.1.2 Definition** Any of the properties 9.1.1 a), 9.1.1 b) is called the *relative inertial condition* (RIC). $u, x$ are the *input*, respectively the *state* (or *output*) and $\mu_r, \delta_r, \mu_f, \delta_f$ are the (*rising, falling*) *inertia parameters*. We say that the tuple $(u, \mu_r, \delta_r, \mu_f, \delta_f, x)$ satisfies RIC.

We call also relative inertial condition the function $Sol_{RIC}^{\mu_r,\delta_r,\mu_f,\delta_f} : S \to P^*(S)$ defined by

$$Sol_{RIC}^{\mu_r,\delta_r,\mu_f,\delta_f}(u) = \{x \mid (u, \mu_r, \delta_r, \mu_f, \delta_f, x) \text{ satisfies RIC}\}$$

**9.1.3 Remarks** The compatibility between RIC and the initial values $u(0-0), x(0-0)$ is given by the fact that for any $t<0$ we have $\overline{x(t-0)} \cdot x(t) = x(t-0) \cdot \overline{x(t)} = 0$, thus RIC is fulfilled trivially.

The word '*relative*' in our terminology refers to the fact that $u, x$ occur both in RIC. We interpret this condition now by recalling some quotations.

[12], [13], see 5.13 a): the inertial delays '*model the fact that the practical circuits will not respond* (at the output) *to two transitions* (on the input) *which are very close together*'. In case that two transitions of $u$ are 'very close together', i.e. if $u$ has a 1-pulse of length $\leq \mu_r$, respectively a 0-pulse of length $\leq \mu_f$, then the function $\bigcap_{\xi \in [t-\delta_r, t-\delta_r+\mu_r]} u(\xi)$, respectively $\bigcap_{\xi \in [t-\delta_f, t-\delta_f+\mu_f]} \overline{u(\xi)}$ is null, thus the function $\overline{x(t-0)} \cdot x(t)$, respectively $x(t-0) \cdot \overline{x(t)}$ is null too.

In this context we rewrite a quotation from [11], see 5.13 b): '*pulses shorter than or equal to the delay magnitude are not transmitted*' in the next manner: '*pulses shorter than or equal to* $\mu_r$ (*respectively* $\mu_f$) *are not transmitted and pulses strictly longer than* $\mu_r$ (*respectively than* $\mu_f$) *may be transmitted*'.

On the other hand 5.14 ii) and 5.16 i) look like RIC with $\delta_r = \delta_f = d_{\min}$, $\mu_r = \mu_f = d_{\min} - 0$ in the latter, point of view that agrees with the one from [12], [13] see Convention 5.3 stating that : '*the transmission delay for transitions is the same as the threshold for cancellation*'. Here $\delta_r, \delta_f$ act as 'transmission delays for transitions' even if they are rather 'minimum transmission delays for transitions' and $\mu_r, \mu_f$ act as 'thresholds for cancellation'. 'The same as' means that the two quantities differ by a small infinitesimal.

Furthermore, let us recall [1], see 5.15 with: '*changes should persist for at least* $l_1$ *time units but propagated after* $l_2, l_2 > l_1$ *time*'. In our formalism, if we accept the rising-falling symmetry, we have: $l_1 = \mu_r = \mu_f, l_2 = \delta_r = \delta_f$ and '*changes should persist for strictly more than* $l_1$ *time units but propagated after more than or equal with* $l_2, l_2 \geq l_1$ *time*'.

**9.1.4 Theorem** (*The relation between RIC and AIC*) Let $0 \leq \mu_r \leq \delta_r, 0 \leq \mu_f \leq \delta_f$ arbitrary. If $\delta_f \geq \delta_r - \mu_r, \delta_r \geq \delta_f - \mu_f$ then $\forall u, Sol_{RIC}^{\mu_r, \delta_r, \mu_f, \delta_f}(u) \subset Sol_{AIC}^{\delta_f - \delta_r + \mu_r, \delta_r - \delta_f + \mu_f}$.

**Proof** Let $d < d'$, $u$ and $x \in Sol_{RIC}^{\mu_r, \delta_r, \mu_f, \delta_f}(u)$ so that
$$\overline{x(d-0)} \cdot x(d) = 1 \text{ and } x(d'-0) \cdot \overline{x(d')} = 1$$
We get
$$\bigcap_{\xi \in [d-\delta_r, d-\delta_r+\mu_r]} u(\xi) = \bigcap_{\xi \in [d'-\delta_f, d'-\delta_f+\mu_f]} \overline{u(\xi)} = 1$$
$$\Rightarrow [d-\delta_r, d-\delta_r+\mu_r] \wedge [d'-\delta_f, d'-\delta_f+\mu_f] = \varnothing$$
$$\Rightarrow d-\delta_r+\mu_r < d'-\delta_f \text{ or } d'-\delta_f+\mu_f < d-\delta_r$$
$$\Rightarrow d'-d > \delta_f - \delta_r + \mu_r \text{ or } d'-d < \delta_f - \mu_f - \delta_r$$
$$\Rightarrow d'-d > \delta_f - \delta_r + \mu_r$$

(The inequality $d'-d < \delta_f - \mu_f - \delta_r$ is false, because the left term is strictly positive and the right term is non-positive).

The proof is similar for the second inequality.

**9.1.5 Definition** In the next property:

$$\forall \varepsilon > 0, \exists d, \exists d', \exists u, \ \exists x \in Sol_{RIC}^{\mu_r, \delta_r, \mu_f, \delta_f}(u) \text{ so that}$$

$$\overline{x(d-0)} \cdot x(d) = 1 \text{ and } x(d'-0) \cdot \overline{x(d')} = 1 \text{ and } |d-d'| < \varepsilon$$

the signal $x$ whose existence is stated is called *Zeno solution of RIC* (expressed by $Sol_{RIC}^{\mu_r, \delta_r, \mu_f, \delta_f}$).

**9.1.6 Theorem** The necessary and the sufficient condition in order that $Sol_{RIC}^{\mu_r, \delta_r, \mu_f, \delta_f}$ has no Zeno solutions is that $\delta_f > \delta_r - \mu_r, \delta_r > \delta_f - \mu_f$.

**Proof** <u>The necessity</u> We suppose against all reason that $\delta_f \leq \delta_r - \mu_r$ and let $u = \chi_{(-\infty, 0)}$ for which

$$\bigcap_{\xi \in [t-\delta_r, t-\delta_r+\mu_r]} \chi_{(-\infty, 0)}(\xi) = \chi_{(-\infty, \delta_r-\mu_r)}(t)$$

$$\bigcap_{\xi \in [t-\delta_f, t-\delta_f+\mu_f]} \chi_{[0, \infty)}(\xi) = \chi_{[\delta_f, \infty)}(t)$$

We have for any $\varepsilon > 0$ that $x = \chi_{[\delta_f - \varepsilon, \delta_f)} \in Sol_{RIC}^{\mu_r, \delta_r, \mu_f, \delta_f}(u)$, contradiction with the hypothesis. The supposition that $\delta_r \leq \delta_f - \mu_f$ is similar.

<u>The sufficiency</u> results from Theorem 9.1.4.

**9.1.7 Remarks** We have for any $a \in B$ that $Sol_{RIC}^{\mu_r, \delta_r, \mu_f, \delta_f}(a) = \{0, 1\} \vee \{\overline{a} \oplus \chi_{[\tau, \infty)} \mid \tau \geq 0\}$.

On the other hand RIC is a constancy condition, in the sense that

$$\overline{x(t-0)} \cdot x(t) \leq u(t-\delta_r)$$
$$x(t-0) \cdot \overline{x(t)} \leq \overline{u(t-\delta_f)}$$

## 9.2 Relative Inertial Delays

**9.2.1 Definition** Let the numbers $0 \leq \mu_r \leq \delta_r, 0 \leq \mu_f \leq \delta_f$ and the DC $i$. If $i$ satisfies the condition $\forall u, i(u) \subset Sol_{RIC}^{\mu_r, \delta_r, \mu_f, \delta_f}(u)$, then it is called *relative inertial delay condition* (RIDC). We also say that $i$ satisfies RIC (represented by $Sol_{RIC}^{\mu_r, \delta_r, \mu_f, \delta_f}$).

**9.2.2 Examples** a) If the DC $i$ fulfills $\forall u, i(u) \wedge Sol_{RIC}^{\mu_r, \delta_r, \mu_f, \delta_f}(u) \neq \varnothing$, then it defines (see Theorem 6.2.4 c)) the RIDC $i \wedge Sol_{RIC}^{\mu_r, \delta_r, \mu_f, \delta_f}$. The extreme situations are expressed by $Sol_{SC}$ and respectively by $I_d$, that define in this manner RIDC's for all $\mu_r, \delta_r, \mu_f, \delta_f$, respectively for $\mu_r = \mu_f = 0, \delta_r = \delta_f = d$.

b) The next functions

$$i(u) = \begin{cases} 1, \text{if } \exists \lim_{t \to \infty} u(t) \text{ and } \lim_{t \to \infty} u(t) = 1 \\ 0, \text{otherwise} \end{cases}$$

$$i(u) = \begin{cases} 0, \text{if } \exists \lim_{t \to \infty} u(t) \text{ and } \lim_{t \to \infty} u(t) = 0 \\ 1, \text{otherwise} \end{cases}$$

and more general the next functions

$$i(u) = \begin{cases} \{1\} \vee \{\chi_{[\tau,\infty)} \mid \tau \geq \delta\}, \delta = \begin{cases} \delta_r + \max \text{supp } Du, \text{supp } Du \neq \emptyset \\ 0, \text{supp } Du = \emptyset \end{cases}, \\ \quad \text{if } \exists \lim_{t \to \infty} u(t) \text{ and } \lim_{t \to \infty} u(t) = 1 \\ 0, \text{otherwise} \end{cases}$$

$$i(u) = \begin{cases} \{0\} \vee \{\chi_{(-\infty,\tau)} \mid \tau \geq \delta\}, \delta = \begin{cases} \delta_f + \max \text{supp } Du, \text{supp } Du \neq \emptyset \\ 0, \text{supp } Du = \emptyset \end{cases}, \\ \quad \text{if } \exists \lim_{t \to \infty} u(t) \text{ and } \lim_{t \to \infty} u(t) = 0 \\ 1, \text{otherwise} \end{cases}$$

are DC's defining RIDC's like at a) for all $\mu_r, \delta_r, \mu_f, \delta_f$.

c) Let $0 \leq m \leq d$. From 4.2.2 we have that the DC $x(t) = \bigcap_{\xi \in [t-d, t-d+m]} u(\xi)$ satisfies RIC

$$\overline{x(t-0)} \cdot x(t) \leq \bigcap_{\xi \in [t-d, t-d+m]} u(\xi)$$

$$x(t-0) \cdot \overline{x(t)} \leq \overline{u(t-d+m)}$$

and dually for $x(t) = \bigcup_{\xi \in [t-d, t-d+m]} u(\xi)$.

d) The DC that is defined by 8.2.5 (1), 8.2.5 (2) satisfies the relative inertial condition

$$\overline{x(t-0)} \cdot x(t) \leq \bigcap_{\xi \in [t-d_r, t-d_r+m_r]} u(\xi)$$

$$x(t-0) \cdot \overline{x(t)} \leq \bigcap_{\xi \in [t-d_f, t-d_f+m_f]} \overline{u(\xi)}$$

**9.2.3 Theorem** Let $0 \leq \mu_r \leq \delta_r, 0 \leq \mu_f \leq \delta_f$.

a) $Sol_{SC} \wedge Sol_{RIC}^{\mu_r, \delta_r, \mu_f, \delta_f}$ is time variable.

b) $Sol_{SC} \wedge Sol_{RIC}^{\mu_r, \delta_r, \mu_f, \delta_f}$ is symmetrical iff $\mu_r = \mu_f, \delta_r = \delta_f$.

**Proof** a) Like at Example 6.5.5 b), $1 = 1 \circ \tau^{-1} \in S$ and $\chi_{[0,\infty)} \in Sol_{SC}(1) \wedge Sol_{RIC}^{\mu_r, \delta_r, \mu_f, \delta_f}(1)$, but $\chi_{[0,\infty)} \circ \tau^{-1} = \chi_{[-1,\infty)} \notin S$.

b) <u>Only if</u> For $u = \chi_{[0,\infty)}$ we compute

$$\bigcap_{\xi \in [t-\delta_r, t-\delta_r+\mu_r]} \chi_{[0,\infty)}(\xi) = \chi_{[\delta_r,\infty)}(t)$$

$$\bigcap_{\xi \in [t-\delta_f, t-\delta_f+\mu_f]} \chi_{(-\infty,0)}(\xi) = \chi_{(-\infty,\delta_f-\mu_f)}(t)$$

$$\bigcap_{\xi \in [t-\delta_r, t-\delta_r+\mu_r]} \chi_{(-\infty,0)}(\xi) = \chi_{(-\infty,\delta_r-\mu_r)}(t)$$

$$\bigcap_{\xi \in [t-\delta_f, t-\delta_f+\mu_f]} \chi_{[0,\infty)}(\xi) = \chi_{[\delta_f,\infty)}(t)$$

The systems of inequalities

$$\begin{cases} \overline{x(t-0)} \cdot x(t) \leq \chi_{[\delta_r,\infty)}(t) \\ x(t-0) \cdot \overline{x(t)} \leq \chi_{(-\infty,\delta_f-\mu_f)}(t) \end{cases} \quad \begin{cases} x(t-0) \cdot \overline{x(t)} \leq \chi_{(-\infty,\delta_r-\mu_r)}(t) \\ \overline{x(t-0)} \cdot x(t) \leq \chi_{[\delta_f,\infty)}(t) \end{cases}$$

are equivalent only if $\delta_r = \delta_f$ and $\mu_r = \mu_f$.

**9.2.4 Remark** A property like the one that was stated for AIC in Theorem 8.2.8 is not true for RIC. Let for this the statement 8.2.8 a)

$$(Sol_{SC} \wedge Sol_{RIC}^{\mu'_r,\delta'_r,\mu'_f,\delta'_f}) \circ (Sol_{SC} \wedge Sol_{RIC}^{\mu_r,\delta_r,\mu_f,\delta_f}) =$$

$$= (Sol_{SC} \circ (Sol_{SC} \wedge Sol_{RIC}^{\mu_r,\delta_r,\mu_f,\delta_f})) \wedge Sol_{RIC}^{\mu'_r,\delta'_r,\mu'_f,\delta'_f} \subset$$

$$\subset (Sol_{SC} \circ Sol_{SC}) \wedge Sol_{RIC}^{\mu''_r,\delta''_r,\mu''_f,\delta''_f} \tag{1}$$

where $\mu''_r = \mu'_r, \delta''_r = \delta'_r, \mu''_f = \mu'_f, \delta''_f = \delta'_f$ or, in any case, $\mu''_r, \delta''_r, \mu''_f, \delta''_f$ are fixed and depend on $\mu_r, \delta_r, \mu_f, \delta_f, \mu'_r, \delta'_r, \mu'_f, \delta'_f$ only. We take $u(t) = \chi_{[t_1,\infty)}(t)$, $x(t) = 1, y(t) = \chi_{[t_0,\infty)}(t), 0 \leq t_0, 0 \leq t_1$ for which

$$y \in Sol_{SC}(x) \text{ and } y \in Sol_{RIC}^{\mu'_r,\delta'_r,\mu'_f,\delta'_f}(x) \text{ and } x \in Sol_{SC}(u) \text{ and } x \in Sol_{RIC}^{\mu_r,\delta_r,\mu_f,\delta_f}(u)$$

but

$$y \in (Sol_{SC} \circ Sol_{SC})(u) \text{ and } y \notin Sol_{RIC}^{\mu''_r,\delta''_r,\mu''_f,\delta''_f}(u)$$

if $t_1 > t_0 - \delta''_r$ and this shows the falsity of (1). In other words, inertia's inertia is not inertia, the inertia's paradox that we have mentioned in our introduction (Section 1).

The explanation of this paradox brings us from 8.2.8 to 6.8.6 a), where the proof for

$$(i \wedge U) \circ j = (i \circ j) \wedge U$$

was the next one: for any $u$

$$((i \wedge U) \circ j)(u) = \{y \mid \exists x, y \in i(x) \text{ and } y \in U \text{ and } x \in j(u)\} = ((i \circ j) \wedge U)(u)$$

and this was true for $U = Sol_{AIC}^{\delta'_r,\delta'_f}$ at 8.2.8. If $U$ is variable however $U = \varphi(\cdot)$ like here, with $\varphi(u) = Sol_{RIC}^{\mu'_r,\delta'_r,\mu'_f,\delta'_f}(u)$ the property 6.8.6 is false because the statements

$$((i \wedge \varphi) \circ j)(u) = \{y \mid \exists x, y \in i(x) \text{ and } y \in \varphi(x) \text{ and } x \in j(u)\}$$
$$((i \circ j) \wedge \varphi)(u) = \{y \mid \exists x, y \in i(x) \text{ and } y \in \varphi(u) \text{ and } x \in j(u)\}$$

are not equivalent. The first equality from (1) is false.

## 9.3 The Consistency Condition

**9.3.1 Theorem** The numbers $0 \leq m_r \leq d_r, 0 \leq m_f \leq d_f, 0 \leq \mu_r \leq \delta_r, 0 \leq \mu_f \leq \delta_f$ are given. The next system

$$\bigcap_{\xi \in [t-d_r, t-d_r+m_r]} u(\xi) \leq x(t) \leq \bigcup_{\xi \in [t-d_f, t-d_f+m_f]} u(\xi)$$

$$\overline{x(t-0)} \cdot x(t) \leq \bigcap_{\xi \in [t-\delta_r, t-\delta_r+\mu_r]} u(\xi)$$

$$x(t-0) \cdot \overline{x(t)} \leq \bigcap_{\xi \in [t-\delta_f, t-\delta_f+\mu_f]} \overline{u(\xi)}$$

where $u, x \in S$ has solutions for any $u$ if and only if one of the next requests is satisfied:

a)  $d_f - m_f \leq \delta_r \leq d_r \leq \delta_r - \mu_r + m_r$
    $d_r - m_r \leq \delta_f \leq d_f \leq \delta_f - \mu_f + m_f$

b)  $d_r - m_r + \mu_r \leq \delta_r \leq d_f - m_f \leq d_r$
    $d_f - m_f + \mu_f \leq \delta_f \leq d_r - m_r \leq d_f$

c)  $d_f - m_f \leq \delta_r \leq d_r - m_r + \mu_r \leq d_r$
    $d_r - m_r \leq \delta_f \leq d_f - m_f + \mu_f \leq d_f$

d)  $\delta_r \leq d_f - m_f \leq \delta_r + m_r - \mu_r \leq d_r$
    $\delta_f \leq d_r - m_r \leq \delta_f + m_f - \mu_f \leq d_f$

**Proof** Solutions exist iff whenever $x$ must have the value 1, respectively the value 0 in $t$ ( $\bigcap_{\xi \in [t-d_r, t-d_r+m_r]} u(\xi)$ switches in $t$ from 0 to 1, respectively $\bigcup_{\xi \in [t-d_f, t-d_f+m_f]} u(\xi)$ switches in $t$ from 1 to 0), RIC gives this possibility ($t'$ exists so that $\bigcap_{\xi \in [t'-\delta_r, t'-\delta_r+\mu_r]} u(\xi) = 1$, respectively so that $\bigcap_{\xi \in [t'-\delta_f, t'-\delta_f+\mu_f]} \overline{u(\xi)} = 1$) in time ($t' \in [t - d_r + d_f - m_f, t]$, respectively $t' \in [t - d_f + d_r - m_r, t]$).

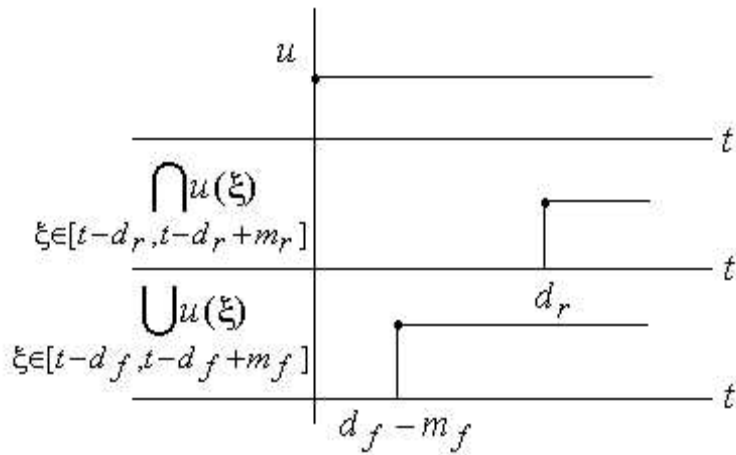

Fig 18

This happens for example in Fig 18 for $t = d_r$. We can write

$$\overline{u(t-d_r-0)} \cdot \bigcap_{\xi \in [t-d_r, t-d_r+m_r]} u(\xi) \leq \bigcup_{t' \in [t-d_r+d_f-m_f, t]} \bigcap_{\xi \in [t'-\delta_r, t'-\delta_r+\mu_r]} u(\xi) \quad (1)$$

inequality that is true for any $u$ [4]. The next statements are all equivalent with the previous one:

$$\forall u, \bigcap_{\xi \in [t-d_r, t-d_r+m_r]} u(\xi) \leq \bigcup_{t' \in [t-d_r+d_f-m_f, t]} \bigcap_{\xi \in [t'-\delta_r, t'-\delta_r+\mu_r]} u(\xi) \quad (2)$$

$((1) \Rightarrow (2))$: Let $0 \leq t_0$ and $u$ so that $\bigcap_{\xi \in [t_0, t_0+m_r]} u(\xi) = 1$ and we define $v(\xi) = \begin{cases} u(\xi), \xi \geq t_0 \\ 0, \xi < t_0 \end{cases}$. From

(1) and from $v(t) \leq u(t)$ we have $\overline{v(t_0 - 0)} \cdot \bigcap_{\xi \in [t_0, t_0+m_r]} v(\xi) = 1$ thus

$\exists t' \in [t_0 + d_f - m_f, t_0 + d_r]$, $1 = \bigcap_{\xi \in [t'-\delta_r, t'-\delta_r+\mu_r]} v(\xi) \leq \bigcap_{\xi \in [t'-\delta_r, t'-\delta_r+\mu_r]} u(\xi)$ thus (2) is true.

$(2) \Rightarrow (1)$: We have

$$\overline{u(t-d_r-0)} \cdot \bigcap_{\xi \in [t-d_r, t-d_r+m_r]} u(\xi) \leq \bigcap_{\xi \in [t-d_r, t-d_r+m_r]} u(\xi) \leq \bigcup_{t' \in [t-d_r+d_f-m_f, t]} \bigcap_{\xi \in [t'-\delta_r, t'-\delta_r+\mu_r]} u(\xi) \quad )$$

$$\exists t' \in [t-d_r+d_f-m_f, t], [t-d_r, t-d_r+m_r] \supset [t'-\delta_r, t'-\delta_r+\mu_r] \quad (3)$$

$$\exists t', \begin{cases} t-d_r+d_f-m_f \leq t' \\ t-d_r+\delta_r \leq t' \end{cases} \text{ and } \begin{cases} t' \leq t \\ t' \leq t-d_r+m_r+\delta_r-\mu_r \end{cases} \quad (4)$$

$$\max(t-d_r+d_f-m_f, t-d_r+\delta_r) \leq \min(t, t-d_r+m_r+\delta_r-\mu_r) \quad (5)$$

one of the next possibilities is true: (6)

j) $-d_r + d_f - m_f \leq -d_r + \delta_r$

$0 \leq -d_r + m_r + \delta_r - \mu_r$

$-d_r + \delta_r \leq 0$

jj) $-d_r + d_f - m_f \geq -d_r + \delta_r$

$0 \leq -d_r + m_r + \delta_r - \mu_r$

$-d_r + d_f - m_f \leq 0$

jjj) $-d_r + d_f - m_f \leq -d_r + \delta_r$

$0 \geq -d_r + m_r + \delta_r - \mu_r$

$-d_r + \delta_r \leq -d_r + m_r + \delta_r - \mu_r$

jv) $-d_r + d_f - m_f \geq -d_r + \delta_r$

$0 \geq -d_r + m_r + \delta_r - \mu_r$

$-d_r + d_f - m_f \leq -d_r + m_r + \delta_r - \mu_r$

It is shown that j), jj), jjj), jv) are equivalent with the first statements of a), b), c), d).

9.3.2 **Definition** The condition

9.3.1 a) *or* 9.3.1 b) *or* 9.3.1 c) *or* 9.3.1 d)

---

[4] The left term of this inequality represents $D_{01} \bigcap_{\xi \in [t-d_r, t-d_r+m_r]} u(\xi)$, see 4.2.2

is called the *consistency condition* (of the bounded relative inertial delay condition) (CC$_{BRIDC}$).

**9.3.3 Theorem** a) Any of 9.3.1 a),...,9.3.1 d) implies $d_f \geq d_r - m_r, d_r \geq d_f - m_f$ in other words CC$_{BRIDC}$ in stronger than CC$_{BDC}$.

b) CC$_{BRIDC}$ implies the next conditions (that are conditions of necessity for the existence of a solution of the system 9.3.1)
$$m_r \geq \mu_r, m_f \geq \mu_f$$
$$\delta_r \leq d_r, \delta_f \leq d_f$$
$$d_f - m_f \leq \delta_r - \mu_r + m_r, d_r - m_r \leq \delta_f - \mu_f + m_f$$

c) The next conditions (that are sufficient for the existence of a solution of the system 9.3.1)
$$d_f - m_f \leq \delta_f - \mu_f \leq \delta_r \leq d_r$$
$$d_r - m_r \leq \delta_r - \mu_r \leq \delta_f \leq d_f$$
imply CC$_{BRIDC}$.

**Proof** b) It is shown that any of 9.3.1 a),...,d) implies these inequalities..

c) It is shown that they imply 9.3.1 a).

**9.3.4 Remarks** (*Special cases of* CC$_{BRIDC}$) If in CC$_{BRIDC}$ $\mu_r = m_r = \mu_f = m_f = 0$, then it takes the form $\delta_r = d_r = \delta_f = d_f = d$ and the system 9.3.1 degenerates in FDC. If $m_r = d_r, m_f = d_f$, $\mu_r = \delta_r, \mu_f = \delta_f$ then CC$_{BRIDC}$ becomes $\delta_r \leq d_r, \delta_f \leq d_f$. If $d_r = \delta_r, d_f = \delta_f, m_r = \mu_r$, $m_f = \mu_f$, then CC$_{BRIDC}$ becomes $d_f \geq d_r - m_r, d_r \geq d_f - m_f$, i.e. CC$_{BDC}$.

**9.4 Bounded Relative Inertial Delays**

**9.4.1 Definition** Let $u, x \in S$ and the numbers $0 \leq m_r \leq d_r, 0 \leq m_f \leq d_f$, $0 \leq \mu_r \leq \delta_r, 0 \leq \mu_f \leq \delta_f$ so that CC$_{BRIDC}$ be satisfied. The system of inequalities

$$\bigcap_{\xi \in [t-d_r, t-d_r+m_r]} u(\xi) \leq x(t) \leq \bigcup_{\xi \in [t-d_f, t-d_f+m_f]} u(\xi)$$

$$\overline{x(t-0)} \cdot x(t) \leq \bigcap_{\xi \in [t-\delta_r, t-\delta_r+\mu_r]} u(\xi)$$

$$x(t-0) \cdot \overline{x(t)} \leq \bigcap_{\xi \in [t-\delta_f, t-\delta_f+\mu_f]} \overline{u(\xi)}$$

is called the *bounded relative inertial delay condition* (BRIDC); We say that the tuple $(u, m_r, d_r, m_f, d_f, \mu_r, \delta_r, \mu_f, \delta_f, x)$ satisfies BRIDC.

We call also bounded relative inertial delay condition the DC $Sol_{BDC}^{m_r, d_r, m_f, d_f} \wedge Sol_{RIC}^{\mu_r, \delta_r, \mu_f, \delta_f}$.

**9.4.2 Remark** BRIDC is obviously a DC, from Theorem 9.3.1. Its meaning results from Fig 19 representing the situation from 9.3.1 c), where we have supposed that $x(0-0) = 0, u(t) = \chi_{[0,\tau)}(t), \tau > m_r$. The functions $\bigcap_{\xi \in [t-d_r, t-d_r-m_r]} u(\xi)$, $\bigcup_{\xi \in [t-d_f, t-d_f+m_f]} u(\xi)$ from BDC give the possibility, respectively state the necessity of a value of $x$, while the functions $\bigcap\limits_{\xi\in[t-\delta_r,t-\delta_r+\mu_r]} u(\xi)$, $\bigcap\limits_{\xi\in[t-\delta_f,t-\delta_f+\mu_f]} \overline{u(\xi)}$ from RIC give the possibility that some switches of $x$ happen.

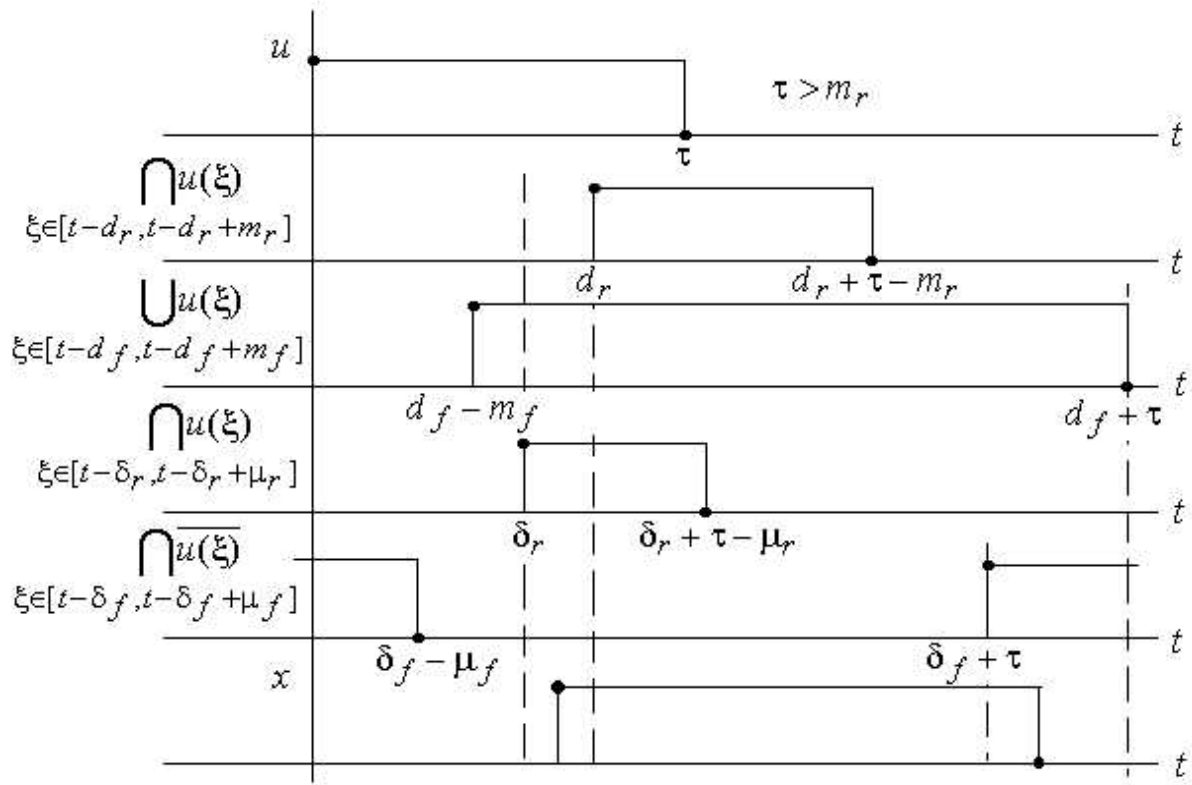

Fig 19

These interpretations may be done in other cases of $CC_{BRIDC}$ (or violations of $CC_{BRIDC}$!), for example 9.3.1 a) from Fig 20, where $x(0-0)=0, u(t)=\chi_{[0,\tau)}(t), \tau>m_r$

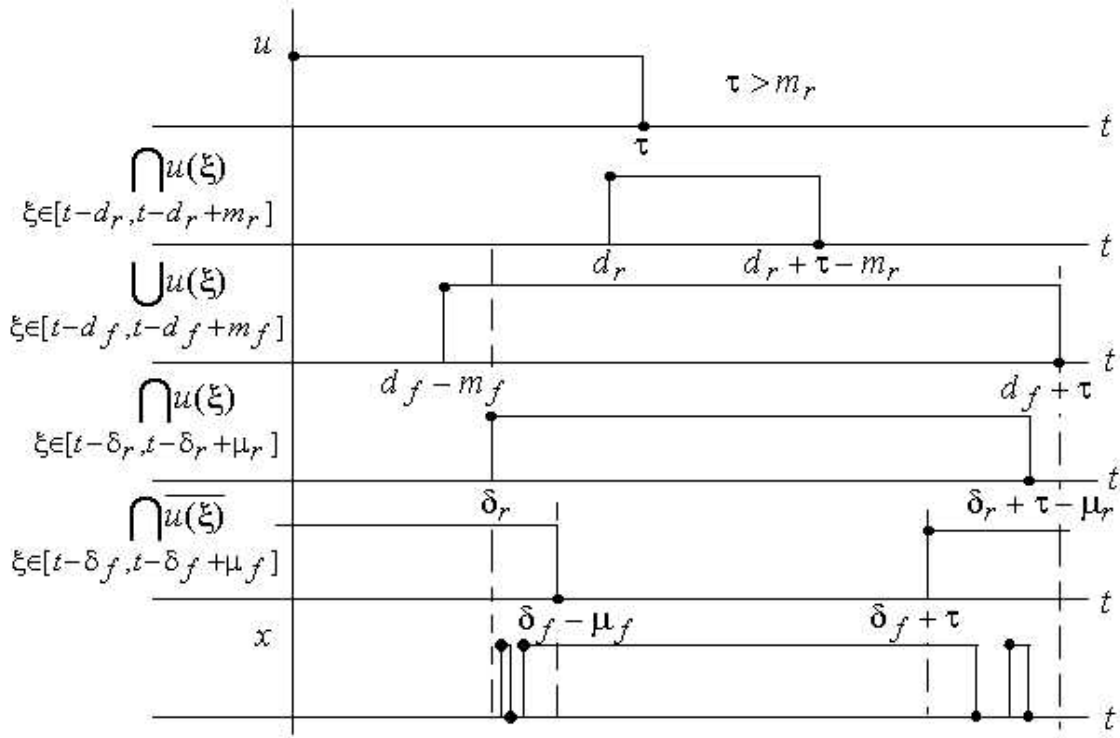

Fig 20

The intervals $[\delta_r, \delta_f - \mu_f]$ and $[\delta_f + \tau, \delta_r + \tau - \mu_r]$ are observed; if they are non-empty, then Zeno solutions of RIC exist.

We refer in Fig 21 to the requests 9.3.3 c); we suppose without loss that $x(0-0) = 0, u(t) = \chi_{[0,\tau)}(t), \tau > m_r$ and we have the next time intervals when $u$ has sufficiently long pulses ($u$ is 1 strictly longer than $m_r$ and then $u$ is 0 strictly longer than $m_f$).

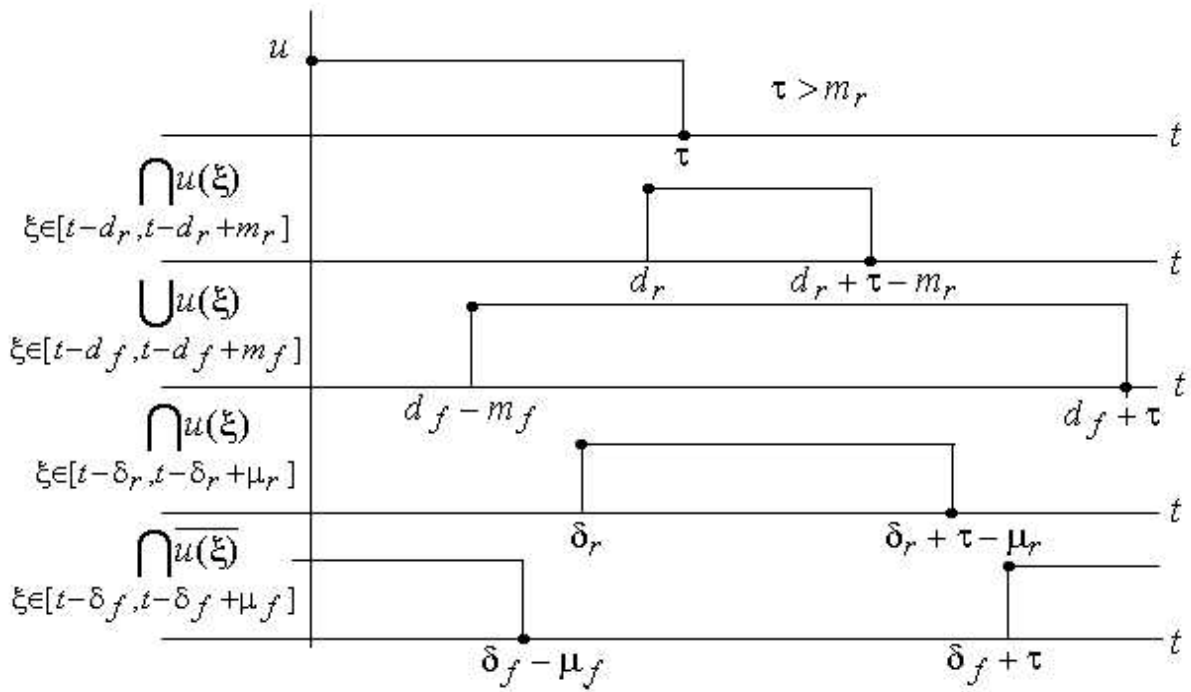

- $t \in (-\infty, d_f - m_f)$; $x(t) = 0$ and the only possible switch -that does not happen- is from 1 to 0
- $t \in [d_f - m_f, \delta_f - \mu_f)$, $x(t) = 0$;. $x(t)$ could be 1, but it is allowed to switch only from 1 to 0 and this does not happen
- $t \in [\delta_f - \mu_f, \delta_r)$, $x(t) = 0$;. $x(t)$ could be 1, but no switch is allowed
- $t \in [\delta_r, d_r]$; $x(t) = 0$ and $x(t) = 1$ are both allowed, switching from 0 to 1 may happen and exactly one such switch takes place
- $t \in (d_r, d_r + \tau - m_r)$; $x(t) = 1$ and the only possible switch -that does not happen- is from 0 to 1
- $t \in [d_r + \tau - m_r, \delta_r + \tau - \mu_r)$, $x(t) = 1$; $x(t)$ could be 0, but it is allowed to switch only from 0 to 1 and this does not happen
- $t \in [\delta_r + \tau - \mu_r, \delta_f + \tau)$, $x(t) = 1$; $x(t)$ could be 0, but no switches are possible in this time interval
- $t \in [\delta_f + \tau, d_f + \tau]$; $x(t) = 1$ and $x(t) = 0$ are both possible, switching from 1 to 0 is allowed and exactly one such switch takes place
- $t \in (d_f + \tau, \infty)$; $x(t) = 0$ and the only possible switch -that does not happen- is from 1 to 0.

This succession of time intervals is repetitive if $u$ is suitable chosen.

**9.4.3 Theorem** The next properties are equivalent in the sense that the arbitrary signals $u, x$ satisfy one of them if and only if they satisfy the other one.

a) $m_r, d_r, m_f, d_f, \mu_r, \delta_r, \mu_f, \delta_f$ are given and the next inequalities are fulfilled (see also 9.3.3 c))

$$0 \leq m_r \leq d_r, 0 \leq m_f \leq d_f \tag{1}$$

$$0 \leq \mu_r \leq \delta_r, 0 \leq \mu_f \leq \delta_f \tag{2}$$

$$d_f - m_f \leq \delta_f - \mu_f \leq \delta_r \leq d_r \tag{3}$$

$$d_r - m_r \leq \delta_r - \mu_r \leq \delta_f \leq d_f \tag{4}$$

$$\bigcap_{\xi \in [t-d_r, t-d_r+m_r]} u(\xi) \leq x(t) \leq \bigcup_{\xi \in [t-d_f, t-d_f+m_f]} u(\xi) \tag{5}$$

$$\overline{x(t-0)} \cdot x(t) \leq \bigcap_{\xi \in [t-\delta_r, t-\delta_r+\mu_r]} u(\xi) \tag{6}$$

$$x(t-0) \cdot \overline{x(t)} \leq \bigcap_{\xi \in [t-\delta_f, t-\delta_f+\mu_f]} \overline{u(\xi)} \tag{7}$$

b) The numbers $m_{r,\min}, d_{r,\min}, m_{r,\max}, d_{r,\max}, m_{f,\min}, d_{f,\min}, m_{f,\max}, d_{f,\max}$ are given and we have

$$0 \leq m_{r,\max} \leq d_{r,\max}, 0 \leq m_{f,\max} \leq d_{f,\max} \tag{8}$$

$$0 \leq m_{r,\min} \leq d_{r,\min}, 0 \leq m_{f,\min} \leq d_{f,\min} \tag{9}$$

$$d_{f,\max} - m_{f,\max} \leq d_{f,\min} - m_{f,\min} \leq d_{r,\min} \leq d_{r,\max} \tag{10}$$

$$d_{r,\max} - m_{r,\max} \leq d_{r,\min} - m_{r,\min} \leq d_{f,\min} \leq d_{f,\max} \tag{11}$$

$$\overline{x(t-0)} \cdot \bigcap_{\xi \in [t-d_{r,\max}, t-d_{r,\max}+m_{r,\max}]} u(\xi) \leq \overline{x(t-0)} \cdot x(t) \leq \tag{12}$$
$$\leq \overline{x(t-0)} \cdot \bigcap_{\xi \in [t-d_{r,\min}, t-d_{r,\min}+m_{r,\min}]} u(\xi)$$

$$x(t-0) \cdot \bigcap_{\xi \in [t-d_{f,\max}, t-d_{f,\max}+m_{f,\max}]} \overline{u(\xi)} \leq x(t-0) \cdot \overline{x(t)} \leq \tag{13}$$
$$\leq x(t-0) \cdot \bigcap_{\xi \in [t-d_{f,\min}, t-d_{f,\min}+m_{f,\min}]} \overline{u(\xi)}$$

**Proof** The next equalities take place

$$m_{r,\min} = \mu_r, d_{r,\min} = \delta_r \tag{14}$$
$$m_{r,\max} = m_r, d_{r,\max} = d_r \tag{15}$$
$$m_{f,\min} = \mu_f, d_{f,\min} = \delta_f \tag{16}$$
$$m_{f,\max} = m_f, d_{f,\max} = d_f \tag{17}$$

under the form 'equal by definition with' in both directions $a) \Rightarrow b)$ and $b) \Rightarrow a)$ resulting that (1),…,(4) and (8),…,(11) coincide.

$a) \Rightarrow b)$ The left inequality of (5) multiplied with $\overline{x(t-0)}$ gives the left inequality of (12) and (6) multiplied with $\overline{x(t-0)}$ gives the right inequality of (12). The rest results by duality.

$b) \Rightarrow a)$ We suppose that $\bigcap_{\xi \in [t-d_r, t-d_r+m_r]} u(\xi) = 1$ and we have the next possibilities

i) $x(t-0) = 0$

Then the left inequality of (12) shows that $x(t) = 1$ and the left inequality of (5) is satisfied.

ii) $x(t-0) = 1$

and the right inequality of (13) becomes

$$\overline{x(t)} \leq \bigcap_{\xi \in [t-\delta_f, t-\delta_f+\mu_f]} \overline{u(\xi)} \tag{18}$$

(10), (11) are written under the form (3), (4) and this implies

$$t - \delta_f + \mu_f \geq t - d_r \tag{19}$$
$$t - d_r + m_r \geq t - \delta_f \tag{20}$$

i.e.

$$[t-d_r, t-d_r+m_r] \wedge [t-\delta_f, t-\delta_f+\mu_f] \neq \emptyset \tag{21}$$

and thus $\bigcap_{\xi \in [t-\delta_f, t-\delta_f+\mu_f]} \overline{u(\xi)} = 0$. From (18) we get $x(t) = 1$ and the left inequality of (5) is satisfied in this case too.

On the other hand, the right inequality of (12) gives

$$\overline{x(t-0)} \cdot x(t) \leq \overline{x(t-0)} \cdot \bigcap_{\xi \in [t-\delta_r, t-\delta_r+\mu_r]} u(\xi) \leq \bigcap_{\xi \in [t-\delta_r, t-\delta_r+\mu_r]} u(\xi) \tag{22}$$

i.e. (6).

The other implications result by duality.

## 9.5 Deterministic Bounded Relative Inertial Delays

**9.5.1 Theorem** Let the real numbers $0 \leq m_r \leq d_r, 0 \leq m_f \leq d_f$ arbitrary with $d_r - m_r \leq d_f$, $d_f - m_f \leq d_r$. The next systems are equivalent, in the sense that if $u, x \in S$ satisfy one of them, then they also satisfy any other.

a)
$$\bigcap_{\xi \in [t-d_r, t-d_r-m_r]} u(\xi) \leq x(t) \leq \bigcup_{\xi \in [t-d_f, t-d_f+m_f]} u(\xi) \quad (1)$$

$$\overline{x(t-0)} \cdot x(t) \leq \bigcap_{\xi \in [t-d_r, t-d_r+m_r]} u(\xi) \quad (2)$$

$$x(t-0) \cdot \overline{x(t)} \leq \bigcap_{\xi \in [t-d_f, t-d_f+m_f]} \overline{u(\xi)} \quad (3)$$

b)
$$\overline{x(t-0)} \cdot x(t) = \overline{x(t-0)} \cdot \bigcap_{\xi \in [t-d_r, t-d_r+m_r]} u(\xi) \quad (4)$$

$$x(t-0) \cdot \overline{x(t)} = x(t-0) \cdot \bigcap_{\xi \in [t-d_f, t-d_f+m_f]} \overline{u(\xi)} \quad (5)$$

c)
$$\bigcap_{\xi \in [t-d_r, t-d_r+m_r]} u(\xi) \leq x(t) \quad (6)$$

$$\bigcap_{\xi \in [t-d_f, t-d_f+m_f]} \overline{u(\xi)} \leq \overline{x(t)} \quad (7)$$

$$\overline{\bigcap_{\xi \in [t-d_r, t-d_r+m_r]} u(\xi)} \cdot \overline{\bigcap_{\xi \in [t-d_f, t-d_f+m_f]} \overline{u(\xi)}} \leq$$
$$\leq \overline{x(t-0)} \cdot \overline{x(t)} \cup x(t-0) \cdot x(t) \quad (8)$$

d)
$$x(t) = \begin{cases} 1, & \bigcap_{\xi \in [t-d_r, t-d_r+m_r]} u(\xi) = 1 \\ 0, & \bigcap_{\xi \in [t-d_f, t-d_f+m_f]} \overline{u(\xi)} = 1 \\ x(t-0), & \text{otherwise} \end{cases} \quad (9)$$

e) $x(t) = \bigcap_{\xi \in [t-d_r, t-d_r+m_r]} u(\xi) \cup x(t-0) \cdot \bigcup_{\xi \in [t-d_f, t-d_f+m_f]} u(\xi) \quad (10)$

f) $Dx(t) = \overline{x(t-0)} \cdot \bigcap_{\xi \in [t-d_r, t-d_r+m_r]} u(\xi) \cup$
$$\cup x(t-0) \cdot \bigcap_{\xi \in [t-d_f, t-d_f+m_f]} \overline{u(\xi)} \quad (11)$$

g) $\overline{x(t-0)} \cdot x(t) \cdot \bigcap_{\xi \in [t-d_r, t-d_r+m_r]} u(\xi) \cup x(t-0) \cdot \overline{x(t)} \cdot \bigcap_{\xi \in [t-d_f, t-d_f+m_f]} \overline{u(\xi)} \cup$
$$\cup \overline{x(t-0)} \cdot \overline{x(t)} \cdot \overline{\bigcap_{\xi \in [t-d_r, t-d_r+m_r]} u(\xi)} \cup$$
$$\cup x(t-0) \cdot x(t) \cdot \overline{\bigcap_{\xi \in [t-d_f, t-d_f+m_f]} \overline{u(\xi)}} = 1 \quad (12)$$

**Proof** In a),...,g) three possibilities exist, due to the satisfaction of CC$_{BDC}$:

i) $$\bigcap_{\xi \in [t-d_r, t-d_r+m_r]} u(\xi) = \bigcup_{\xi \in [t-d_f, t-d_f+m_f]} u(\xi) = 0$$

ii) $$\bigcap_{\xi \in [t-d_r, t-d_r+m_r]} u(\xi) = 0, \quad \bigcup_{\xi \in [t-d_f, t-d_f+m_f]} u(\xi) = 1$$

iii) $$\bigcap_{\xi \in [t-d_r, t-d_r+m_r]} u(\xi) = \bigcup_{\xi \in [t-d_f, t-d_f+m_f]} u(\xi) = 1$$

<u>Case i)</u> a) gives $0 \leq x(t) \leq 0$ from (1), thus $x(t) = 0$. By taking into account the fact that $\overline{\bigcup_{\xi \in [t-d_f, t-d_f+m_f]} u(\xi)} = \bigcap_{\xi \in [t-d_f, t-d_f+m_f]} \overline{u(\xi)} = 1$, b) is

$$x(t-0) \cdot \overline{x(t)} = 0$$
$$x(t-0) \cdot \overline{x(t)} = x(t-0)$$

whose unique solution is $x(t) = 0$. c) shows, from (7), that $1 \leq \overline{x(t)}$, that is $x(t) = 0$. d) and e) give $x(t) = 0$ too. f) becomes $x(t-0) \oplus x(t) = x(t-0)$, in other words $x(t) = 0$. Because $\overline{\bigcap_{\xi \in [t-d_r, t-d_r+m_r]} u(\xi)} = 1$, g) is in this case

$$x(t-0) \cdot \overline{x(t)} \cup \overline{x(t-0)} \cdot \overline{x(t)} = (x(t-0) \cup \overline{x(t-0)}) \cdot \overline{x(t)} = \overline{x(t)} = 1$$

i.e. $x(t) = 0$.

The other two cases are similar, with $x(t) = x(t-0)$ for ii) and $x(t) = 1$ for iii). $\square$

**9.5.2 Remarks** In any of the equivalent conditions 9.5.1 a),...,9.5.1 g), CC$_{BRIDC}$ and CC$_{BDC}$ coincide, as we have mentioned at 9.3.4.

The implications of the violation of this condition in 9.5.1 a),...,9.5.1 g) are the following. Let $u$ and $t'$ so that

$$[t'-d_r, t'-d_r+m_r] \wedge [t'-d_f, t'-d_f+m_f] = \emptyset$$
$$\forall \xi \in [t'-d_r, t'-d_r+m_r], u(\xi) = 1$$
$$\forall \xi \in [t'-d_f, t'-d_f+m_f], u(\xi) = 0$$

Let us suppose, in order to make a choice, that $t'-d_r+m_r < t'-d_f$. From the last two equations and from the right continuity of $u$ in $t'-d_r+m_r, t'-d_f+m_f$ we get the existence of $\varepsilon > 0$ so that $t'-d_r+m_r+\varepsilon < t'-d_f$ and

$$\forall \xi \in [t'-d_r, t'-d_r+m_r+\varepsilon], u(\xi) = 1$$
$$\forall \xi \in [t'-d_f, t'-d_f+m_f+\varepsilon], u(\xi) = 0$$

9.5.1 a) shows that $Sol_{BDC}^{m_r, d_r, m_f, d_f}(u) = \emptyset$

9.5.1 b) becomes for any $t \in [t', t'+\varepsilon]$

$$x(t-0) \cdot x(t) = \overline{x(t-0)}$$
$$x(t-0) \cdot \overline{x(t)} = x(t-0)$$

and the system accepts two possibilities $x(t) = 0, x(t-0) = 1$ and $x(t) = 1, x(t-0) = 0$, meaning that $x$ should continuously switch in this interval; no signal satisfies such requests. 9.5.1 c) has no solution either, because 9.5.1 (6), 9.5.1 (7) show that $x(t) = 0$ and $x(t) = 1$ are both true for $t \in [t', t'+\varepsilon]$ and this is the case of 9.5.1 d) also, where $x(t)$ is not a well defined function for $t \in [t',t'+\varepsilon]$. 9.5.1 e) gives $x(t) = 1$, $t \in [t',t'+\varepsilon]$. 9.5.1 f) and 9.5.1 g) become both
$$Dx(t) = 1, t \in [t',t'+\varepsilon]$$
and this equation represents a nonsense similar with the one from 9.5.1 b), because the set $\{t \mid t \in [t',t'+\varepsilon], Dx(t) = 1\}$ should be finite ($x$ has resulted to be 'everywhere discontinuous' in $[t',t'+\varepsilon]$).

On the other hand, Theorem 8.2.5 characterizes, via 9.5.1 b), all of 9.5.1 a),…,9.5.1 g).

**9.5.3 Definition** For $u, x \in S$ and $0 \leq m_r \leq d_r, 0 \leq m_f \leq d_f$ so that CC$_{BDC}$ is satisfied, any of the equivalent properties 9.5.1 a), ..., 9.5.1 g) is called the *deterministic bounded relative inertial delay condition* (DBRIDC). We say that the tuple $(u, m_r, d_r, m_f, d_f, x)$ satisfies DBRIDC.

We call also deterministic bounded relative inertial delay condition the function $Sol_{DBRIDC}^{m_r, d_r, m_f, d_f} : S \to P*(S)$ defined by
$$Sol_{DBRIDC}^{m_r, d_r, m_f, d_f}(u) = Sol_{BDC}^{m_r, d_r, m_f, d_f}(u) \wedge Sol_{RIC}^{m_r, d_r, m_f, d_f}(u)$$

# 10 Alternative Definitions. Symmetrical Deterministic Upper Bounded, Lower Unbounded Relative Inertial Delays

## 10.1 Alternative Definitions

We just mention the possibility of replacing
- in BDC the functions $\bigcap_{\xi \in [t-d_r, t-d_r+m_r]} u(\xi)$, $\bigcup_{\xi \in [t-d_f, t-d_f+m_f]} u(\xi)$ with $\bigcap_{\xi \in [t-d_r, t)} u(\xi)$, $\bigcup_{\xi \in [t-d_f, t)} u(\xi)$; new notation BDC'

- in AIC the functions $\bigcap_{\xi \in [t, t+\delta_r]} x(\xi), \bigcap_{\xi \in [t, t+\delta_f]} \overline{x(\xi)}$ with $\bigcap_{\xi \in [t, t+\delta_r)} x(\xi), \bigcap_{\xi \in [t, t+\delta_f)} \overline{x(\xi)}$; new notation AIC'

- in RIC the functions $\bigcap_{\xi \in [t-\delta_r, t-\delta_r+\mu_r]} u(\xi)$, $\bigcap_{\xi \in [t-\delta_f, t-\delta_f+\mu_f]} \overline{u(\xi)}$ with $\bigcap_{\xi \in [t-\delta_r, t)} u(\xi), \bigcap_{\xi \in [t-\delta_f, t)} \overline{u(\xi)}$; new notation RIC'.

In BDC' $d_r > 0, d_f > 0$ is a consistency condition, but in AIC' and RIC' any of $\delta_r, \delta_f$ can be null, resulting trivial conditions. Such substitutions are natural, but BDC' and RIC' are suggested also by the 'common form of implementation…', from Convention 5.3.

None of the six new functions are signals, however, because they are not right continuous.

## 10.2 Symmetrical Deterministic Upper Bounded, Lower Unbounded Relative Inertial Delays

**10.2.1 Lemma** Let $d_r > 0, d_f > 0$ and $u \in S$. The next formulas are true:
$$\bigcap_{\xi \in [t-d_r, t)} u(\xi) = u(t-0) \cdot \overline{\bigcup_{\xi \in (t-d_r, t)} Du(\xi)}$$

$$\bigcap_{\xi \in [t-d_f,t)} \overline{u(\xi)} = \overline{u(t-0)} \cdot \overline{\bigcup_{\xi \in (t-d_f,t)} Du(\xi)}$$

**Proof** We prove the first of these two relations and let $t$ arbitrary, fixed. We have:
$$\bigcap_{\xi \in [t-d_r,t)} u(\xi) = 1 \Leftrightarrow u(t-0) = 1 \text{ and } u_{|[t-d_r,t)} \text{ is constant}$$
$$\Leftrightarrow u(t-0) = 1 \text{ and } u_{|(t-d_r,t)} \text{ is constant}$$

(because $u$ is right continuous in $t - d_r$)
$$\Leftrightarrow u(t-0) = 1 \text{ and } \forall \xi \in (t-d_r,t), Du(\xi) = 0$$
$$\Leftrightarrow u(t-0) \cdot \overline{\bigcup_{\xi \in (t-d_r,t)} Du(\xi)} = 1$$

Because $t$ was arbitrary, the equation is proved.

**10.2.2 Remarks** The idea from 10.1 has implications in the way of understanding 9.5.1, for example 9.5.1 f) can be thought under the form
$$Dx(t) = \overline{x(t-0)} \cdot \bigcap_{\xi \in [t-d_r,t)} u(\xi) \cup x(t-0) \cdot \bigcap_{\xi \in [t-d_f,t)} \overline{u(\xi)} \qquad (1)$$

The natural notation of equation (1) following the previous patterns is DBRIDC'.

In the rising-falling symmetric version, when $d_r = d_f = d > 0$:
$$\overline{x(t-0)} \cdot \bigcap_{\xi \in [t-d,t)} u(\xi) \cup x(t-0) \cdot \bigcap_{\xi \in [t-d,t)} \overline{u(\xi)} =$$
$$= \overline{x(t-0)} \cdot u(t-0) \cdot \overline{\bigcup_{\xi \in (t-d,t)} Du(\xi)} \cup x(t-0) \cdot \overline{u(t-0)} \cdot \overline{\bigcup_{\xi \in (t-d,t)} Du(\xi)} = \quad \text{(from 10.2.1)}$$
$$= (\overline{x(t-0)} \cdot u(t-0) \cup x(t-0) \cdot \overline{u(t-0)}) \cdot \overline{\bigcup_{\xi \in (t-d,t)} Du(\xi)} =$$
$$= (x(t-0) \oplus u(t-0)) \cdot \overline{\bigcup_{\xi \in (t-d,t)} Du(\xi)}$$

and equation (1) becomes
$$Dx(t) = (x(t-0) \oplus u(t-0)) \cdot \overline{\bigcup_{\xi \in (t-d,t)} Du(\xi)} \qquad (2)$$

In [23], the equation of the inertial delay circuit with null initial conditions was written under the form
$$Dx(t) = (x(t-0) \oplus u(t-0)) \cdot \overline{\bigcup_{\xi \in (t-d,t)} Du(\xi)} \cdot \chi_{[d,\infty)}(t) \qquad (3)$$

The context from there demanded that the signals $f$ satisfy by definition the non-restrictive request $supp\, f \subset [0,\infty)$, thus the general form of equation (3) is
$$Dx(t) = (x(t-0) \oplus u(t-0)) \cdot \overline{\bigcup_{\xi \in (t-d,t)} Du(\xi)} \cdot \chi_{[d,\infty)}(t) \oplus x^0 \cdot \chi_{\{0\}}(t) \qquad (3')$$

where the initial value $x^0 \in \boldsymbol{B}$ stands for $x(0-0)$ in our present work.

In the next drawing

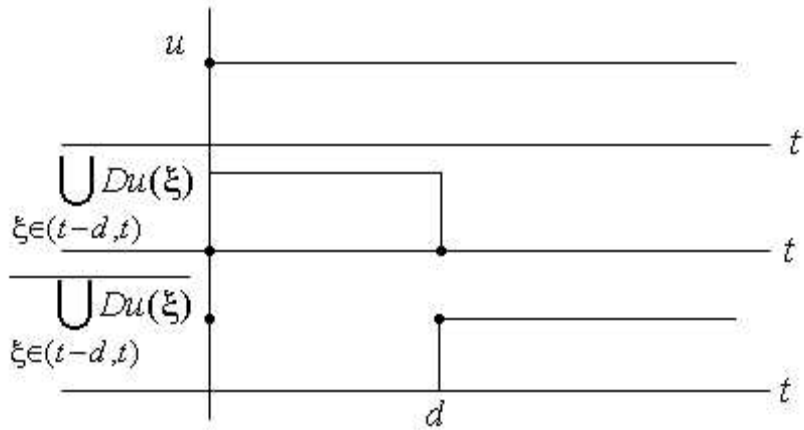

Fig 22

we have shown that, because $x(0-0) \oplus u(0-0) = 0$ (a slightly modified version of Theorem 7.2.4 is valid here) both equations (2) and (3) satisfy $\forall u, \forall t < d, Dx(t) = 0$ thus they are equivalent. The abbreviation is SDBRIDC'.

## 11. Other Examples and Applications

### 11.1 A Delay Line for the Falling Transitions Only

The proposed circuit that is reproduced from [11] has the gates and the wires governed by delays

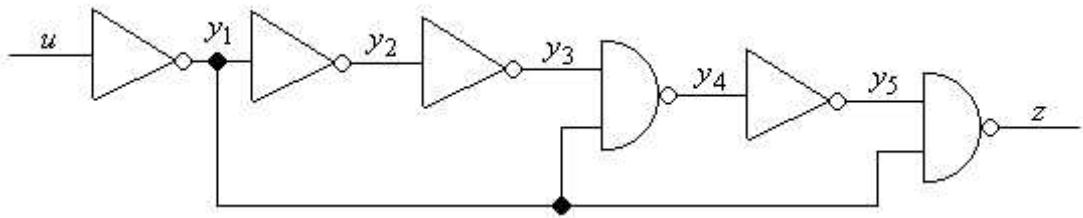

Fig 23

From the static point of view, $u, y_1, ..., y_5, z \in \mathbf{B}$ and we observe that

$$z = \overline{y_1 \cdot y_5} = \overline{y_1 \cdot \overline{y_4}} = \overline{y_1 \cdot \overline{(y_3 \cdot y_1)}} = \overline{y_1 \cdot \overline{y_3}} = \overline{y_1 \cdot \overline{\overline{y_2}}} = \overline{y_1 \cdot \overline{\overline{\overline{y_1}}}} = \overline{\overline{y_1}} = u \quad (1)$$

i.e. the Boolean function that this circuit computes is the identity.

The model is offered by the next circuit where all the variables that occur are signals, the gates and the wires have no delays and the delays are concentrated in the delay circuits. The system of equations and inequalities is:

$$y_1(t) = \overline{u(t)} \quad (2)$$
$$y_2(t) = \overline{x_1(t)} \quad (3)$$
$$y_3(t) = \overline{x_2(t)} \quad (4)$$
$$y_4(t) = \overline{x_3(t) \cdot x_1(t)} \quad (5)$$
$$y_5(t) = \overline{x_4(t)} \quad (6)$$
$$z(t) = \overline{x_5(t) \cdot x_1(t)} \quad (7)$$
$$\bigcap_{\xi \in [t-d_r, t)} y_i(\xi) \leq x_i(t) \leq \bigcup_{\xi \in [t-d_f, t)} y_i(\xi), i = \overline{1,5} \quad (8)$$
$$\bigcap_{\xi \in [t-d_r, t)} z(\xi) \leq w(t) \leq \bigcup_{\xi \in [t-d_f, t)} z(\xi) \quad (9)$$

thus we use BDC' with $d_r > 0, d_f > 0$ the parameters that characterize the six delay circuits. The compatibility with the initial conditions is supposed to be satisfied under the form

$$u(0-0) = \overline{y_1(0-0)} = x_1(0-0) = \overline{y_2(0-0)} = x_2(0-0) = \overline{y_3(0-0)} = x_3(0-0) =$$
$$= \overline{y_4(0-0)} = x_4(0-0) = \overline{y_5(0-0)} = x_5(0-0) = z(0-0) = w(0-0)$$

that simplifies (2),...,(7). For example (5) was written in the next manner:

$$y_4(t) = y_4(0-0) \cdot \chi_{(-\infty,0)}(t) \oplus \overline{x_3(t) \cdot x_1(t)} \cdot \chi_{[0,\infty)}(t) =$$
$$= \overline{x_3(0-0) \cdot x_1(0-0)} \cdot \chi_{(-\infty,0)}(t) \oplus \overline{x_3(t) \cdot x_1(t)} \cdot \chi_{[0,\infty)}(t) = \overline{x_3(t) \cdot x_1(t)}$$

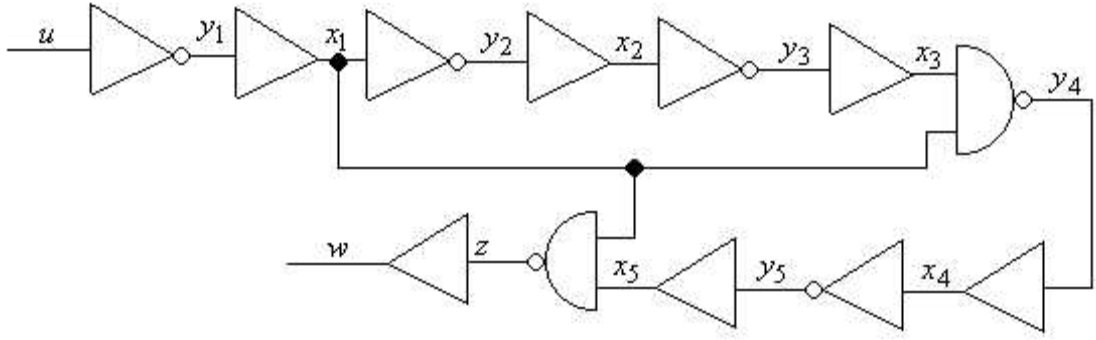

Fig 24

We have

$$\bigcap_{\xi \in [t-d_r, t)} y_1(\xi) \leq x_1(t) \leq \bigcup_{\xi \in [t-d_f, t)} y_1(\xi) \quad \text{(equation (8))} \quad (10)$$

$$\bigcap_{\xi \in [t-d_r, t)} \overline{u(\xi)} \leq x_1(t) \leq \bigcup_{\xi \in [t-d_f, t)} \overline{u(\xi)} \quad \text{(from (2) and (10))} \quad (11)$$

$$\bigcap_{\xi \in [t-d_r, t)} y_3(\xi) \leq x_3(t) \leq \bigcup_{\xi \in [t-d_f, t)} y_3(\xi) \quad \text{(equation (8))} \quad (12)$$

$$\bigcap_{\xi \in [t-d_r, t)} \overline{x_2(\xi)} \leq x_3(t) \leq \bigcup_{\xi \in [t-d_f, t)} \overline{x_2(\xi)} \quad \text{(from (4) and (12))} \quad (13)$$

$$\bigcap_{\xi \in [t-d_r,t)} y_2(\xi) \le x_2(t) \le \bigcup_{\xi \in [t-d_f,t)} y_2(\xi) \qquad \text{(equation (8))} \qquad (14)$$

$$\bigcap_{\xi \in [t-d_f,t)} \overline{y_2(\xi)} \le \overline{x_2(t)} \le \bigcup_{\xi \in [t-d_r,t)} \overline{y_2(\xi)} \qquad \text{(from (14))} \qquad (15)$$

$$\bigcap_{\xi \in [t-d_f,t)} x_1(\xi) \le \overline{x_2(t)} \le \bigcup_{\xi \in [t-d_r,t)} x_1(\xi) \qquad \text{(from (3) and (15))} \qquad (16)$$

$$\bigcap_{\xi \in [t-d_r-d_f,t)} x_1(\xi) \le x_3(t) \le \bigcup_{\xi \in [t-d_r-d_f,t)} x_1(\xi) \qquad \text{(from (13) and (16))} \qquad (17)$$

$$\bigcap_{\xi \in [t-2d_r-d_f,t)} \overline{u(\xi)} \le x_3(t) \le \bigcup_{\xi \in [t-d_r-2d_f,t)} \overline{u(\xi)} \qquad \text{(from (11) and (17))} \qquad (18)$$

$$\overline{y_4(t)} = x_3(t) \cdot x_1(t) \qquad \text{(from (5))} \qquad (19)$$

$$\bigcap_{\xi \in [t-2d_r-d_f,t)} \overline{u(\xi)} \cdot \bigcap_{\xi \in [t-d_r,t)} \overline{u(\xi)} \le \overline{y_4(t)} \le \bigcup_{\xi \in [t-d_r-2d_f,t)} \overline{u(\xi)} \cdot \bigcup_{\xi \in [t-d_f,t)} \overline{u(\xi)} \qquad (20)$$

from (18), (11) and (19). But $[t-2d_r-d_f,t) \supset [t-d_r,t), [t-d_r-2d_f,t) \supset [t-d_f,t)$ imply

$$\bigcap_{\xi \in [t-2d_r-d_f,t)} \overline{u(\xi)} \le \bigcap_{\xi \in [t-d_r,t)} \overline{u(\xi)}, \qquad \bigcup_{\xi \in [t-d_r-2d_f,t)} \overline{u(\xi)} \ge \bigcup_{\xi \in [t-d_f,t)} \overline{u(\xi)}$$

$$\bigcap_{\xi \in [t-2d_r-d_f,t)} \overline{u(\xi)} \cdot \bigcap_{\xi \in [t-d_r,t)} \overline{u(\xi)} = \bigcap_{\xi \in [t-2d_r-d_f,t)} \overline{u(\xi)}$$

$$\bigcup_{\xi \in [t-d_r-2d_f,t)} \overline{u(\xi)} \cdot \bigcup_{\xi \in [t-d_f,t)} \overline{u(\xi)} = \bigcup_{\xi \in [t-d_f,t)} \overline{u(\xi)}$$

from where (20) becomes

$$\bigcap_{\xi \in [t-2d_r-d_f,t)} \overline{u(\xi)} \le \overline{y_4(t)} \le \bigcup_{\xi \in [t-d_f,t)} \overline{u(\xi)} \qquad (21)$$

Furthermore

$$\bigcap_{\xi \in [t-d_r,t)} y_5(\xi) \le x_5(t) \le \bigcup_{\xi \in [t-d_f,t)} y_5(\xi) \qquad \text{(equation (8))} \qquad (22)$$

$$\bigcap_{\xi \in [t-d_r,t)} \overline{x_4(\xi)} \le x_5(t) \le \bigcup_{\xi \in [t-d_f,t)} \overline{x_4(\xi)} \qquad \text{(from (6) and (22))} \qquad (23)$$

$$\bigcap_{\xi \in [t-d_f,t)} \overline{y_4(\xi)} \le \overline{x_4(t)} \le \bigcup_{\xi \in [t-d_r,t)} \overline{y_4(\xi)} \qquad \text{(similar with (15))} \qquad (24)$$

$$\bigcap_{\xi \in [t-d_r-d_f,t)} \overline{y_4(\xi)} \le x_5(t) \le \bigcup_{\xi \in [t-d_r-d_f,t)} \overline{y_4(\xi)} \qquad \text{(from (23) and (24))} \qquad (25)$$

$$\bigcap_{\xi \in [t-3d_r-2d_f,t)} \overline{u(\xi)} \le x_5(t) \le \bigcup_{\xi \in [t-d_r-2d_f,t)} \overline{u(\xi)} \qquad \text{(from (21) and (25))} \qquad (26)$$

$$\overline{z(t)} = x_5(t) \cdot x_1(t) \qquad \text{(from (7))} \qquad (27)$$

$$\bigcap_{\xi \in [t-3d_r-2d_f,t)} \overline{u(\xi)} \cdot \bigcap_{\xi \in [t-d_r,t)} \overline{u(\xi)} \le \overline{z(t)} \le \bigcup_{\xi \in [t-d_r-2d_f,t)} \overline{u(\xi)} \cdot \bigcup_{\xi \in [t-d_f,t)} \overline{u(\xi)} \qquad (28)$$

from (27), (26) and (11). With arguments like those around (20) we infer from (28)

$$\bigcap_{\xi\in[t-3d_r-2d_f,t)}\overline{u(\xi)} \leq \overline{z(t)} \leq \bigcup_{\xi\in[t-d_f,t)}\overline{u(\xi)} \tag{29}$$

$$\bigcap_{\xi\in[t-d_f,t)} u(\xi) \leq z(t) \leq \bigcup_{\xi\in[t-3d_r-2d_f,t)} u(\xi) \tag{30}$$

From (9) and (30) we get

$$\bigcap_{\xi\in[t-d_r-d_f,t)} u(\xi) \leq w(t) \leq \bigcup_{\xi\in[t-3d_r-3d_f,t)} u(\xi) \tag{31}$$

The conclusion expressed by (31) is that the circuit increases the one gate rising delay from $d_r$ to $d_r+d_f$ and respectively the one gate falling delay from $d_f$ to $3d_r+3d_f$, i.e. the growth of the falling delay is bigger than the growth of the rising delay. This justifies the title of the paragraph.

## 11.2 Example of Circuit with Tranzient Oscillations

We reproduce in Fig 25 an example of circuit from [3].

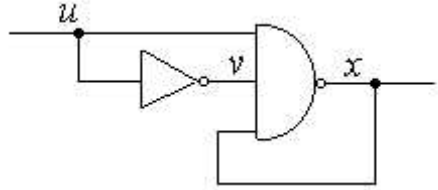

Fig 25

Even if the static analyzis, when $u,v,x \in \boldsymbol{B}$ of such a circuit is not appropriate due to the feedback loop we remark that the proposed circuit computes the 1 Boolean function because

$$x = \overline{u \cdot v \cdot x} = \overline{u \cdot \overline{u} \cdot x} = \overline{0} = 1$$

Modeling is done like in the next drawing where $u,v,x,y,z \in S$ and after solving the system we must obtain $\lim_{t\to\infty} z(t) = 1$ independent on the type of DC's that we choose. The delays of the gates and of the wires from the original circuit have been concentrated in the two delay circuits from Fig 26, where the gates and the wires have no delays. We choose that the delays associated to the two logical gates be FDC's. The equations are:

$$v(t) = v(0-0) \cdot \chi_{(-\infty,0)}(t) \oplus \overline{u(t)} \cdot \chi_{[0,\infty)}(t) \tag{1}$$

$$y(t) = v(t-d) \tag{2}$$

$$x(t) = x(0-0) \cdot \chi_{(-\infty,0)}(t) \oplus \overline{u(t) \cdot y(t) \cdot z(t)} \cdot \chi_{[0,\infty)}(t) \qquad (3)$$
$$z(t) = x(t-d') \qquad (4)$$

resulting

$$y(t) = v(0-0) \cdot \chi_{(-\infty,0)}(t-d) \oplus \overline{u(t-d)} \cdot \chi_{[0,\infty)}(t-d)$$
$$= v(0-0) \cdot \chi_{(-\infty,d)}(t) \oplus \overline{u(t-d)} \cdot \chi_{[d,\infty)}(t) \qquad \text{(from (1), (2))} \qquad (5)$$
$$x(t) = x(0-0) \cdot \chi_{(-\infty,0)}(t) \oplus \qquad \text{(from (3), (4), (5))} \qquad (6)$$
$$\oplus \overline{u(t) \cdot (v(0-0) \cdot \chi_{(-\infty,d)}(t) \oplus \overline{u(t-d)} \cdot \chi_{[d,\infty)}(t)) \cdot x(t-d')} \cdot \chi_{[0,\infty)}(t)$$

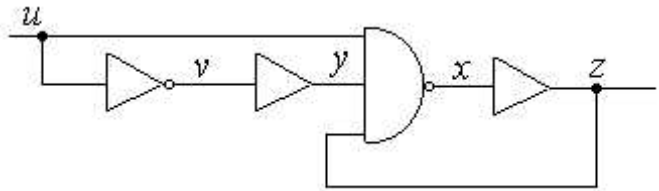

Fig 26

We solve (6) in the special case when $x(0-0) = v(0-0) = 1$ and $u(t) = 1$, when it becomes

$$x(t) = \chi_{(-\infty,0)}(t) \oplus \overline{\chi_{(-\infty,d)}(t) \cdot x(t-d')} \cdot \chi_{[0,\infty)}(t) \qquad (7)$$

The solution of (7) is the following

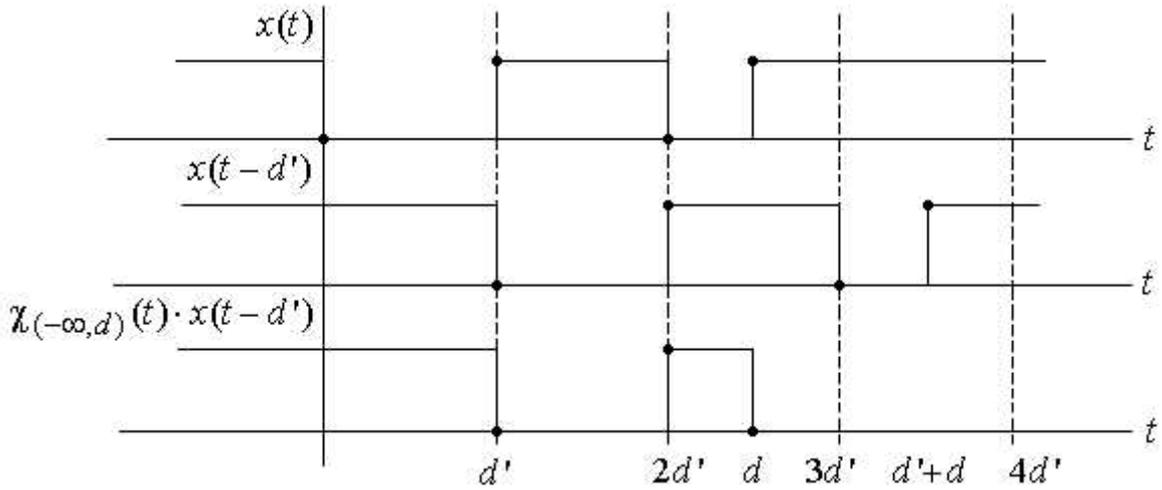

Fig 27

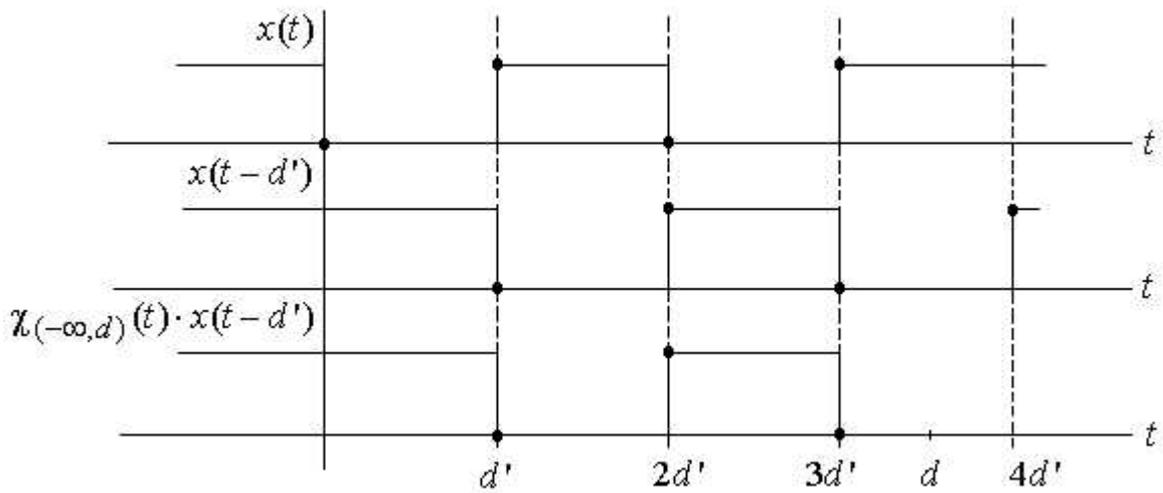

Fig 28

$2kd' < d \leq (2k+1)d'$ implies
$$x(t) = \begin{cases} \chi_{(-\infty,0)}(t) \oplus \chi_{[d',2d')}(t) \oplus ... \oplus \chi_{[(2k-1)d',2kd')}(t) \oplus \chi_{[d,\infty)}(t), k \geq 1 \\ \chi_{(-\infty,0)}(t) \oplus \chi_{[d,\infty)}(t), k = 0 \end{cases}$$

$(2k+1)d' < d \leq (2k+2)d'$ implies
$$x(t) = \begin{cases} \chi_{(-\infty,0)}(t) \oplus \chi_{[d',2d')}(t) \oplus ... \oplus \chi_{[(2k-1)d',2kd')}(t) \oplus \chi_{[(2k+1)d',\infty)}(t), k \geq 1 \\ \chi_{(-\infty,0)}(t) \oplus \chi_{[d',\infty)}(t), k = 0 \end{cases}$$

where $k = 0, 1, 2, ...$ We have represented in Fig 27 and Fig 28 these two formulas for $k = 1$.

The output $z(t)$ of the circuit is obtained from (4).

The idea of solving the equation (6) in other cases as well as the behavior of the circuit from Fig 25 are obvious now.

## 11.3 Example of C Gate. Generalization

The circuit from Fig 29 where the gates and the wires have delays is modeled like in Fig 30

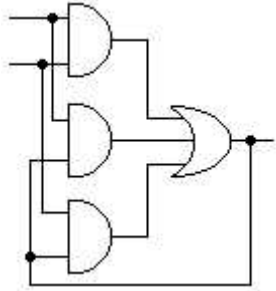

Fig 29

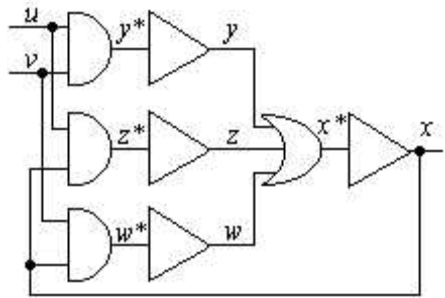

Fig 30

where the Boolean functions are computed instantaneously:
$$y^*(t) = u(t) \cdot v(t) \qquad (1)$$
$$z^*(t) = u(t) \cdot x(t) \qquad (2)$$
$$w^*(t) = v(t) \cdot x(t) \qquad (3)$$
$$x^*(t) = y(t) \cup z(t) \cup w(t) \qquad (4)$$

and the delays on gates and wires are localized in the four delay circuits. We have supposed for simplicity that all the signals have the same initial values $u(0-0) = v(0-0) = ... = x(0-0)$ and for example equation (1) has resulted in the next way:
$$y^*(t) = y^*(0-0) \cdot \chi_{(-\infty,0)}(t) \oplus u(t) \cdot v(t) \cdot \chi_{[0,\infty)}(t) =$$

$$= u(0-0) \cdot v(0-0) \cdot \chi_{(-\infty,0)}(t) \oplus u(t) \cdot v(t) \cdot \chi_{[0,\infty)}(t) = u(t) \cdot v(t)$$

a) The bounded delay model

$$\bigcap_{\xi \in [t-d_r, t-d_r+m_r]} y^*(\xi) \leq y(t) \leq \bigcup_{\xi \in [t-d_f, t-d_f+m_f]} y^*(\xi) \tag{5}$$

$$\bigcap_{\xi \in [t-d_r, t-d_r+m_r]} z^*(\xi) \leq z(t) \leq \bigcup_{\xi \in [t-d_f, t-d_f+m_f]} z^*(\xi) \tag{6}$$

$$\bigcap_{\xi \in [t-d_r, t-d_r+m_r]} w^*(\xi) \leq w(t) \leq \bigcup_{\xi \in [t-d_r, t-d_r+m_r]} w^*(\xi) \tag{7}$$

$$\bigcap_{\xi \in [t-D_r, t-D_r+M_r]} x^*(\xi) \leq x(t) \leq \bigcup_{\xi \in [t-D_f, t-D_f+M_f]} x^*(\xi) \tag{8}$$

with $0 \leq m_r \leq d_r$, $0 \leq m_f \leq d_f$, $0 \leq M_r \leq D_r$, $0 \leq M_f \leq D_f$ and the consistency conditions are fulfilled under the form: $d_r \geq d_f - m_f$, $d_f \geq d_r - m_r$ respectively $D_r \geq D_f - M_f$, $D_f \geq D_r - M_r$. We have considered that the three AND gates are identical. We eliminate the intermediary variables $y^*, z^*, w^*, y, z, w, x^*$:

$$x(t) \overset{(4),(8)}{\geq} \bigcap_{\xi \in [t-D_r, t-D_r+M_r]} (y(\xi) \cup z(\xi) \cup w(\xi)) \geq \bigcap_{\xi \in [t-D_r, t-D_r+M_r]} y(\xi) \geq$$

$$\overset{(5)}{\geq} \bigcap_{\xi \in [t-D_r, t-D_r+M_r]} \bigcap_{\omega \in [\xi-d_r, \xi-d_r+m_r]} y^*(\omega) = \bigcap_{\xi \in [t-d_r-D_r, t-d_r-D_r+m_r+M_r]} y^*(\xi) =$$

$$\overset{(1)}{=} \bigcap_{\xi \in [t-d_r-D_r, t-d_r-D_r+m_r+M_r]} (u(\xi) \cdot v(\xi)) \tag{9}$$

$$x(t) \overset{(4),(8)}{\leq} \bigcup_{\xi \in [t-D_f, t-D_f+M_f]} (y(\xi) \cup z(\xi) \cup w(\xi)) =$$

$$= \bigcup_{\xi \in [t-D_f, t-D_f+M_f]} y(\xi) \cup \bigcup_{\xi \in [t-D_f, t-D_f+M_f]} z(\xi) \cup \bigcup_{\xi \in [t-D_f, t-D_f+M_f]} w(\xi) \leq$$

$$\overset{(5),(6),(7)}{\leq} \bigcup_{\xi \in [t-D_f, t-D_f+M_f]} \bigcup_{\omega \in [\xi-d_f, \xi-d_f+m_f]} y^*(\omega) \cup$$

$$\cup \bigcup_{\xi \in [t-D_f, t-D_f+M_f]} \bigcup_{\omega \in [\xi-d_f, \xi-d_f+m_f]} z^*(\omega) \cup \bigcup_{\xi \in [t-D_f, t-D_f+M_f]} \bigcup_{\omega \in [\xi-d_f, \xi-d_f+m_f]} w^*(\omega) =$$

$$= \bigcup_{\xi \in [t-d_f-D_f, t-d_f-D_f+m_f+M_f]} y^*(\xi) \cup \bigcup_{\xi \in [t-d_f-D_f, t-d_f-D_f+m_f+M_f]} z^*(\xi) \cup$$

$$\cup \bigcup_{\xi \in [t-d_f-D_f, t-d_f-D_f+m_f+M_f]} w^*(\xi) =$$

$$\overset{(1),(2),(3)}{=} \bigcup_{\xi \in [t-d_f-D_f, t-d_f-D_f+m_f+M_f]} u(\xi) \cdot v(\xi) \cup \bigcup_{\xi \in [t-d_f-D_f, t-d_f-D_f+m_f+M_f]} u(\xi) \cdot x(\xi) \cup$$

$$\cup \bigcup_{\xi \in [t-d_f-D_f, t-d_f-D_f+m_f+M_f]} v(\xi) \cdot x(\xi) =$$

$$= \bigcup_{\xi \in [t-d_f-D_f, t-d_f-D_f+m_f+M_f]} (u(\xi) \cdot v(\xi) \cup (u(\xi) \cup v(\xi)) \cdot x(\xi))$$

$$\leq \bigcup_{\xi \in [t-d_f-D_f, t-d_f-D_f+m_f+M_f]} (u(\xi) \cdot v(\xi) \cup u(\xi) \cup v(\xi)) =$$

$$= \bigcup_{\xi \in [t-d_f-D_f, t-d_f-D_f+m_f+M_f]} (u(\xi) \cup v(\xi)) \tag{10}$$

thus, by cumulating (9), (10)

$$\bigcap_{\xi \in [t-d_r-D_r, t-d_r-D_r+m_r+M_r]} (u(\xi) \cdot v(\xi)) \leq x(t) \leq \bigcup_{\xi \in [t-d_f-D_f, t-d_f-D_f+m_f+M_f]} (u(\xi) \cup v(\xi))$$

we have obtained a system that is very much similar with BDC.

  b) The deterministic model

We ask that (9), (10) be fulfilled together with

$$\overline{x(t-0)} \cdot x(t) \leq \bigcap_{\xi \in [t-d_r-D_r, t-d_r-D_r+m_r+M_r]} (u(\xi) \cdot v(\xi)) \tag{11}$$

$$x(t-0) \cdot \overline{x(t)} \leq \bigcap_{\xi \in [t-d_f-D_f, t-d_f-D_f+m_f+M_f]} \overline{u(\xi)} \cdot \overline{v(\xi)} \tag{12}$$

The system (9), (10), (11), (12) represents a deterministic model, similar with 9.5.1 a) and it is equivalent with the next one, that is similar with 9.5.1 b):

$$\overline{x(t-0)} \cdot x(t) = \overline{x(t-0)} \cdot \bigcap_{\xi \in [t-d_r-D_r, t-d_r-D_r+m_r+M_r]} (u(\xi) \cdot v(\xi))$$

$$x(t-0) \cdot \overline{x(t)} = x(t-0) \cdot \bigcap_{\xi \in [t-d_f-D_f, t-d_f-D_f+m_f+M_f]} \overline{u(\xi)} \cdot \overline{v(\xi)}$$

etc. The C gate, also called Muller C element generalizes the delay circuit, in the sense that when the two inputs are equal, it becomes a delay circuit. In the same manner, the systems that we have obtained generalize our systems BDC, respectively DBRIDC and may be called 2-delays: 2-BDC, respectively 2-DBRIDC. The general case of n-BDC and n-DBRIDC is obvious now.

## Bibliography

[1] E. Asarin, O. Maler, A. Pnueli, On Discretization of Delays in Timed Automata and Digital Circuits, in R. de Simone and D. Sangiorgi (Eds), Proc. Concur '98, 470-484, LNCS 1466, Springer, 1998

[2] M. Bozga, Hou Jianmin, O. Maler, S. Yovine, Verification of Asynchronous Circuits using Timed Automata, Rejected from ASCD'01

[3] J. A. Brzozowski, C-J. H. Seger, Advances in Asynchronous Circuit Theory, Part I: Gate and Unbounded Inertial Delay Models, Bulletin of the European Association for Theoretical Computer Science, Number 42, pp. 198-249, February 1990

[4] J. A. Brzozowski, C-J. H. Seger, Advances in Asynchronous Circuit Theory, Part II: Bounded Inertial Delay Models, MOS Circuits, Design Techniques, Bulletin of the European Association for Theoretical Computer Science, Number 43, pp. 199-263, February 1991

[5] Supratik Chakraborty, Polynomial-Time Techniques for Approximate Timing Analysis of Asynchronous Systems, PhD Thesis, August, 1998

[6] Graham Clark, George Taylor, The Verification of Asynchronous Circuits Using CCS, Technical Report ECS-LCFS-97-369, Department of Computer Science, University of Edinburgh, October 23, 1997

[7] Al Davis, Steven Mark Nowick, An Introduction to Asynchronous Circuit Design, UUCS-97-013, 1997

[8] Anton Dumitriu, The Logical Mechanism of Mathematics (in Romanian), Editura Academiei Republicii Socialiste Romania, Bucuresti, 1968

[9] Michael Kishinevsky, Luciano Lavagno, P. Vanbekbergen, The Systematic Design of Asynchronous Circuits, ICCAD '95, Tutorial

[10] William Kwei-Cheung Lam, Algebraic Methods for Timing Analysis and Testing in the High Performance Designs, PhD Thesis, Univ. of California at Berkeley, 1993

[11] Luciano Lavagno, Synthesis and Testing of Bounded Wire Delay Asynchronous Circuits from Signal Transition Graphs, PhD Thesis, Electrical Engineering and Computer Sciences, Univ. of California at Berkeley, 1992

[12] M. J. Liebelt, Progress Report on Research on the Testing of Asynchronous Circuits, The University of Adelaide, Department of Electrical and Electric Engineering, internal report HPCA-ECS-95/03, 29 December 1995

[13] Michael Liebelt, A Proposal for Research on the Testability of Asynchronous Circuits, The University of Adelaide, Department of Electrical and Electric Engineering, Internal Report HPCA-ECS-96/01, August 1998

[14] O. Maler, A. Pnueli, Timing Analysis of Asynchronous Circuits Using Timed Automata, in P. E. Camurati, H. Eveking (Eds), Proc. CHARME'95, 189-205, LNCS 987, Springer, 1995

[15] Steven Mark Nowick, Automatic Synthesis of Burst-Mode Asynchronous Controllers, Technical Report CSL-TR-95-686, 1995 (revised version of PhD dissertation, March 1993)

[16] Alexander Rabinovich, Automata over Continuous Time, Dept. of Computer Science, Tel-Aviv University, June 26, 1997

[17] Polly Sara Kay Siegel, Automatic Technology Mapping for Asynchronous Designs, Department of Electrical Enginering, Stanford University, PhD Thesis, 1995

[18] Serdar Tasiran, Rajeev Alur, Robert P. Kurshan, Robert K. Brayton, Verifying Abstractions of Timed Systems, In Proc. of 7th International Conference on Concurrency Theory, CONCUR '96, LNCS 1119, Pisa, Italy, 1996, pp. 546-562.

[19] Serdar Tasiran, Yuji Kukimoto, Robert K. Brayton, Computing Delay with Coupling Using Timed Automata, In Proc. of IEEE/ACM Intl. Workshop on Timing Issues in the Specification and Synthesis of Digital Systems, TAU '97, Austin, TX, 1997.

[20] Serdar Tasiran, Compositional and Hierarchical Techniques for the Formal Verification of Real-Time Systems, PhD Thesis, Univ. of California at Berkeley, 1998

[21] S. Tasiran, S.P.Khatri, S. Yovine, R.K. Brayton, A. Sangiovanni-Vincentelli, A Timed Automaton-Based Method for Accurate Computation of Circuit Delay in the Presence of Cross-Talk, In Proc. of the 2$^{nd}$ Intl. Conf. on Formal Methods in Computer-aided Design, FMCAD '98, LNCS 1522, Palo Alto, CA, 1998, pp. 149-166.

[22] Serban E. Vlad  $R \to B$  Differentiable Functions, Analele Universitatii din Oradea, Fasciclola Matematica, TOM V, 1995-1996

[23] Serban E. Vlad, Selected Topics in Asynchronous Automata , Analele Universitatii din Oradea, Fascicola matematica, Tom VII, 1999

[24] Serban E. Vlad, On Timed Automata: the Inertial Delay Buffer, the 9-th Symposium of Mathematics and Its Applications, Timisoara-Romania, November 1-4, 2001

[25] Serban E. Vlad, Towards a Mathematical Theory of the Delays of the Asynchronous Circuits, Analele Universitatii din Oradea, Fascicola matematica, Tom IX, 2002